\documentclass[prb,twocolumn,floatfix]{revtex4}
\usepackage{amsmath}
\usepackage{graphicx}
\usepackage{latexsym}
\usepackage{amsfonts}
\usepackage{amssymb}
\usepackage{bm}
\usepackage{longtable}

\begin{document}
\title{Optical Probe Diffusion in Polymer Solutions }
\author{George D. J. Phillies}
\email[To whom inquiries should be sent ] {phillies@wpi.edu}
\affiliation{Department of Physics, Worcester Polytechnic Institute, Worcester, MA 01609}

\begin{abstract}

The experimental literature on the motion of mesoscopic probe particles through polymer solutions is systematically reviewed.  The primary focus is the study of diffusive motion of small probe particles.  Comparison is made with measurements of solution viscosities. A coherent description was obtained, namely that the probe diffusion coefficient generally depends on polymer concentration as $D_{p} = D_{p0} \exp(-\alpha c^{\nu})$.  One finds that $\alpha$ depends on polymer molecular weights as $\alpha \sim M^{0.8}$, and $\nu$ appears to have large-$M$ and small-$M$ values with a crossover linking them. The probe diffusion coefficient does not simply track the solution viscosity; instead, $D_{p}\eta$ typically increases markedly with increasing polymer concentration and molecular weight.  In some systems, e.g., hydroxypropylcellulose:water, the observed probe spectra are bi- or tri-modal.  Extended analysis of the full probe phenomenology implies that hydroxypropylcellulose solutions are characterized by a single, concentration-independent, length scale that is approximately the size of a polymer coil.  In a very few systems, one sees re-entrant or low-concentration-plateau behaviors of uncertain interpretation; from their rarity, these behaviors are reasonably interpreted as corresponding to specific chemical effects.  True microrheological studies examining the motion of mesoscopic particles under the influence of a known external force are also examined. Viscosity from true microrheological measurements is in many cases substantially smaller than the viscosity measured with a macroscopic instrument.

\end{abstract} 

\maketitle

\section{Introduction}

This review treats probe diffusion and related methods of investigating polymer dynamics.  In a probe diffusion experiment, a dilute dispersion of mesoscopic probe particles is mixed into a polymer solution.  The motions or relative motions of the probe particles are then measured.   In most experiments discussed here, probe motions arise from diffusion; the small literature on driven motions of mesoscopic probes through polymer solutions is also reviewed here.  In some systems, probe motions involve multiple relaxations whose time dependences can be independently determined. In others, a single relaxation determined a probe diffusion coefficient $D_{p}$.  $D_{p}$ is sensitive to the probe radius $R$, polymer molecular weight $M$ and concentration $c$, solution viscosity $\eta$, solvent viscosity $\eta_{s}$, and other variables.  The dependence of $D_{p}$ on these and other variables is used to infer how polymers move in solution.  

The remainder of this Section presents a historical background.  Section II remarks briefly on the theory underlying major experimental methods for studying probe diffusion.  Section III presents the experimental phenomenology.  Section IV discusses the systematics of that phenomenology.

The literature on probe diffusion studies of polymer solutions tends to be divided into three parts, namely (i) {\em optical probe diffusion} studies, largely with quasi-elastic light scattering spectroscopy (QELSS), fluorescence recovery after photobleaching (FRAP), and Forced Rayleigh Scattering (FRS), of the thermal motion of dilute probe particles, (ii) {\em microrheology} studies in which an inferred mean-square particle displacement and a generalized Stokes-Einstein relation are used to compute dynamic moduli of the solution, and (iii) \emph{particle-tracking} studies in which the detailed motions of individual particles are recorded. 

Historically, Brown used particle tracking to study the motion now called {\em Brownian}.  Optical probe diffusion as a method for studying polymer solutions dates back to Turner and Hallett\cite{turner1976aDp}, who in 1976 examined polystyrene spheres diffusing through dextran solutions.  The microrheology approach to interpreting probe diffusion stems from early work on Diffusing Wave Spectroscopy (DWS), e.g., the 1993 chapter by Weitz and Pines\cite{weitz1993aDp}.  Particle tracking methods were early applied to observing the motion of labelled tags in cell membranes as reviewed, e.g., by Saxton and Jacobson\cite{saxton1997aDp}. The experimental literatures in these three areas are less than entirely communicating.

\subsection{Other Reviews}

The optical probe diffusion literature has not recently been reviewed systematically.  In 1985, Phillies, et al.\cite{phillies1985bDp} re-examined extant optical probe studies of solutions of bovine serum albumin, polyethylene oxide, and polyacrylic acid, primarily from their own laboratory.  A uniform stretched-exponential concentration dependence $\exp(- \alpha c^{\nu})$ of $D_{p}$ was found.  A dependence $\alpha \sim M^{0.8 \pm 0.1}$ was noted over a range of matrix species.  $\alpha$ for probes in BSA solutions complies with this dependence if BSA is assigned an effective $M$ corresponding to its radius.

Phillies and Streletzky\cite{phillies2001aDp} presented in 2001 a short review (38 references) of the literature on optical probe diffusion.  They treat primarily systems studied with QELSS.  In some cases $D_{p}$ follows the Stokes-Einstein equation evaluated using the solution viscosity.  In other cases, diffusion is appreciably faster than would be expected from the solution viscosity.  In a few cases one finds re-entrant behavior in which the Stokes-Einstein equation fails but only over a narrow band of concentrations.  Finally (taking probes in hydroxypropyl-cellulose as an examplar), in some cases the spectral mode structure is too complicated to be characterized with a single diffusion coefficient.  Phillies and Streletzky's review does not contact the microrheology literature.

A series of reviews treat microrheology: 

Diffusing Wave Spectroscopy was reviewed (49 references) by Harden and Viasnoff\cite{harden2001aDp}.  The primary emphases are on two-cell light scattering measurements, in which light scattered by the sample of interest is rescattered by a second scattering cell before being collected, and on CCD methods, which allow one to collect simultaneously the light scattered into a substantial number of coherence areas.

Solomon and Lu\cite{solomon2001aDp} review (54 references) {\em microrheology} studies of probe diffusion in complex fluids, in particular uses of the general Stokes-Einstein equation and its range of validity, correlations in the motion of pairs of large particles, data analysis methods, and possible paths for extending their methods to smaller probe particles.  The reference list has very limited contact with the optical probe diffusion literature.

MacKintosh and Schmidt\cite{mackintosh1999aDp} discuss (60 references) studies on the diffusion of microscopic probe particles through polymer solutions by means of particle tracking and diffusing wave spectroscopy, as performed under the cognomen \emph{microrheology}, as well as studies of viscoelasticity using atomic force microscopy and the driven motion of mesoscopic particles. The excellent list of references lacks contact with the QELSS/FRAP-based optical probe diffusion literature.

Mukhopadhay and Granick\cite{mukhopaday2001aDp} review (41 references) experimental methods of driving and measuring the displacement of mesoscopic particles, such as DWS and optical tweezer techniques.  They obtain the complex modulus of the fluid is via a generalized Stokes-Einstein equation.  The discussion and references do not contact the optical probe diffusion literature.

Among reviews of particle tracking methods, note: 

Saxton and Jacobson\cite{saxton1997aDp} treat (105 references) experimental methods for tracking single particles as they move in cell membranes.  Motion of membrane proteins and other probes is complex, because the motion may be diffusive, obstructed, include nondiffusive convective motion, or involve trapping. Correlations between single particle tracking and FRAP are considered, including the FRAP-determined nominal fraction of immobile particles.  

Tseng, et al.\cite{tseng2002aDp} review (52 references) particle tracking methods as applied to solutions of F-actin and other cytoskeletal proteins.  While mention is made of measuring mean-square displacements of diffusing particles, this is in the context of a technique that actually does measure particle positions and displacements.  

A review (56 references) of microscopy techniques by Habdas and Weeks\cite{habdas2002aDp} emphasizes video microscopy, imaging crystals, glasses, and phase transitions, and measurement of particle interactions with laser tweezer methods.  The focus is the application of these techniques to colloidal particles in solution.

\section{Theory Underlying Experimental Methods}

Many methods have been applied to study the motion of mesoscopic probe particles through polymer solutions, including QELSS, FRS, FRAP, DWS, particle tracking, and interferometry.  Optical probe methods refer to a special case of scattering from a general three-component probe:matrix polymer:solvent system.  In the special case, one of the components, the \emph{probe}, is dilute yet nonetheless dominates the scattering process, even though the other solute component, the \emph{matrix}, may be far more concentrated than the probe.  Most of these methods, in particular QELSS, FRS, and FRAP, have also been used to study motion of dilute labelled polymer chains in polymer solutions, leading to measurements of polymer self- and tracer- diffusion coefficients. 

Light scattering spectroscopy is sensitive to the dynamic structure factor $S(q,\tau)$ of the scattering particles. The theoretical basis for applying QELSS, FRS, and FRAP to three-component solutions appears in our prior review\cite{phillies2004xDp} and need not be repeated here in detail.  For dilute probe particles diffusing in a non-scattering matrix, $S(q,\tau)$ reduces to 
\begin{equation}
    S(q,\tau) = \langle \sum_{i=1}^{N} \exp(i {\bf q} \cdot \Delta {\bf r}_{i}(\tau)) \rangle.
    \label{eq:s1qtaudefDp}
\end{equation}
Here $i$ labels the $N$ probe particles whose locations at time $t$ are the ${\bf r}_{i}(t)$.  $S(q,\tau)$ is determined by the particle displacements $\Delta {\bf r}_{i}(\tau) = {\bf r}_{i}(t + \tau)  - {\bf r}_{i}(t)$ during $\tau$.  

The scattering vector ${\bf q}$, with
\begin{equation}
     q = \frac{4 \pi n}{\lambda} \sin(\theta/2),
   \label{eq:qmagdefDp}
\end{equation}
determines the distance a particle must diffuse to have a significant effect on $S(q,\tau)$.  Here $n$ is the solution index of refraction, $\lambda$ is the illuminating wavelength {\em in vacuo}, and $\theta$ is the scattering angle. In observing a polymer coil with QELSS, complications arise if the polymer coil radius $R_{g}$ is comparable with $q$, because coil internal modes enter the spectrum if $qR_{g}$ is not $\ll 1$.  However, the rigid probes considered here have no significant internal modes, so light scattering spectroscopy measures rigid-probe center-of-mass motion regardless of $q R_{g}$. 

For dilute monodisperse rigid probes \emph{in a simple Newtonian solvent}, probe motion is described to good approximation by the Langevin equation. In this case, the dynamic structure factor reduces to a simple exponential\cite{berne1976aDp}
\begin{equation}
    S(q,\tau) = \exp(- 2 D_{p} q^{2} \tau)
    \label{eq:s1qtausimpleDp}
\end{equation}    
in which $D_{p}$ is the probe diffusion coefficient.  In one dimension, $D_{p}$ is related to particle displacement by \begin{equation}
        \langle (\Delta x(\tau))^{2} \rangle = 2 D_{p} \tau.
        \label{eq:DDeltarDp}
\end{equation}
The Stokes-Einstein equation relates $D_{p}$ for spheres of radius $R$ to other parameters, namely
\begin{equation}
     D_{p}=\frac{k_{B}T}{6 \pi \eta R}.
     \label{eq:SEeqDp}
\end{equation}
Here $k_{B}$ is Boltzmann's constant, $T$ is the absolute temperature, and  $\eta$ is the experimentally-measured macroscopic solution viscosity.  When $D$ and $R$ or $\eta$ are known, this equation may be inverted to yield a {\em microviscosity} $\eta_{\mu}$ or an {\em apparent hydrodynamic radius} $r_{H}$, respectively, viz.
\begin{equation}
   r_{H} = \frac{k_{B} T}{6 \pi \eta D_{p}}
   \label{eq:rhdefinition}
\end{equation}
and
\begin{equation}
  \eta_{\mu}  = \frac{k_{B} T}{6 \pi R D_{p}}.
   \label{eq:etamudefinition}
\end{equation}
For probes in polymer solutions, $\eta_{\mu}$ may differ markedly from the macroscopically measured viscosity $\eta$, typically with $\eta_{\mu}/\eta \ll 1$.  The $r_{H}$ may be much less than the experimentally-determined physical radius $R$.  We denote $\eta_{\mu} \approx \eta$ and $r_{H} \approx R$ as {\em Stokes-Einsteinian} behavior.  The contrary case ($\eta_{\mu} \neq \eta$ and $r_{H} \neq R$) we describe as {\em non-Stokes-Einsteinian} behavior.

In more complex systems, the dynamic structure factor is non-exponential.  $S(q,\tau)$ remains monotonic, with Laplace transform 
\begin{equation}
        S(q,\tau) = \int_{0}^{\infty} d \Gamma A(\Gamma) \exp(- \Gamma t).
        \label{eq:SqtLaplaceDp}
\end{equation}
Here $A(\Gamma) \geq 0$ is the relaxation distribution function. In some cases, $A(\Gamma)$ is qualitatively well-described by a sum of relatively separated peaks, commonly termed 'modes'.  This terminology does not imply that there is necessarily a 1-to-1 correspondence between individual modes and individual physical relaxation processes, though such a correspondence often exists.

An arbitrary physical $A(\Gamma)$ can be generated by scattering light from an appropriately-chosen mixture of dilute Brownian probes in a simple fluid.  For each Brownian species, the mean-square displacement increases linearly in time. Correspondingly, every physical $S(q,\tau)$ corresponds to a system in which the mean-square particle displacement increases linearly in time.  However, a non-exponential $S(q,\tau)$ can also arise from systems in which $\langle \Delta x(\tau))^{2} \rangle$ has more complex behavior.  It is impossible to distinguish these possibilities by examining spectra taken at a single scattering angle or from spectra uniformly averaged over a range of $q$.  Thus, it is impossible to determine $ \langle (\Delta x(\tau))^{2} \rangle$ from one light scattering spectrum of an unknown system.

Recently, a misinterpretation of the historical literature on light scattering spectra has emerged.  The starting point is the excellent book of Berne and Pecora\cite{berne1976aDp}, in particular their treatment of light scattering spectra of Brownian particles that follow the Langevin equation. Berne and Pecora's treatment is without error, but the restrictions on the range of validity of their results are not uniformly recognized. 

The Langevin equation is a model for an isolated particle floating in a simple solvent.  The particle is assumed to be subject to two forces, namely a drag force ${\bf F}_{D}(t) = - f {\bf v}(t)$ and a 'random' force ${\cal F}(t)$.  Here $f$ is the drag coefficient and ${\bf v}(t)$ is the time-dependent particle velocity.  The random force is uncorrelated with the drag force.  Also, the random forces at any distinct pair of times are uncorrelated, i.e., $\langle {\cal F}(t) \cdot {\cal F}(t+\tau)\rangle = 0$ for $\tau\neq 0$.  The particle's motion follows from Newton's Second Law.  The model only specifies statistical properties of ${\cal F}(t)$, so only statistical properties of particle displacements can be obtained from the Langevin model. 

Under these conditions, Doob\cite{doob1942aDp} showed in 1942 that: (i)  $P(\Delta {\bf r}, \tau)$ is a Gaussian in $\Delta {\bf r}$, with mean-square displacement in one dimension satisfying $\langle \Delta x^{2}(t) \rangle = 2 D t$; (ii) the spectrum reduces to
\begin{equation}
     S(q,\tau) = \exp(- q^{2} \langle  \Delta x^{2}(t) \rangle)
     \label{eq:sqtaureducedDp}
\end{equation}
and (iii) as an absolutely rigorous mathematical consequence, $S(q,\tau)$ is a single exponential, namely $S(q,\tau) = \exp(-2 D_{p} q^{2} t)$.  These results describe, e.g., the light scattering spectrum of dilute polystyrene spheres in pure water, which was an early target for experimental studies. 

Spectra of probe particles in viscoelastic fluids commonly are not simple exponentials.  In the literature misinterpretation, eq.\ \ref{eq:sqtaureducedDp} is incorrectly claimed--under the cognomen 'gaussian assumption' or 'gaussian approximation'--to be uniformly applicable to light scattering spectra, in particular to spectra that do not show a single-exponential decay.  Is this realistic? In a viscoelastic liquid, the elastic moduli are frequency-dependent.  Correspondingly, the stress fluctuations in the liquid create random forces with non-zero correlation times, i.e., $\langle {\cal F}(t) \cdot {\cal F}(t+\tau)\rangle \neq 0$ for a range of $\tau \neq 0$. Correspondingly, the displacement distribution function is not a Gaussian, and $\langle (\Delta x(\tau))^{2} \rangle$ shows non-diffusive behavior $(\nsim t^{1})$ that is not characterized by a diffusion coefficient.  A contrapositive statement of Doob's theorem shows that non-exponential spectra correspond necessarily to particles whose motion is not described by eq.\ \ref{eq:sqtaureducedDp}.  

An explicit calculation that correctly expresses $S(q,\tau)$ in terms of all moments $\langle \Delta x^{2n}(t) \rangle$, $n \geq 1$, has been made\cite{phillies2005aDp}. $S(q,\tau)$ reflects all moments $\langle \Delta x^{2n} \rangle$  of $P(\Delta {\bf r}, \tau)$.  Except under special conditions not satisfied by probe particles in complex fluids, the higher moments $n>1$ make non-trivial contributions to $S(q,\tau)$.  $S(q,\tau)$ and QELSS single- and multiple-scattering spectra are in general not determined by $\langle \Delta x^{2} \rangle$.  Equivalently, there are pairs of systems having the same $S(q,\tau)$, but very different time dependences for $\langle (\Delta x)^{2} \rangle$. Articles based on the gaussian approximation are considered in Section III.F.

$S(q,\tau)$ is the spatial Fourier transform of the probability distribution $P(\Delta {\bf r}, \tau)$ for finding a probe displacement $\Delta {\bf r}$ during $\tau$.  While a single measurement of $S(q,\tau)$ cannot be used to compute $P(\Delta {\bf r}, \tau)$, determination of the complete functional dependence of $S(q,\tau)$ on $q$ could in principle be used to determine  $P(\Delta {\bf r}, \tau)$.  To the authors knowledge, this has only been done in the trivial case of simple Brownian motion. Particle tracking methods yield directly $P(\Delta {\bf r}, \tau)$, and can be used to compute complex many-time displacement cross-correlations, such as $P(\Delta {\bf r}(t, t+\tau), \Delta {\bf r}(t+\tau, t+\tau+\theta))$, where $\Delta {\bf r}(t, t+\tau)$ is the probe displacement between $t$ and $t+\tau$. These cross-correlations, which have not yet been intensively exploited, are substantially inaccessible to conventional light scattering spectroscopy.

\section{Phenomenology}

This section describes experiments on the diffusion of rigid probes through polymer solutions and related systems, and related measurements using other techniques, including particle tracking measurements and true microscopic rheological measurements.  The major subsections of Section III are (A) probe diffusion in polymer solutions, (A.1) probe diffusion in hydroxypropylcellulose (HPC) solutions, (B) rotational diffusion of probes, (C) particle tracking methods, (D) true microrheological measurements, (E) probes in polyacrylamide gels, protein gels, and in vivo and in other structured systems, and (F) probe spectra interpreted with the gaussian displacement assumption. For simplicity, within each section descriptions of the primary literature are sorted alphabetically, except that studies on HPC:water are ordered chronologically. Measurements on probes in solutions of rigid-rod polymers and in colloidal systems will be discussed elsewhere.

In generating the results of this Section, experimental findings were taken from tabulated numbers in the original papers, or were recovered from the original papers by scanning original figures and digitizing the images.  Fitting curves were generated by non-linear least-squares analysis based on the simplex method, leading to parameters in Tables I-V and the smooth curves in the Figures.  Figures were generated \emph{de novo} from digitized numbers and our fitting curves.  Most smooth curves represent fits to stretched exponentials in concentration, namely 
\begin{equation}
    D_{p} = D_{p0} \exp(- \alpha c^{\nu}).
    \label{eq:seeqxxx}
\end{equation}
Here $\nu$ is a scaling exponent, $\alpha$ is a scaling prefactor, and $D_{p0}$ reflect probe behavior at very low polymer concentration.

\subsection{Probe Diffusion in Polymer Solutions}

This section treats experiments that measured probe translational diffusion.


\begin{figure}[tb] 
\includegraphics{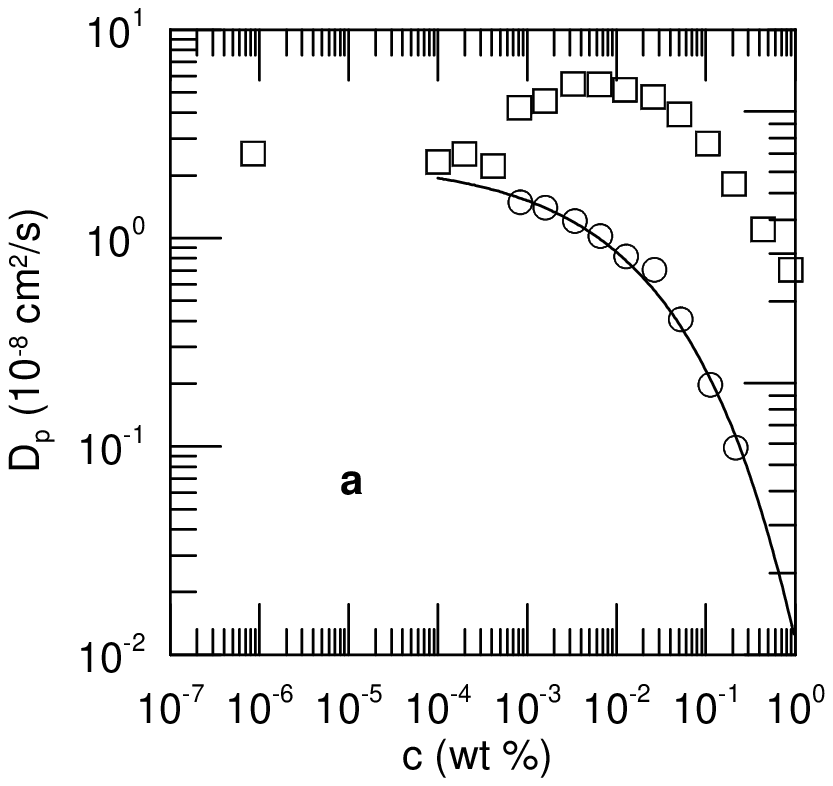}
\includegraphics{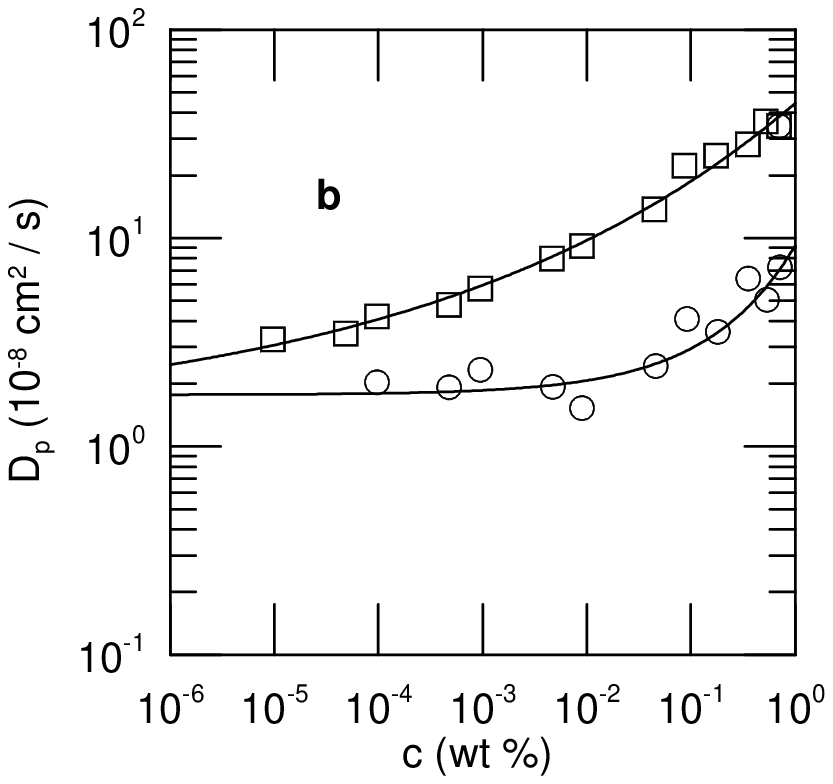}
\caption{\label{figurebremm2001aDp} 
Fast and slow modes in spectra of (a) 100 nm sulfate latex spheres in polyacrylamide solution, and (b) 33 nm hematite particles in sodium polyacrylate solutions, pH 10, 0.1M NaNO$_{3}$, both as functions of polymer concentration, after Bremmell, et al.\cite{bremmell2001aDp}}
\end{figure}

Bremmell, et al.\cite{bremmell2001aDp} used QELSS to examine the diffusion of positively-charged 34 nm and negatively-charged 100 nm diameter polystyrene latex (PSL) spheres and 33 nm diameter hematite particles.  Matrix solutions included water: glycerol, aqueous 3 Mda polyacrylamide (PAAM), and aqueous high-molecular weight sodium polyacrylate (NaPAA).  Spectra were fit to a sum of exponentials. PSL spheres in water:glycerol showed single-exponential relaxations whose $D$ scaled linearly with $T/\eta$ as the temperature and glycerol concentration were varied. Polyacrylamide solutions show marked shear thinning. 

As seen in Figure \ref{figurebremm2001aDp}, in polyacrylamide solutions: Except at very low polymer concentrations, PSL sphere spectra are bimodal, usefully characterized by fast and slow diffusion coefficients $D_{f}$ and $D_{s}$.  With increasing polyacrylamide concentration, $D_{s}$ of the 200 nm spheres falls 30-fold while $\eta$ increases by only 20-fold.  $D_{s}$ does not depend strongly on $q$.  $D_{f}$ shows re-entrance: it first increases with increasing polymer $c$ and then decreases to below its zero-$c$ value.  At elevated $c$, $D_{f}$ increases profoundly, and more rapidly than linearly, with increasing $q^{2}$.  

With 68 nm spheres, $D_{f}$ at smaller $c$ increases with increasing polymer $c$, though less than the increase with 200 nm spheres, while $D_{s}$ falls with increasing $c$.  With hematite particles in sodium polyacrylate solutions, with concentrations increasing up to 1 wt\%, $D_{f}$ and $D_{s}$ \emph{both} increase, $D_{f}$ by ten-fold and $D_{s}$ by at least three-fold.  Bremmell, et al.\cite{bremmell2001aDp} refer to speculation that elasticity (solution viscoelasticity) is related to the observed hyperdiffusivity.

\begin{figure}[tb] 
\includegraphics{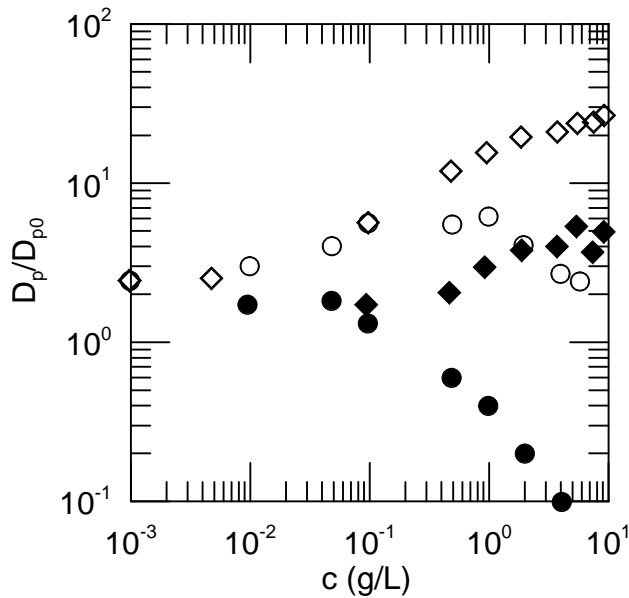}
\caption{\label{figurebremm2002aDp} Fast (open points) and slow (filled points) diffusion coefficients of 100 nm polystyrene sulfate spheres in solutions of polyacrylic acid-co-acrylamide in the presence of 0.001 ($\bigcirc$) or 1.0 ($\lozenge$) M NaNO$_{3}$, from data of Bremmell and Dunstan\cite{bremmell2002aDp}.}
\end{figure}

Bremmel and Dunstan\cite{bremmell2002aDp} examined the diffusion of 100 nm radius PSL spheres in 3 MDa poly(acrylic acid-co-acrylamide) at a range of ionic strengths and polymer concentrations.  Inverse Laplace transformation of QELSS spectra found a bimodal distribution of relaxation rates. Figure \ref{figurebremm2002aDp} shows representative results from the smallest and largest ionic strengths examined. In 1 M NaNO$_{3}$, $D_{f}$ and $D_{s}$ both increase with increasing $c$.  In 0.001M NaNO$_{3}$, $D_{f}$ shows re-entrant behavior, while $D_{s}$ simply decreases with increasing $c$. $D_{f}$ and $D_{s}$ both show a complex ionic-strength-dependent dependence on $q^{2}$.  

Brown and Rymden\cite{brown1987bDp} used QELSS to examine the diffusion of PSL spheres through carboxymethylcellulose (CMC) solutions.  The primary interest was to treat adsorption by this polymer to the latex spheres.  Brown and Rymden conclude that CMC goes down on the surface in a relatively flat conformation.  CMC is a weak polyelectrolyte.  The extend of its binding to polystyrene latex is complexly influenced by factors including salt concentration, pH, and probe size and surface chemistry.

\begin{figure}[tb] 
\includegraphics{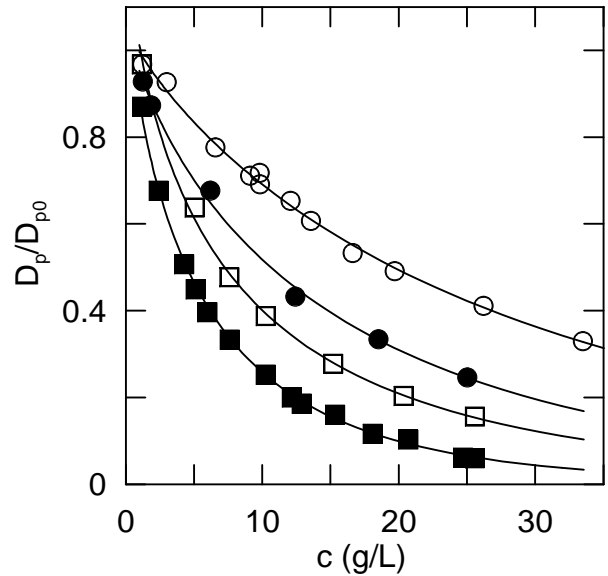}
\caption{\label{figurebrown1988c1} 
$D_{p}$ of 160 nm silica spheres through solutions (top to bottom) of 101, 163, 268, and 445 kDa PMMA,  as obtained 
with QELSS by Brown and Rymden\cite{brown1988cDp}, and fits to stretched exponentials in $c$.}
\end{figure}

Brown and Rymden\cite{brown1988cDp} used QELSS to determine the diffusion of 160 nm radius silica spheres and polystyrene random-coil-polymer fractions through polymethylmethacrylate (PMMA) in CHCl$_{3}$, as seen in Fig.\ \ref{figurebrown1988c1}.  They also measured the solution viscosity.  PMMA molecular weights for the sphere diffusion study were 101, 163, 268, and 445 kDa. QELSS spectra were uniformly quite close to single exponentials.  For silica spheres, $D_{p}/D_{po}$ was to good approximation a universal function of $c[\eta]$.  The product $D_{p} \eta$ for spheres was independent from $c$, even though $D_{p}$ changes by three orders of magnitude over this concentration interval. For chain probes, $D_{s}\eta$ increases by a few to 70\% with increasing $c$, the increase in $D_{s} \eta$ increasing with increasing matrix molecular weight.  Brown and Rymden found that concentration dependence of the single particle diffusion coefficient, for either chain or sphere probes, is to good accuracy given by a stretched exponential in polymer concentration. 

\begin{figure}[hbt] 
\includegraphics{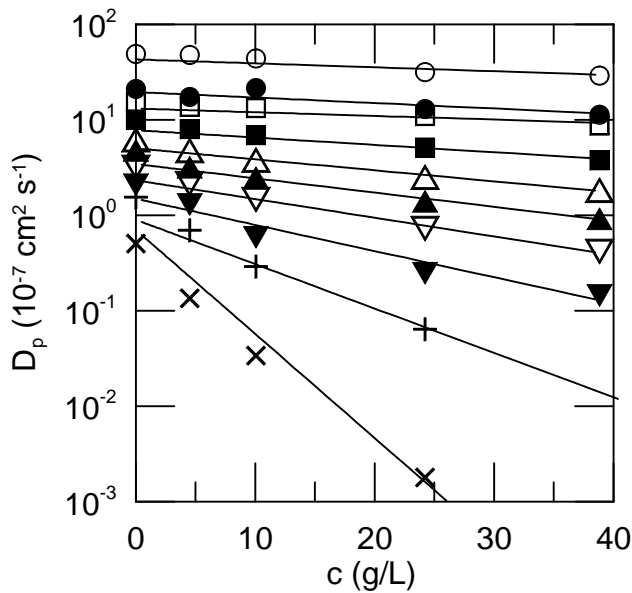}
\caption{\label{figurebu1994aDp1} 
$D_{p}$ of fluorescein (0.5 nm), 1.3, 1.7, 2.8, 4.5, 5.8, 8.8, 13.3, and 17.9 nm hydrodynamic radius dextrans, and 55 nm polystyrene spheres in aqueous 300 kDa HPC, as obtained 
with FRAP by Bu and Russo\cite{bu1994aDp}, and fits to exponentials in $c$.}
\end{figure}

Bu and Russo\cite{bu1994aDp} used FRAP to measure the diffusion of 55 nm PSL spheres, fluorescein, and eight fluorescently labeled dextrans (molecular weights 3.9 kDa to 2 MDa) through HPC solutions.  Nominal matrix molecular weights were 60, 300, and 1000 kDa.  Detailed results were presented for the 300 kDa matrix as seen in Figure \ref{figurebu1994aDp1}. Probe particles diffused more rapidly than expected from the macroscopic solution viscosity, with $\eta_{\mu}/\eta$ increasing toward 1 as the probe radius is increased.  Comparison was made by Bu and Russo with the Langevin-Rondelez equation\cite{langevin1978aDp}, which describes this data well. $D_{p}(c)$ for each probe fits well to a simple exponential, except for the largest probe for which more data appears needed.

\begin{figure}[tb] 
\includegraphics{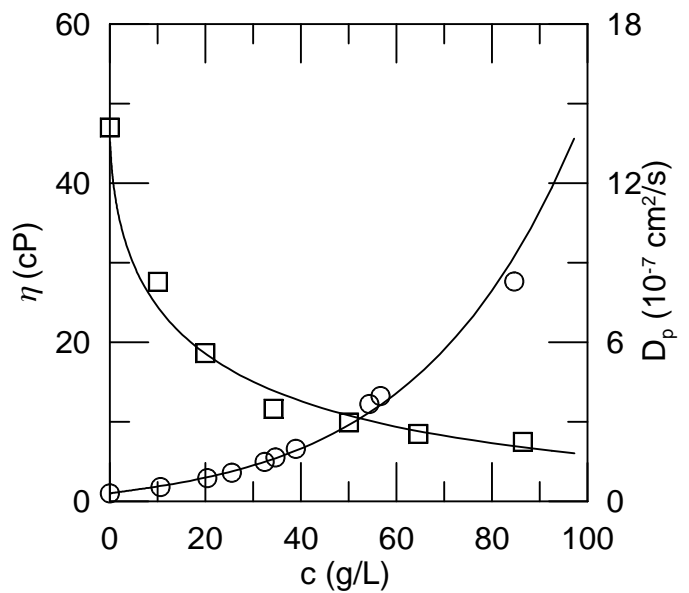}
\caption{\label{figurebusch2000aDp3} 
Viscosity of DNA solutions and $D_{p}$ of green fluorescent protein in those solutions, based on data of Busch, et al.\cite{busch2000aDp}, and fits to stretched exponentials in $c$.}
\end{figure}

\begin{figure}[tb] 
\includegraphics{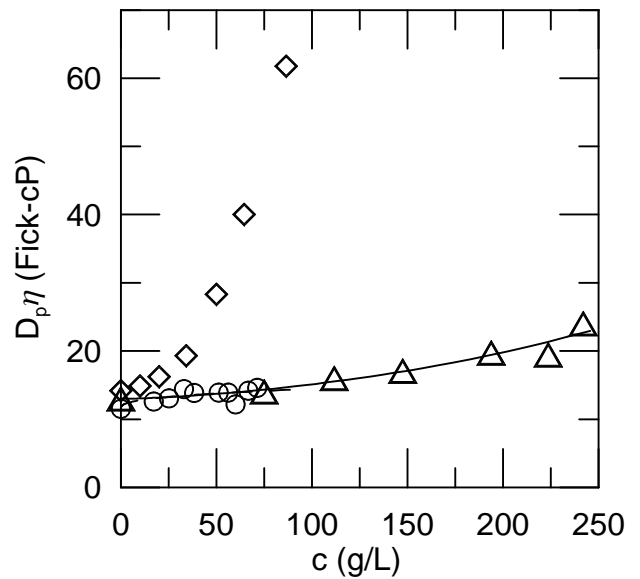}
\caption{\label{figurebusch2000aDpnc} 
$D_{p} \eta$ of green fluorescent protein\cite{busch2000aDp} in solutions of glycerol($\bigcirc$), ficoll 70 ($\vartriangle$), and DNA ($\lozenge$), showing near-Stokes-Einstein behavior in glycerol and ficoll 70 solutions, and large deviations from Stokes-Einstein behavior in DNA solutions.}
\end{figure}

Busch, et al.\cite{busch2000aDp} used FRAP to study diffusion of green fluorescent protein (GFP) through glycerol, ficoll, and 160 bp calf thymus DNA solutions.  QELSS was used to measure the scattering spectrum of the DNA itself; macroscopic viscosities were obtained using capillary viscometers. Representative measurements on DNA solutions appear as Figure \ref{figurebusch2000aDp3}.  In water:glycerol, the probes followed the Stokes-Einstein equation, as seen in Figure \ref{figurebusch2000aDpnc}.  In ficoll solutions the probes diffused moderately more rapidly than expected from the solution viscosity.  In DNA solutions, at elevated $c$, $D_{p}$ of the GFP was much larger than expected from the solution viscosity.

\begin{figure}[tb] 
\includegraphics{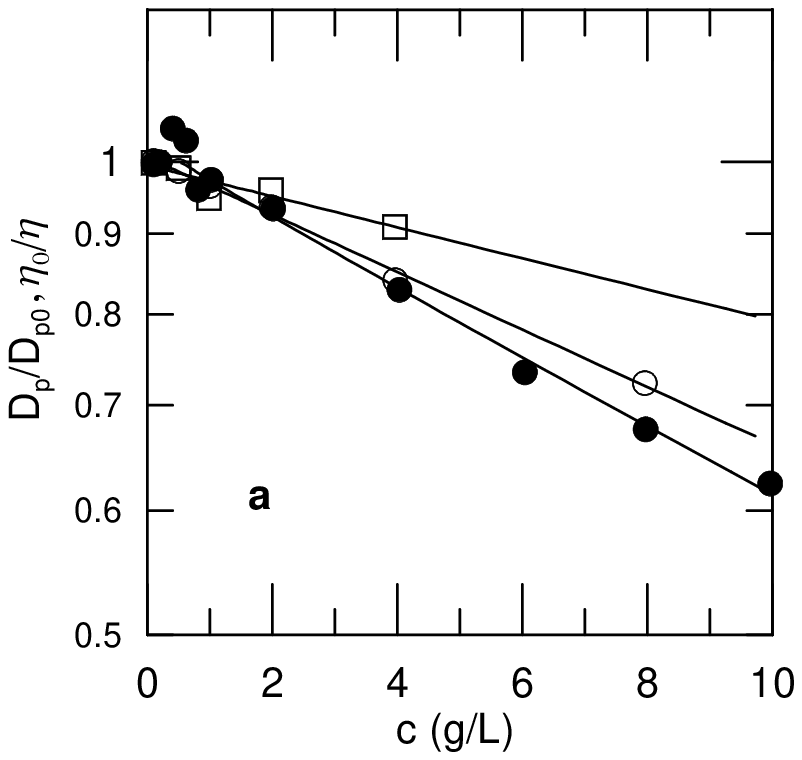}
\includegraphics{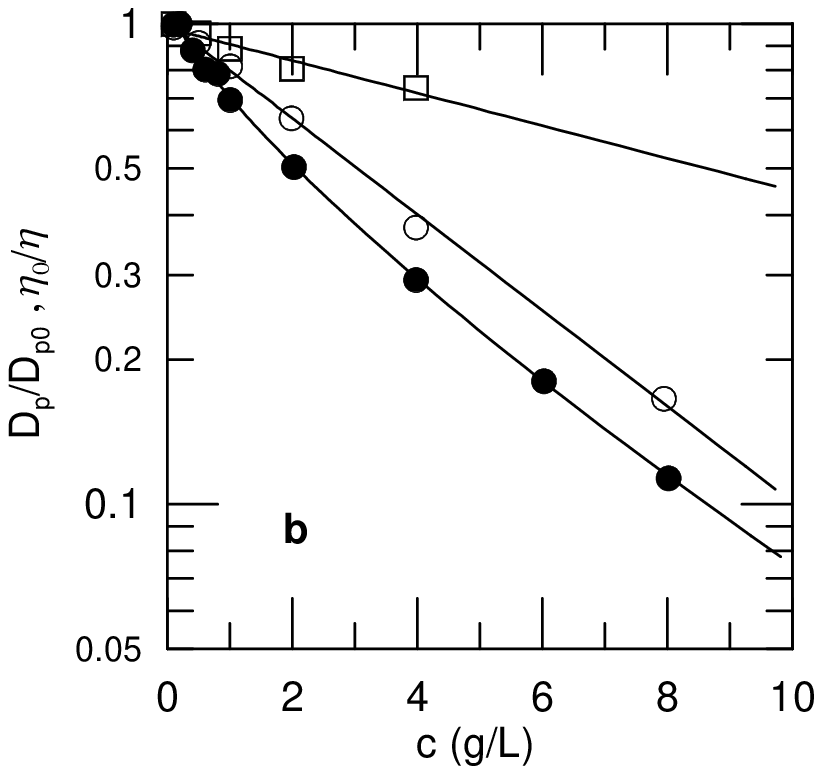}
\caption{\label{figurecao1997aDp2} $D_{p}/D_{p0}$ of 64 nm polystyrene latex spheres ($\bigcirc$) and 65 nm radius unilamellar lipid vesicles ($\square$), and normalized fluidity $\eta_{0}/\eta$ ($\bullet$) in (a) 65 kDa and (b) 1 MDa polyacrylamide solutions, after Cao, et al.\cite{cao1997aDp}.  }
\end{figure}

Cao, et al.\cite{cao1997aDp} used QELSS to measure $D_{p}$ of 32 and 54 nm radius polystyrene spheres, phospholipid/cholesterol vesicles, and multilamellar vesicles diffusing in aqueous non-crosslinked 65 and 1000 kDa polyacrylamides.  $D_{p}$ arose from a relaxation rate that was accurately linear in $q^{2}$; its concentration dependence was  a stretched exponential in $c$. As seen in Figure \ref{figurecao1997aDp2}, comparison of $D_{p}$ with the solution fluidity $\eta^{-1}$ finds that both polyacrylamides were more effective at increasing the solution viscosity than at retarding probe diffusion of polystyrene latex spheres, and were more effective at retarding the motion of  spheres than at retarding unilamellar vesicles of the same radius.  The difference between the vesicle and sphere diffusive mobilities was interpreted as arising from the flexibility of the vesicles.

\begin{figure}[tb] 
\includegraphics{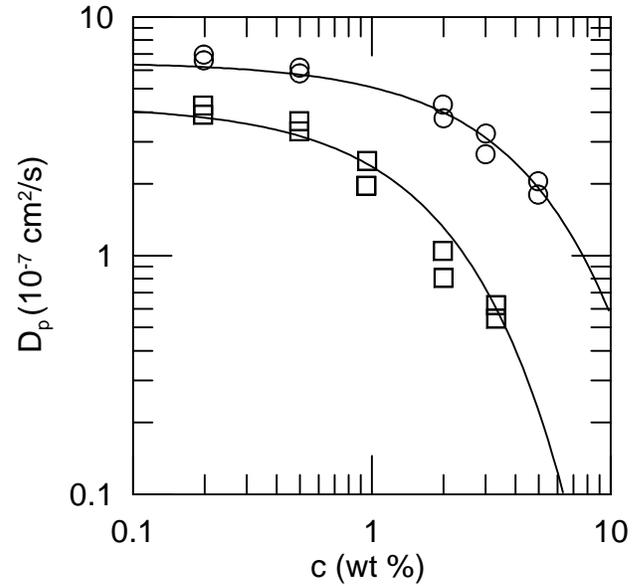}
\caption{\label{figurecheng2002aDp6} $D_{p}$ of 20 and 70 kDa dextrans in 2MDa  aqueous guar solutions as functions of guar concentration, and simple-exponential fits.  The larger probe is more effectively retarded by the solution, after Cheng, et al.\cite{cheng2002aDp}.  }
\end{figure}

\begin{figure}[bth] 
\includegraphics{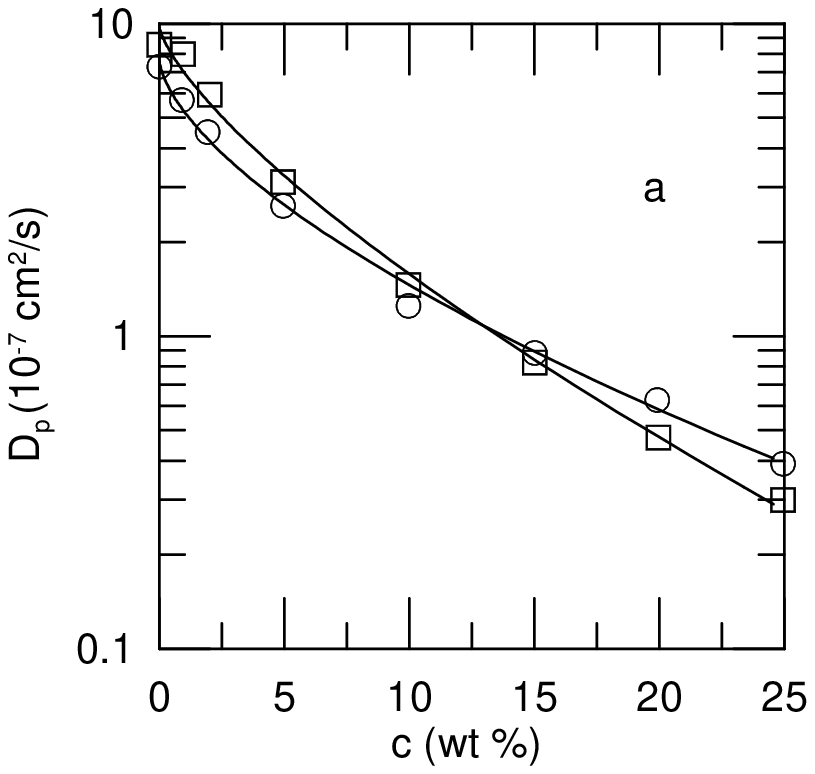}
\includegraphics{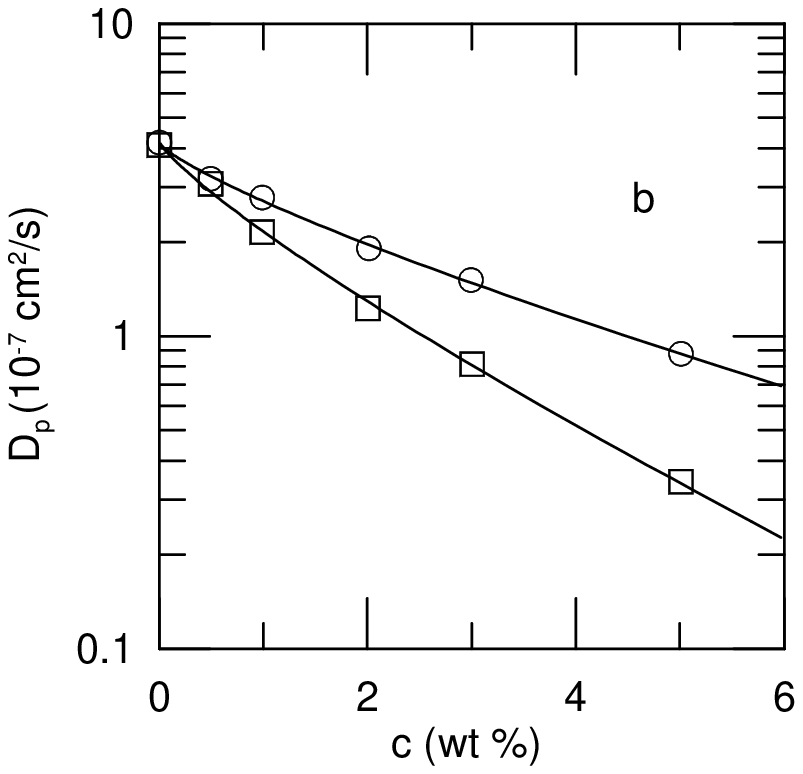}
\caption{\label{figurecheng2002aDp8} $D_{p}$ of (a) 20 kDa dextran ($\bigcirc$) and ovalbumin ($\square$) and (b) 70 kDa dextran ($\bigcirc$) and dendrimer G8 ($\square$) in (a) 40 and (b) 200 kDa PEO solutions, and stretched-exponential fits, after Cheng, et al.\cite{cheng2002aDp}.  }
\end{figure}

Cheng, et al.\cite{cheng2002aDp} used FRAP to monitor the diffusion of extremely small probes ($1 \leq R \leq 20$ nm) in solutions of guar galactomannan and polyethylene oxide.  The probes were labelled dextrans (20, 40, 70, and 500 kDa), ovalbumin, and 6th and 8th generation starburst dendrimers.  PEO fractions had molecular weights 40, 200, and 600 kDa; the guar had molecular weight 2MDa. $D_{p}$ of various probe:matrix combinations and $\eta$ of one polymer solution were obtained. Figure \ref{figurecheng2002aDp6} shows $D_{p}$ of 20 and 70 kDa dextrans through 2 MDa guar solutions, the larger dextran being more effectively retarded by the guar.  Figures \ref{figurecheng2002aDp8}a and \ref{figurecheng2002aDp8}b compare the diffusion of (a) 20 kDa dextran and ovalbumin in 40 kDa PEO, and (b) 70 kDa dextran and dendrimer G8 in 200 kDa PEO. For various probes, $D_{p}$ falls by factors of 5 to 30 over the observed concentration range.   PEO is roughly equally effective at retarding the diffusion of the 20 kDa dextran and ovalbumin, but is more effective at retarding the diffusion of G8 dendrimer than at retarding the motion of 70 kDa dextran, even though G8 dendrimer and 70 kDa dextran have the same $D_{p0}$.

\begin{figure}[tb] 
\includegraphics{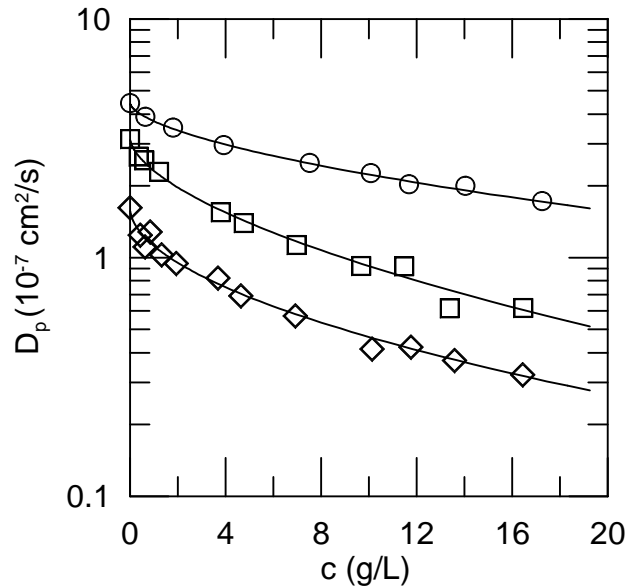}
\caption{\label{figuredesmedt1994aDp1} $D_{p}$ of 71.2 ($\bigcirc$), 147.8 ($\square$), and 487 ($\lozenge$) kDa dextrans in $M_{w} = 680$ kDa hyaluronic acid solutions as functions of hyaluronic acid concentration, after De Smedt, et al.\cite{desmedt1994aDp}.  }
\end{figure}

De Smedt, et al.\cite{desmedt1994aDp} used FRAP to examine diffusion of fluorescein-labelled dextrans and polystyrene latex spheres through hyaluronic acid solutions. Dextrans had molecular weights 71, 148, and 487 kDa.  The hyaluronic acid had $M_{n}$ and $M_{w}$ of 390 and 680 kDa.  The dextran diffusion coefficients depend on matrix polymer $c$ as stretched exponentials in $c$, as seen in Figure \ref{figuredesmedt1994aDp1}.  Hyaluronic acid solutions are somewhat more effective at retarding the larger dextran probes. Viscosities for these solutions were reported by De Smedt, et al.\cite{desmedt1993aDp}.  The concentration dependence of $\eta$ is stronger than the concentration dependence of $D_{p}$ of the polystyrene spheres, which is in turn stronger than the concentration dependence of $D_{p}$ of the dextrans.  Spheres and dextrans both diffuse more rapidly expected from the solution viscosity and the Stokes-Einstein equation.

\begin{figure}[t] 
\includegraphics{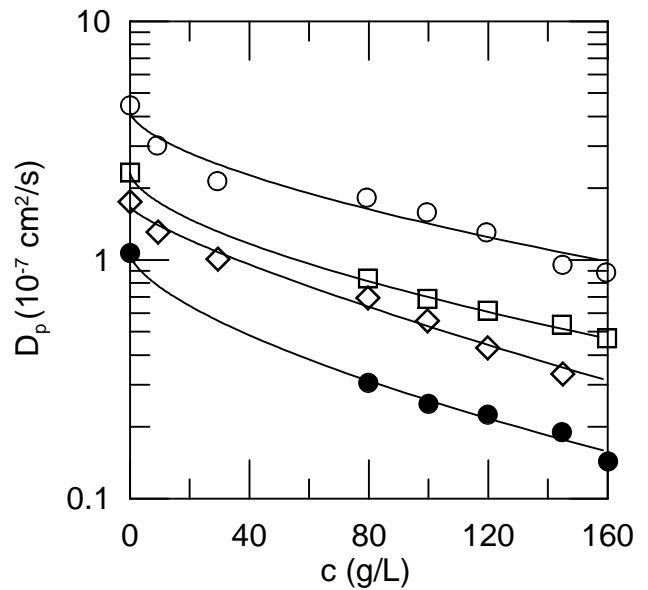}
\caption{\label{figuredesmedt1997aDp1} $D_{p}$ of 19 ($\bigcirc$), 51 ($\square$), 148($\lozenge$), and 487 ($\bullet$) kDa dextrans in dextran methacrylate solutions as functions of dextran methacrylate concentration, after De Smedt, et al.\cite{desmedt1997aDp}.  }
\end{figure}

De Smedt, et al.\cite{desmedt1997aDp} used confocal laser scanning microscopy (CSLM) and FRAP to study the diffusion of fluorescein-labelled dextran molecules ($19 \leq M_{w} \leq 487$ kDa) through dextran-methacrylate, comparing $D_{p}$ of probes in solutions with probes in swollen and relaxed cross-linked gels.  The ability of the crosslinked gels to retard probe motion increased with gel matrix concentration and increasing probe molecular weight, with the smallest probe diffusing at a rate independent of the matrix concentration.  $\eta$, and $D_{p}$ of each probe, all have stretched-exponential concentration dependences, as seen in Figure \ref{figuredesmedt1997aDp1}.  All probes diffuse more rapidly than expected from the solution viscosity.  The larger probes are slightly more effectively retarded by the solution. The concentration dependence of $D_{p}$ approaches the concentration dependence of $\eta$ more closely as the molecular weight of the probes is increased.  $D_{p}$ of probes in gels was always less than or equal to $D_{p}$ of the same probe in a noncrosslinked solution at the same concentration.  Swelling a gel increased $D_{p}$, but only at lower levels of gel crosslinking.

\begin{figure}[bth] 
\includegraphics{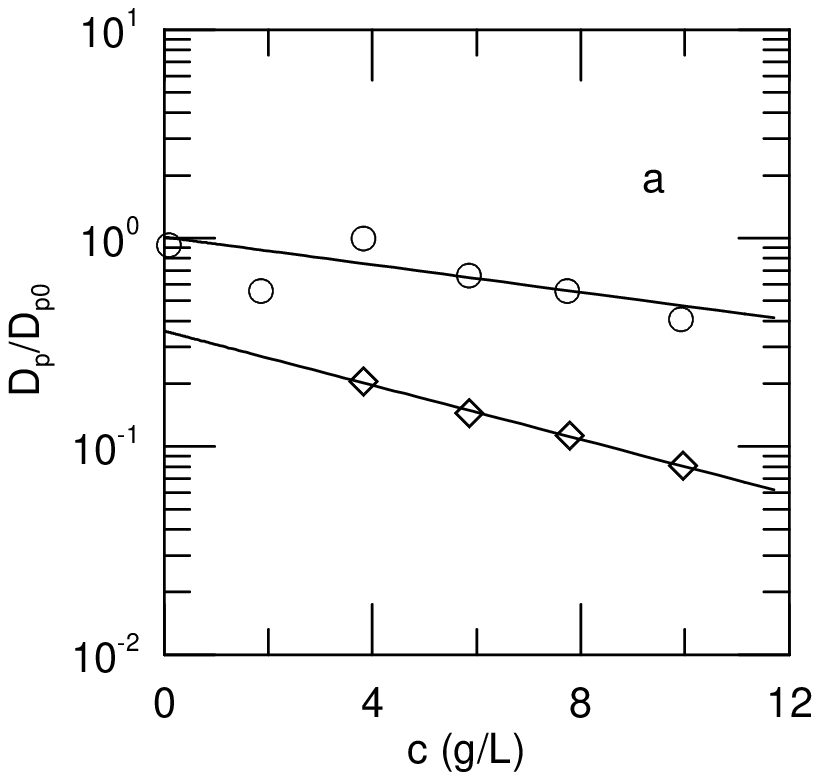}
\includegraphics{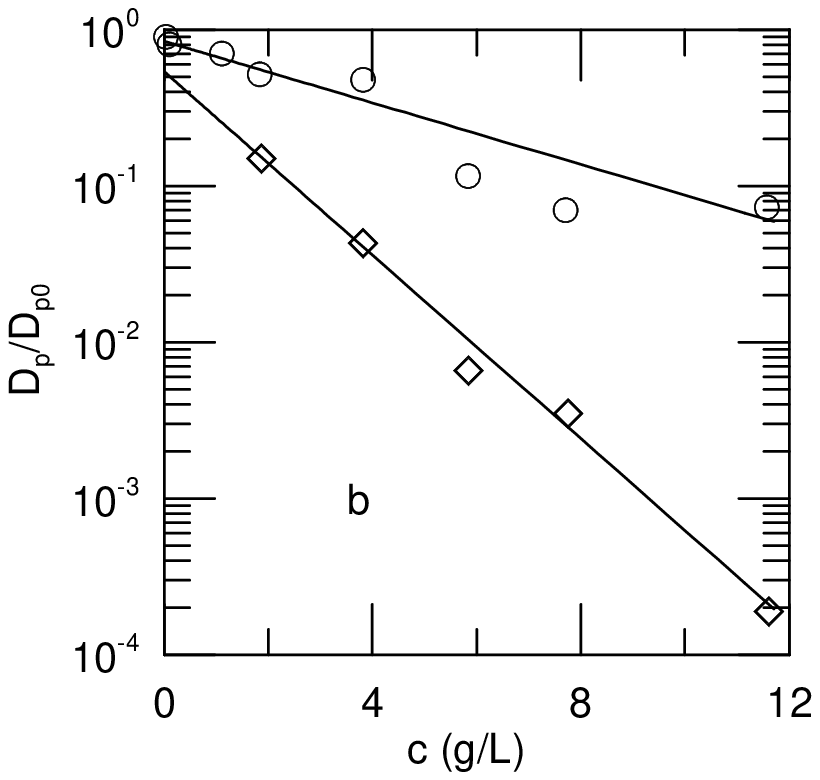}
\caption{\label{figuredelfino2005aDpa} Fast ($\bigcirc$) and slow ($\lozenge$) mode $D_{p}/D_{p0}$ for (a) 14, (b) 47 and (c) 102 nm polystyrene sphere probes in aqueous 700 kDa carboxymethylcellulose, after Delfino, et al.\cite{delfino2005aDp}  }
\end{figure}

\begin{figure}[bth] 
\includegraphics{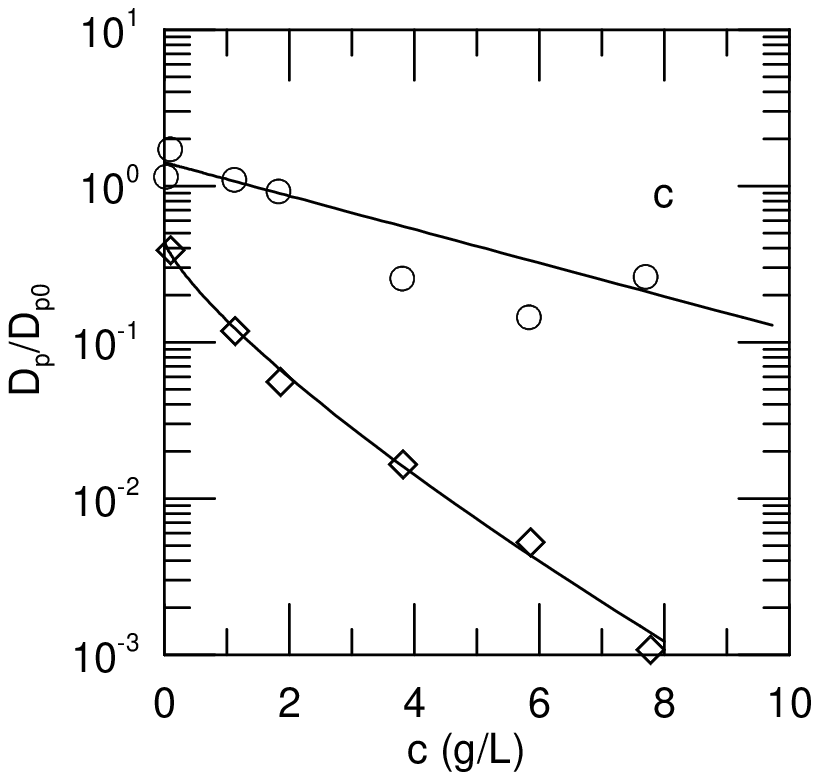}
\end{figure}

Delfino, et al.\cite{delfino2005aDp} used QELSS to study probes in aqueous solutions of carboxymethylcellulose, nominal molecular weight 700 kDa, at concentrations 0.2-11.7 g/L. The probe particles were 14, 47, and 102 nm radius PSL spheres.  Ubbelohde viscometers and a concentric-cylinder Couette viscometer were used to determine rheological properties of the solutions.  From the intrinsic viscosity $[\eta]$, the overlap concentration was $c^{*}=1/[\eta] = 0.73$ g/L, while from the linear slope of $\eta_{\rm SP}$, $R_{g} \approx 50$ nm was found. Probe spectra were fit to an exponential or a sum of two pure exponentials, a fast mode and a slow mode.  For the 14 and 47 nm probes, spectra were unimodal at smaller $c$ and bimodal at larger $c$. With 14 nm probes, the fast mode dominated at all concentrations.  For the 47 nm probes, the fast mode dominated the spectrum at small $c$ and the slow mode dominated at large $c$, while for the 102 nm probes, the slow mode always dominated. As seen in Figure \ref{figuredelfino2005aDpa}, except for the slow mode of the largest spheres, which show a weakly stretched-exponential concentration dependence, $D_{p}$ of either mode falls approximately exponentially with increasing polymer $c$. Increasing $c$ is more effective at retarding the relaxation of the slow mode than at retarding the relaxation of the fast mode.  Increasing $c$ is more effective at retarding the motion of the larger 47 and 102 nm spheres than at retarding the motion of the small 14nm spheres.

Dunstan and Stokes\cite{dunstan2000aDp} employ QELSS to measure diffusion of PSL spheres through glycerol and polyacrylamide solutions.  In glycerol-water, $D_{p}$ tracked the solution viscosity.  In polyacrylamide solutions, spectra were fit to a sum of two exponentials.  For 100 nm radius sulfate-modified spheres: $D_{s}$ of the slow exponential fell with increasing polymer $c$.  $D_{f}$ of the fast exponential showed re-entrant behavior, with a maximum in $D_{f}$ at an intermediate polymer concentration considerably larger than the overlap concentration, followed at still larger concentrations by a decline until $D_{f}$ was less than a third of its zero-concentration value. $\eta_{\mu}$ from the slow exponential is greater than $\eta$ from macroscopic measurements, while $\eta_{\mu}$ from the fast relaxation is less than $\eta$. $D_{s}$ is substantially independent from $q$, while $D_{f}$ of probes in more concentrated polyacrylamide solutions increases strongly with increasing $q$.  More limited measurements on 34 nm radius amidine-modified spheres appear to yield the same pattern of concentration dependences for $D_{s}$ and $D_{f}$.

\begin{figure}[thb] 
\includegraphics{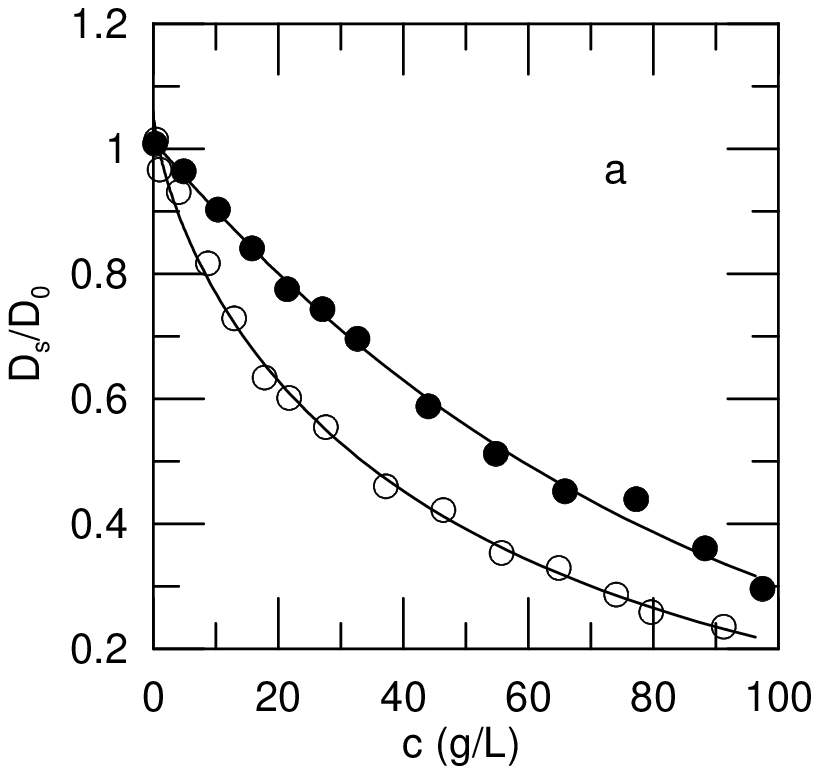}
\includegraphics{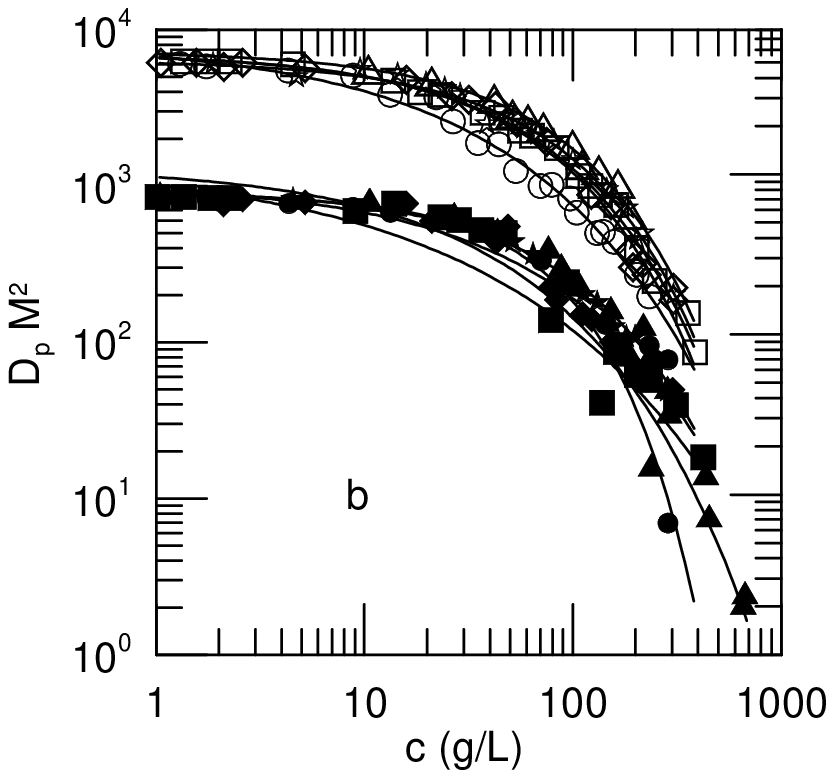}
\caption{\label{figurefurukawa1991aDp1} (a) $D_{s}$ of ($\bullet$) 40 and ($\circ$) 150 kDa dextrans and (b) $D_{p}$ of 40 (open points) and 150 (filled points) kDa dextrans through solutions of 2000 ($\bigcirc$), 500($\lozenge$), 110($\square$), 83($\bigstar$), and 40($\vartriangle$) kDa dextrans, after Furukawa, et al.\cite{furukawa1991aDp}, and fits to stretched exponentials in $c$.}
\end{figure}

\begin{figure}[bht] 
\includegraphics{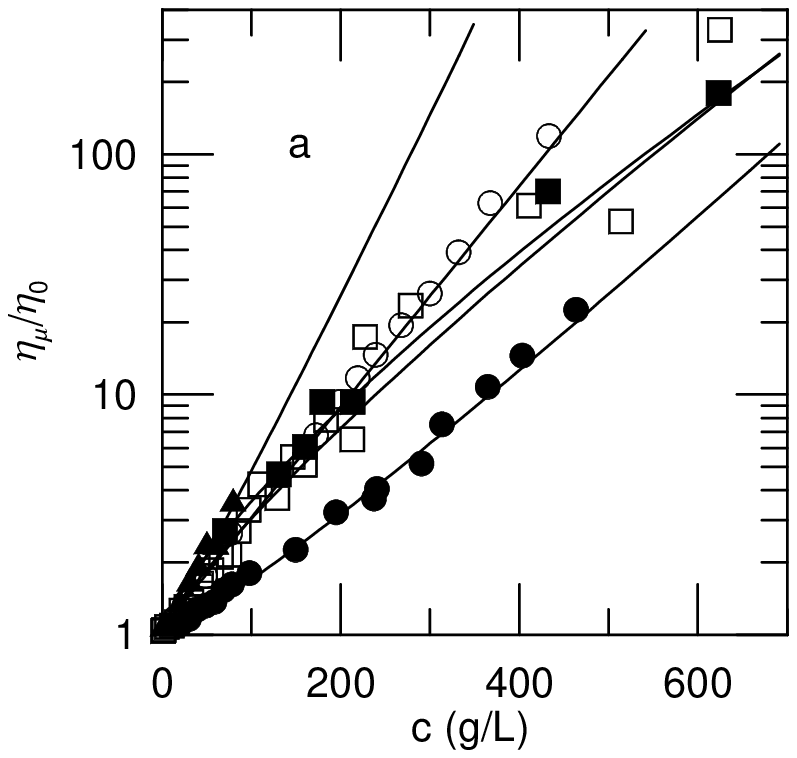}
\includegraphics{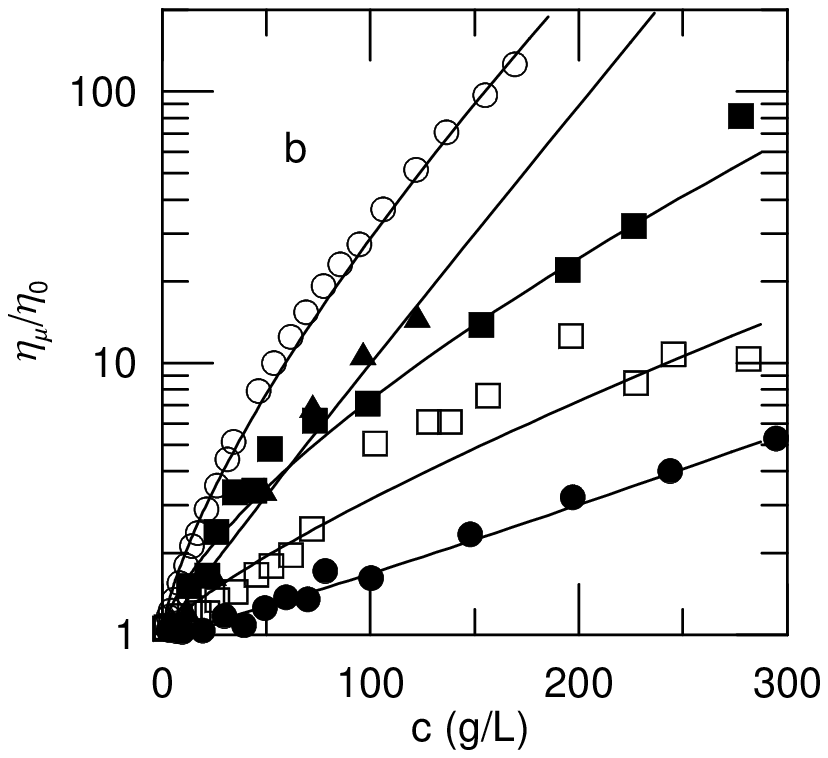}
\caption{\label{figurefurukawa1991aDp5} $\eta/\eta_{0}$ for viscosity from macroscopic measurements ($\circ$) and $\eta_{\mu}/\eta_{0}$ from probe diffusion using 20 nm radius polystyrene latex sphere ($\blacktriangle$), 150 kDa dextran ($\blacksquare$), 40 kDa dextran ($\square$), and fluorescein ($\bullet$) probes in (a) 40 and (b) 150 kDa dextrans, after Furukawa, et al.\cite{furukawa1991aDp}, and fits to stretched exponentials in $c$.}
\end{figure}

Furukawa, et al.\cite{furukawa1991aDp} used FRAP to measure the diffusion of 19 nm radius polystyrene spheres, 40 and 150 kDa dextran, and fluorescein dye through dextran matrix solutions having $40 \leq M_{w} \leq 2000$ kDa and concentrations up to 300 or 600 g/L. $D_{p}$ of dextran chains diffusing through dextran solutions consistently has a stretched exponential concentration dependence; cf.\ Figure \ref{figurefurukawa1991aDp1}.  In 40 kDa dextran solutions, probes other than fluorescein diffuse at very nearly the rate expected from the solution viscosity.  In solutions of 150 kDa dextran, all probes diffuse more rapidly than expected from the solution viscosity, but (cf.\  Figure \ref{figurefurukawa1991aDp5}a) ,with increasing probe size $\eta_{\mu}$ tends toward the macroscopic $\eta$.   

\begin{figure}[bth] 
\includegraphics{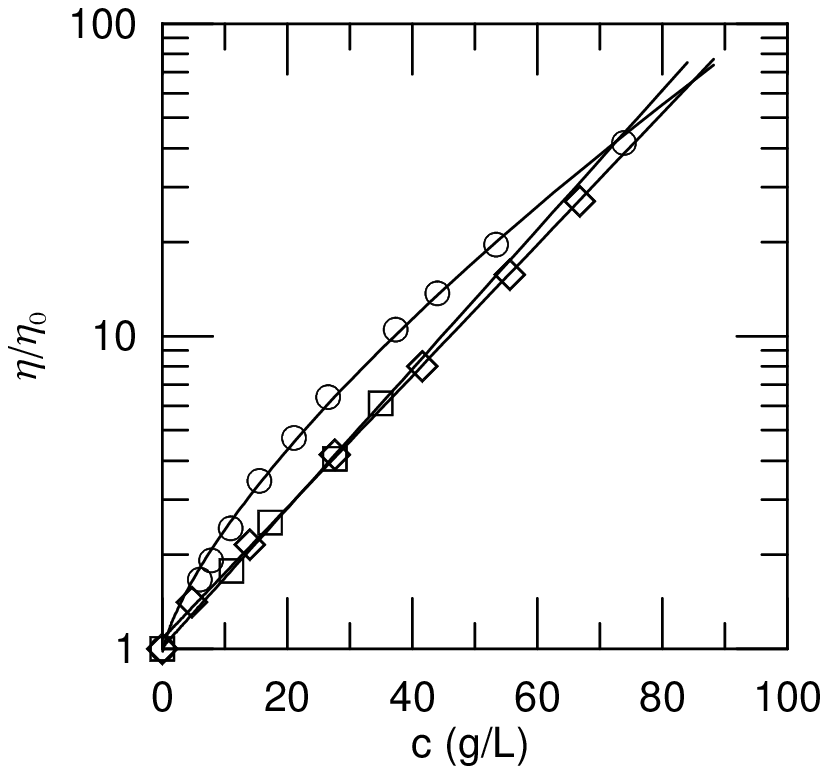}
\includegraphics{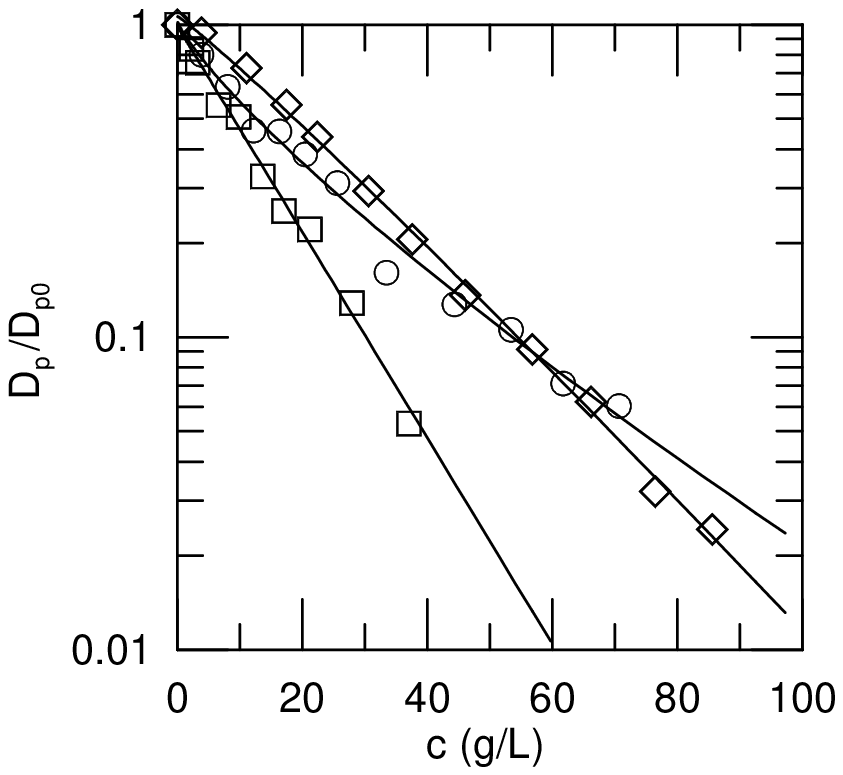}
\caption{\label{figuregold1996aDp1} Reduced viscosities and probe diffusion coefficients for solutions of 350 kDa polymethylmethacrylate.  Solvents are tetrahydrofuran($\bigcirc$), N,N-dimethylformamide($\lozenge$), and dioxane-water($\square$), after Gold, et al.\cite{gold1996aDp}.}
\end{figure}

Gold, et al.\cite{gold1996aDp} studied the diffusion of highly cross-linked polystyrene latex spheres through solutions of 350 kDa polymethylmethacrylate in the good solvents tetrahydrofuran (THF) and N,N-dimethylformamide (DMF), and the theta solvent (at 25$^{o}$ C) dioxane:water.  Spectra were analyzed with Laplace inversion.  For probes in good solvents, cumulant analysis was also applied.  In good solvents, a unimodal decay distribution was observed.  Under Laplace inversion, spectra of probes in dioxane-water often showed a bimodal distribution of decay rates. With dioxane:water as the solvent: The slow mode increased as the sample aged.  The amplitude of the slow mode varied from small up to equal to the amplitude of the fast mode.  The slow mode was more prominent at low angles. Further analysis was based primarily on the relaxation rate of the fast mode.  Diffusion coefficients were inferred from slopes of $\Gamma$ against $q^{2}$, with typical standard errors in the slope of $1-10$\% leading to results of Figure \ref{figuregold1996aDp1}.  For probes in THF and DMF, diffusion is more rapid than expected from the solution viscosities and the Stokes-Einstein equation, the Stokes-Einstein equation failing by 50\% or a factor of 2 at elevated polymer concentration. For probes in dioxane:water, probe motion was slower than anticipated from the Stokes-Einstein equation, an effect attributed by Gold, et al.\ to polymer adsorption by the probe spheres.

\begin{figure}[thb] 
\includegraphics{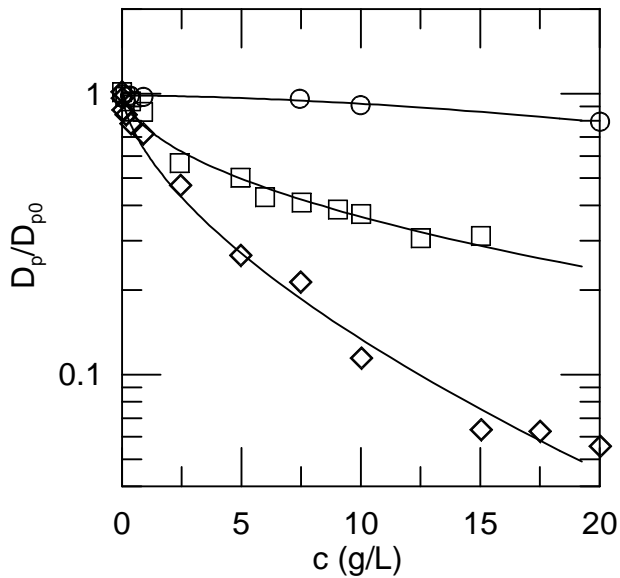}
\caption{\label{figuregorti1985aDp2} $D_{p}$ of fluorescein ($\bigcirc$), bovine serum albumin ($\square$), and 19 nm radius PSL ($\lozenge$) in 500 kDa aqueous NaPSS: 10 mM phosphate buffer, and fits to stretched exponentials, after Gorti and Ware\cite{gorti1985aDp}.  }
\end{figure}

\begin{figure}[bht] 
\includegraphics{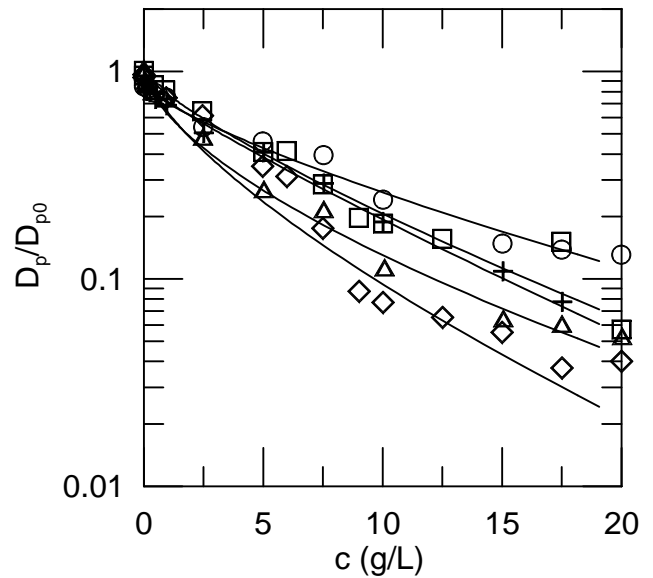}
\caption{\label{figuregorti1985aDp1} $D_{p}$ of 19 nm radius PSL in 500 kDa aqueous polystyrene sulfonate solutions
at various ionic strengths: 0.001 ($\bigcirc$), 0.005 ($\square$), 0.01 ($\bigtriangleup$), 0.02 ($+$), and 0.05 ($\lozenge$) M phosphate ion, and fits of $D_{p}$ to stretched exponentials in $c$, after Gorti and Ware\cite{gorti1985aDp}.  }
\end{figure}

Gorti and Ware\cite{gorti1985aDp} made an extensive study of probe diffusion in polyelectrolyte solutions, using FRAP as the primary method, as seen in figure \ref{figuregorti1985aDp2}.  Probe particles included fluorescein, 19 nm radius polystyrene spheres, and bovine serum albumin. The matrix polymers were 70 and 500 kDa polystyrene sulphonates. Solution viscosities were obtained using Ubbelohde viscometers.  The dependence of $D_{p}$ on polymer concentration, molecular weight, degree of neutralization, solution ionic strength, and probe size was observed. Polymer solutions were more effective at retarding the diffusion of larger probes.  At fixed polymer concentration, probe diffusion was slowed by increasing the polymer molecular weight.  Measurements covered both the nominal dilute and the nominal semidilute regimes.  At low ionic strength, polystyrene sulphonate solutions experience an ordinary-extraordinary transition in their QELSS spectra\cite{lin1978aDp}. For the 1 and 5 mM salt solutions, the transitions occur at 2 and 11 g/L PSS.  Probe diffusion measurements taken above and below the transition appear to lie on the same line; the ordinary-extraordinary transition does not affect the probe diffusion coefficient. Also, $D_{p}$ does not track the solution fluidity $\eta^{-1}$.  Solution ionic strength has a modest effect on $D_{p}$, an increase in $I$ tending to reduce $D_{p}$, as seen in Figure \ref{figuregorti1985aDp1}. 

An early application of probe diffusion was presented by Jamieson, et al.\cite{jamieson1982aDp}, who in the course of reporting on 2.2 MDa xanthan in water, giving measurements of flow birefringence, light scattering spectra, and shear viscosity, also report limited measurements of $D_{p}$ of 100 nm polystyrene spheres as polymer $c$ is increased.  $D_{p}$ fell more than thirty-fold over the allowed range of concentrations, though only for a few data points was $D_{p}$ significantly less than $D_{o}$.  At the highest concentrations studied, $\eta/\eta_{0}$ attained values as large as $10^{4}$.  The microviscosity $\eta_{\mu}$ inferred from $D_{p}$ was thus as much as two orders of magnitude less than $\eta$.

Johanssson, et al.\cite{johansson1991aDp} examined the diffusion of low-molecular-weight (0.3-4 kDa) poly-ethylene glycol fractions in $\kappa$-carrageenan crosslinked gels and in solutions of 500 kDa sodium polystyrene sulphonate.  The probes were radioactively labelled; boundary spreading was used to determine $D_{p}$ of dilute probes.  Even for these very small probes, carrageenan solutions and gels differed substantially in their effect on probe diffusion, gels being substantially more effective at retarding probe motion than were solutions having the same polymer concentration.  Johansson, et al.\ also found that the effect of a fixed concentration of polystyrene sulphonate on probe diffusion decreases with increasing ionic strength.  They proposed that this difference arises because relatively rigid polymers are more able to affect probe motion than are flexible polymers.

\begin{figure}[bth] 
\includegraphics{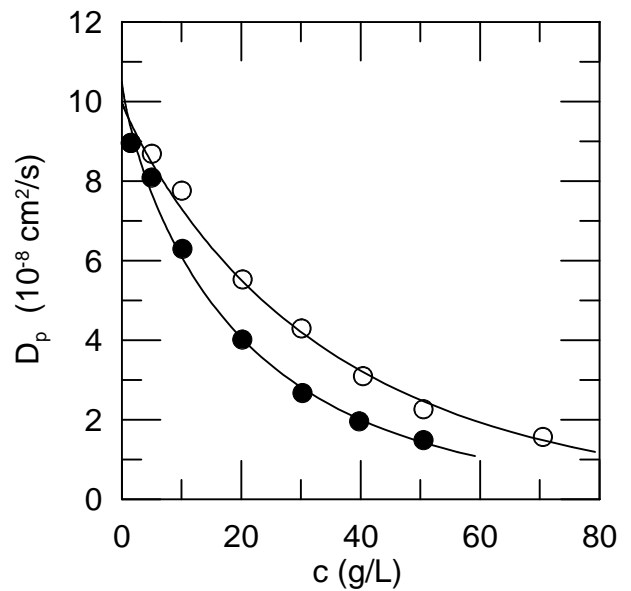}
\caption{\label{figurekonat1982aDp2} $D_{p}$ of 23 nm radius Kraton G-1650 micelles in solutions of 110 ($\bigcirc$) and 200 ($\square$) kDa polystyrenes in dioxane, after data of Konak, et al.\cite{konak1982aDp}. }
\end{figure} 

Konak, et al.\cite{konak1982aDp} report on the diffusion of block copolymer micelles through solutions of linear polystyrene, as shown in Figure \ref{figurekonat1982aDp2}.  The block copolymer was a 74 kDa Kraton G-1650 polystyrene-hydrogenated polybutadiene-polystyrene that forms 4.8 MDa 23 nm radius micelles.  The matrix polymers were 110 and 200 kDa polystyrenes having $M_{w}/M_{n}=1.4$.  The micelle diffusion coefficient $D_{p}$ and the viscosity $\eta$ were reported. At intermediate polystyrene concentrations, $D_{p}$ was larger than expected from $\eta$ and $D_{o}$, but at the largest $c$ examined $D_{p}\eta$ returned to its low-concentration value.  The largest-$c$ re-entrant behavior was only observed at a single point, so it is impossible to say whether $D_{p} \eta/D_{p0}\eta_{0} \approx 1$ represents the general largest-matrix-concentration behavior, or whether at concentrations larger than those studied $D_{p}$ is smaller than expected from $\eta$. 

\begin{figure}[t] 
\includegraphics{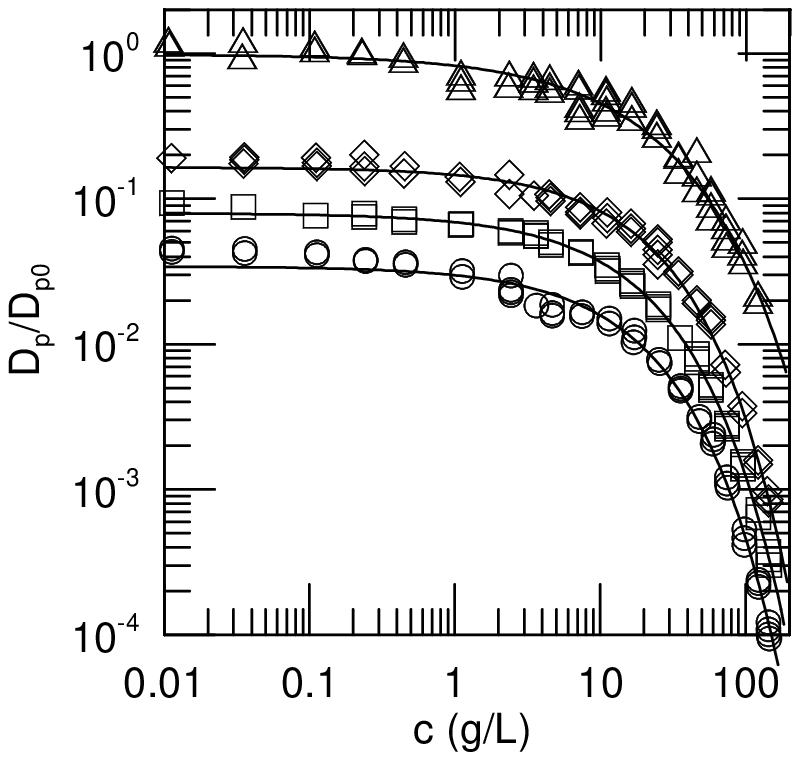}
\includegraphics{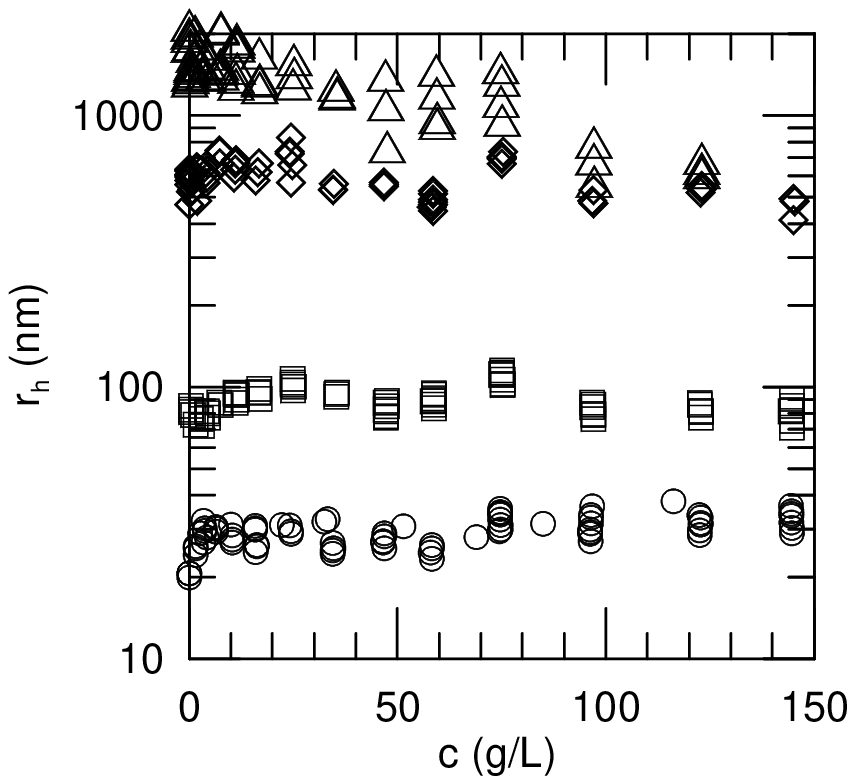}
\caption{\label{figurelin1984bDp6} $D_{p}/D_{p0}$ and $r_{h}$ of 20 ($\bigcirc$), 80 ($\square$), 620 ($\lozenge$) and 1500 ($\bigtriangleup$) nm radius spheres in aqueous 300 kDa non-neutralized polyacrylic acid, after Lin and Phillies\cite{lin1984bDp}.  $D_{p}/D_{p0} \rightarrow 1$ at small $c$; data and fits were shifted vertically for clarity. }
\end{figure}

In a series of three papers, Lin and Phillies\cite{lin1982aDp,lin1984aDp,lin1984bDp} determine diffusion of polystyrene spheres through aqueous solutions of non-neutralized poly-acrylic acid (PAA). Lin and Phillies\cite{lin1982aDp} report on the diffusion of nominal 38 nm carboxylate-modified spheres in solutions of 300 kDa PAA at polymer concentrations $0.37 \leq c \leq 171$ g/L.  They measured $D_{p}$ as a function of temperature at 11 polymer concentrations and $\eta$ of PAA solutions at various polymer concentrations. At each polymer concentration $D_{p}$ tracked $T/\eta$ with reasonable accuracy.  These preliminary measurements were extended in a further study. Lin and Phillies\cite{lin1984bDp} compared the diffusion of probe spheres (radii 20.4-1500 nm) in 300 kDa PAA, for solution concentrations 0.0113-145 g/L.  As seen in Figure \ref{figurelin1984bDp6}, in each solution $D_{p}/D_{po}$ fell markedly, for large spheres by as much as three orders of magnitude, with increasing polymer concentration.  For each sphere size, $D_{p}(c)$ has a stretched-exponential concentration dependence.  

The apparent hydrodynamic radius $r_{H}$ (Figure \ref{figurelin1984bDp6}) of the smaller spheres grew modestly with increasing polymer concentration, $r_{H}$ of the largest spheres fell, and $r_{H}$ of intermediate spheres is nearly independent from $c$.  Lin and Phillies interpreted their results as arising from polymer adsorption (to form a layer 10-20 nm thick) on the spheres, coupled with a failure of the Stokes-Einstein equation (spheres diffuse faster than expected from the solution viscosity), especially for the larger spheres.

\begin{figure}[t] 
\includegraphics{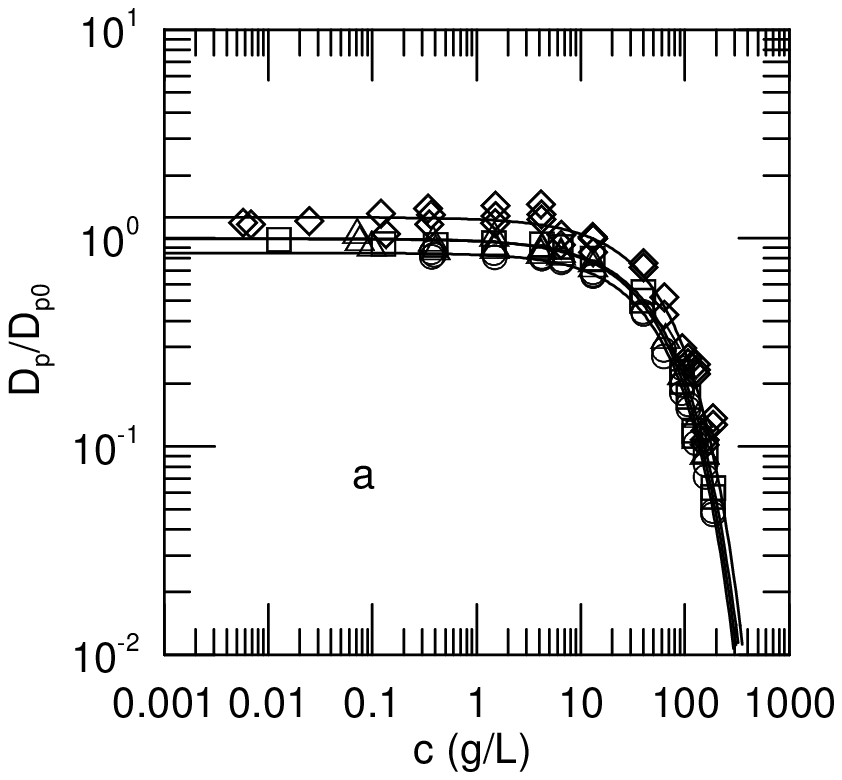}
\includegraphics{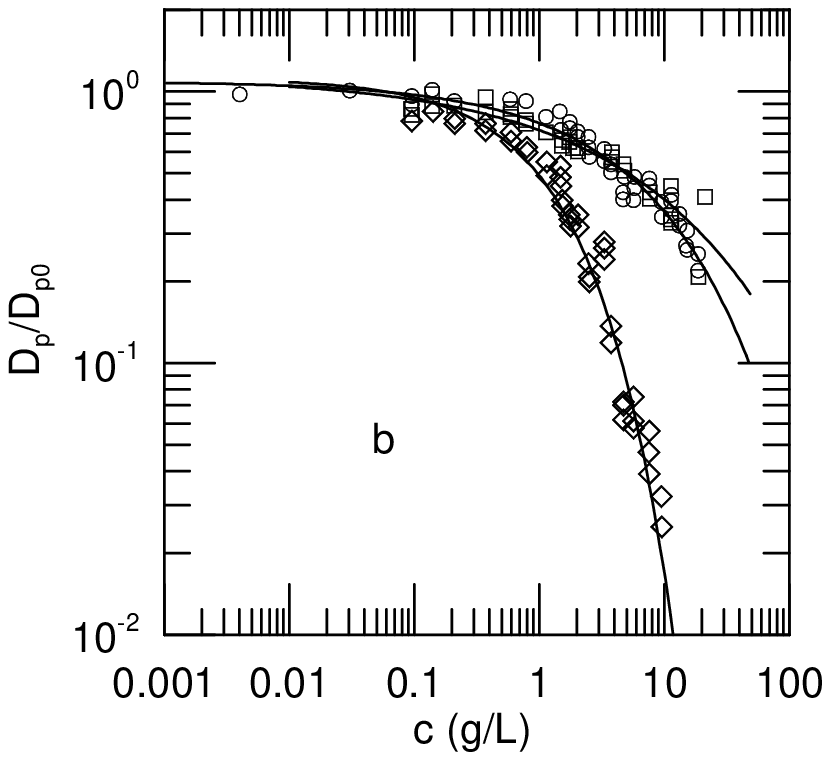}
\caption{\label{figurelin1984aDp5} $D_{p}/D_{p0}$ of 20 ($\bigcirc$), 80 ($\square$), 620 ($\bigtriangleup$), and 1500 ($\lozenge$) nm radius carboxylate-modified polystyrene spheres in solutions of (a) 50 kDa, and (b) 1 MDa non-neutralized polyacrylic acid, and fits of $D_{p}$ to stretched exponentials in $c$, after Lin and Phillies\cite{lin1984aDp}}.
\end{figure} 

\begin{figure}[b] 
\includegraphics{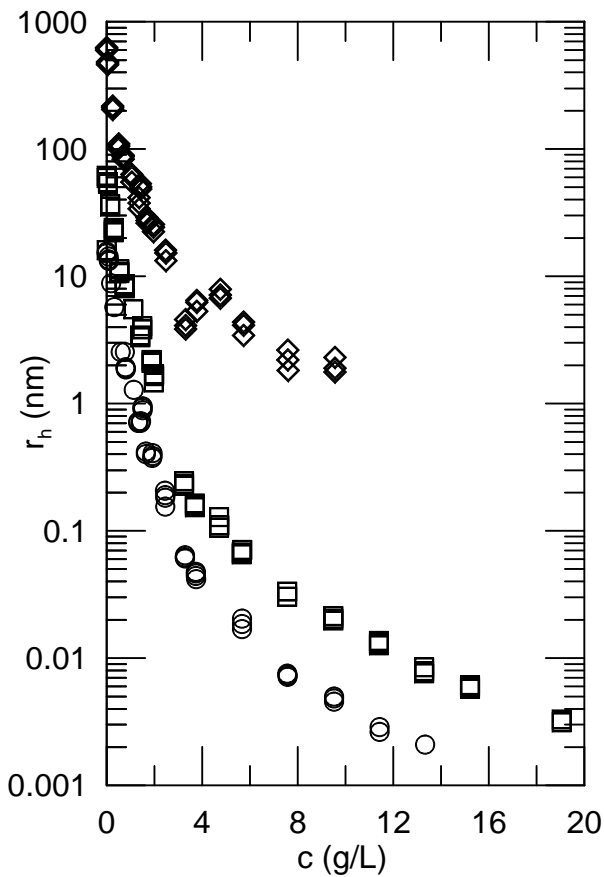}
\caption{\label{figurelin1984aDp1} Hydrodynamic radius of 20 ($\bigcirc$), 80 ($\square$), and 620 ($\lozenge$) nm radius spheres in aqueous 1 MDa polyacrylic acid, after Lin and Phillies\cite{lin1984aDp}}.
\end{figure} 

Finally, Lin and Phillies\cite{lin1984aDp} examined the diffusion of 20.4, 80, 620, and 1500 nm radius carboxylate-modified polystyrene spheres in solutions of 50 kDa and 1 MDa polyacrylic acid.  They also measured the macroscopic solution viscosity $\eta$.   In 50 kDa PAA, $\eta$ follows a stretched exponential in $c$.  1 MDa PAA shows a viscometric transition, $\eta$ changing from a stretched-exponential concentration dependence to a power-law concentration dependence at $c \approx 1.4$ g/L PAA. In 50 kDa (Figure \ref{figurelin1984aDp5}a) and 1 MDa PAA (Figure \ref{figurelin1984aDp5}b), $D_{p}(c)$ for each probe size follows a stretched exponential in $c$. In 50 kDa PAA solutions, $D_{p}$ and $\eta$ are related by the Stokes-Einstein equation. $D_{p}$ in 1 MDa HPC solutions does not show Stokes-Einsteinian behavior.  $D_{p} \eta /D_{po} \eta_{o}$ increases by four orders of magnitude with increasing $c$, as seen in Figure \ref{figurelin1984aDp1}.

Extended searches of the spectrum on the time scale on which the spectrum would have decayed, if the Stokes-Einstein equation were correct, found no sign of a spectral relaxation; the spectrum had already decayed to the baseline.  Lin and Phillies proposed that that the discrepancy between $D_{p}$ and the macroscopic $\eta$ arises from shear thinning {\em in the probe diffusion process}, namely the microviscosity found on the time and distance scales probed by the polystyrene spheres is far less than the macroscopic viscosity $\eta$. 

In 50 kDa HPC, the second cumulant of $S(q,t)$ was substantially independent of $c$.   In 1 MDa PAA, the spectral second cumulant was nearly constant for $c < c^{+}$, but at $c > 1.2$ g/L (for the 620 nm spheres) or $c > 2.5$ g/l (for the smaller spheres) the second cumulant increased substantially.  This very early work was done with a 64-channel linear correlator; it was apparent that at large $c$ the lineshape was changing dramatically, but revelation of the details had to wait for more modern instrumentation.  The near-simultaneous changes at $c^{+}$ in the functional form of $\eta(c)$ and in the qualitative probe lineshape represent the first observations of the solutionlike--meltlike viscosity transition and a spectral correlate.   
 
Lin\cite{lin1986aDp} used QELSS to study diffusion of 155 and 170 nm nominal radius titania spheres as probe particles in a melt of 7500 Da polyethylene oxide, at temperatures from 85 to 160 C.  Over this temperature range, $D_{p}$ changes by nearly two orders of magnitude.  Comparison was made with the shear-dependent viscosity obtained using a cone-and-plate viscometer.  The observed microviscosities were substantially less than the measured viscosity.  An extrapolation procedure based on the apparent activation energy inferred from the temperature dependence of $\eta$ was used to estimate an effective shear rate for probe diffusion $\geq 10^{4}$ s$^{-1}$, corresponding via $\dot{\gamma} \sim D/L^{2}$ to probe diffusion over atomic distances.  Diffusion over an atomic distance is rationally fundamental if one supposes that the basic unit step for probe diffusion in this system corresponds to the displacement of a single layer of polymer chains lying along the probe surfaces.

Mangenot, et al.\cite{mangenot2003aDp} used FCS to measure diffusion of nucleosome core particles (NCP) in water and DNA solutions.  Manipulation of the solution ionic strength was shown to change the size of the core particles.
The particles' surface structure can be manipulated via trypsinization, which removes the histone tails that are ordinarily wrapped about the particles.  The diffusion of trypsinized NCP is reduced by increasing DNA concentration but is almost independent of salt concentration.  In contrast, the histone tails of nontrypsinized particles unfold at elevated ionic strength, changing the particles' interactions with DNA in solution, so $D_{p}$ of nontrypsinized particles in DNA solutions is sensitive to both salt and DNA concentrations.

\begin{figure}[hbt] 
\includegraphics{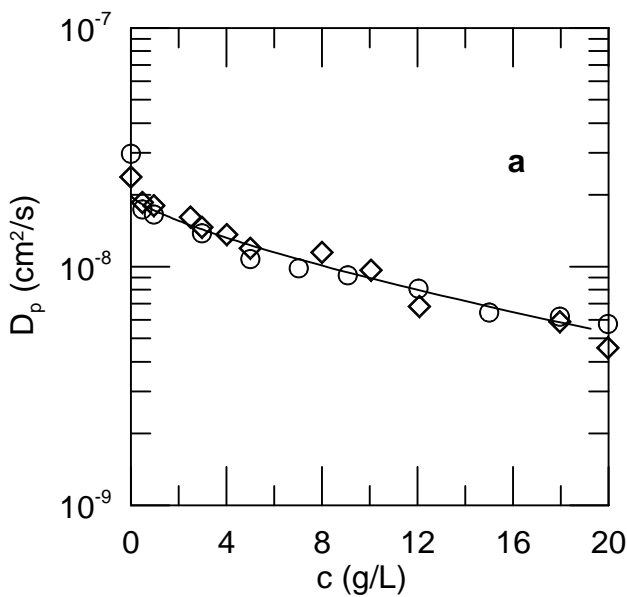}
\includegraphics{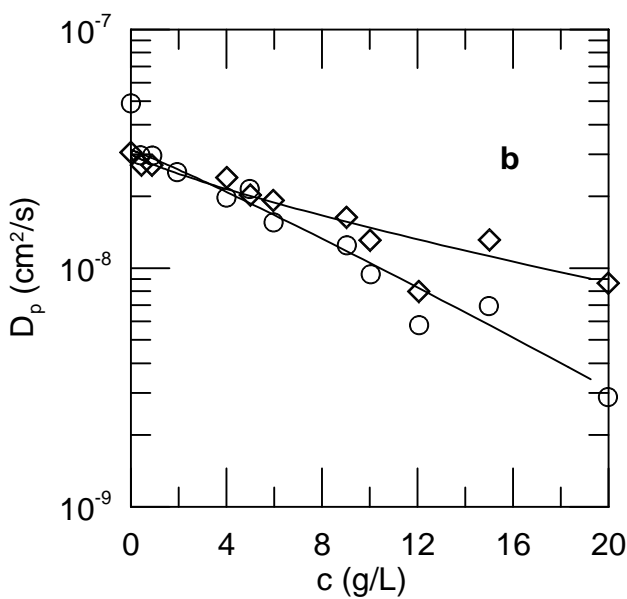}
\caption{\label{figurenehme1989aDp} $D_{p}$ from QELSS ($\bigcirc$) and inferred ($\lozenge$) from the friction factor from the sedimentation coefficient of (a) 80 and (b) 40 nm PSL in solutions of 90 kDa poly-L-lysine, after Nehme, et al.\cite{nehme1989aDp}}.
\end{figure} 

Nehme, et al.\cite{nehme1989aDp} used QELSS and ultracentrifugation to study the motion of 49 and 80 nm radius probe particles in poly-L-lysine ($52 \leq M_{w} \leq 353$ kDa) solutions, leading to Figure \ref{figurenehme1989aDp}.  In most of these experiments, the probe particles were somewhat or substantially larger than the nominal polymer solution mesh size $\xi$. For the smaller probes and the less concentrated larger polymers, at small $c$ the probe $R$ was comparable to $\xi$.  From the ultracentrifuge experiments, the polymer solutions are approximately equally effective at retarding the sedimentation of smaller and larger probes.  For the larger probes, $D_{p}/D_{o}$ and $s_{o}/s$ have nearly the same concentration dependences; the polymers retard sedimentation and probe diffusion to approximately the same extent.  For the smaller probes, the polymers may be more effective at retarding probe diffusion than at retarding probe sedimentation, though only for a few concentrations is the effect large.  

Nehme, et al., interpreted this behavior as a time scale effect: When probe motion is adequately rapid, the polymer is less able to move out of the way of the probes.  The sedimentation coefficients came close to being a universal function of the overlap concentration $c^{*}$.  Polymer adsorption in this system is substantial; the thickness of the adsorbed polymer layer on each probe was estimated as 20 nm.  $D_{p}$ from QELSS consistently tracked the solution $\eta$, but was about 15\% slower than expected from $\eta$ in a concentration-independent way.

\begin{figure}[t] 
\includegraphics{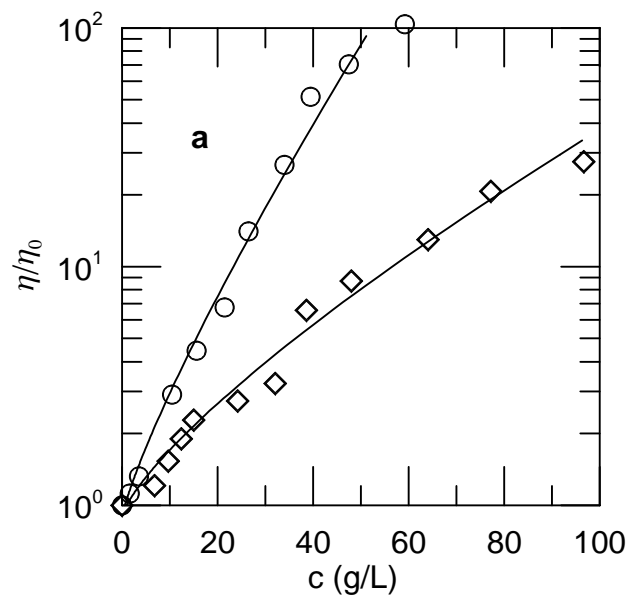}
\includegraphics{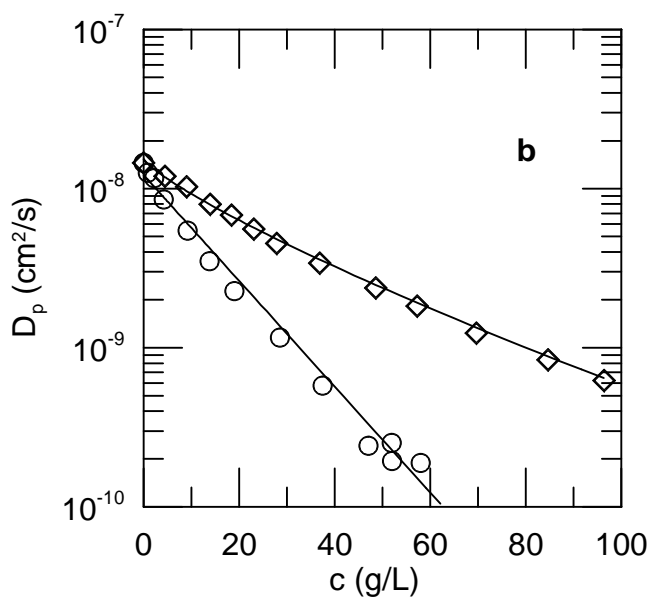}
\caption{\label{figureonyen1993aDp1} (a) $\eta/\eta_{0}$ and (b) $D_{p}$ of 200 nm spheres in solutions of ($\lozenge$) 215 kDa and ($\bigcirc$) 1100 kDa polystyrene in N,N-dimethylformamide, after Onyenemezu, et al.\cite{onyenemezu1993aDp}.}
\end{figure} 

\begin{figure}[t] 
\includegraphics{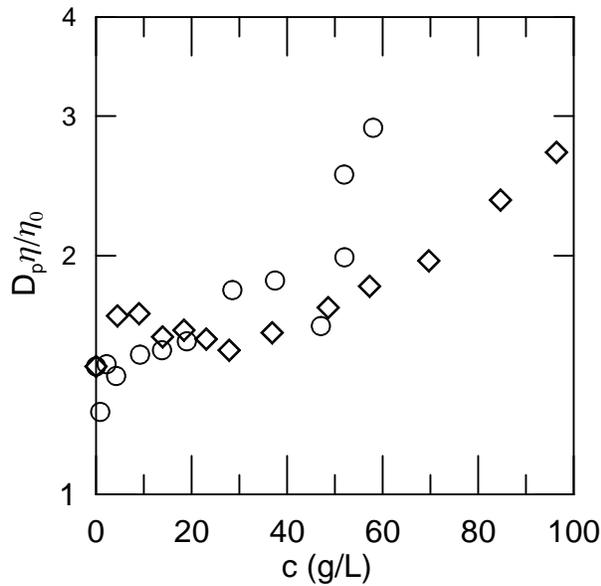}
\caption{\label{figureonyen1993aDp2} $D_{p}\eta/\eta_{0}$ of 200 nm spheres in solutions of ($\lozenge$) 215 kDa and ($\bigcirc$) 1100 kDa polystyrene in N,N-dimethylformamide, after Onyenemezu, et al.\cite{onyenemezu1993aDp}.}
\end{figure} 


Onyenemezu, et al.\cite{onyenemezu1993aDp} studied viscosity and probe diffusion (209 nm radius divinylbenzene-styrene spheres) in polystyrene: N,N-dimethylformamide solutions.  Solution viscosities were measured with Cannon-Ubbelohde viscometers.  Shear rate effects were studied rheometrically; solutions showed Newtonian behavior at the shear rates found in the viscometers. QELSS spectra were obtained at three scattering angles, and fit using cumulant expansions and inverse Laplace transform (CONTIN) methods, all giving good agreement. Polymer molecular weights were 215 and 1100 kDa, with polydispersities 1.05 and 1.07, respectively.  Polymer concentrations approaching 100 and 60 g/L, respectively, were attained, corresponding (cf.\ Figure \ref{figureonyen1993aDp1}a) to $\eta/\eta_{o}$ as large as 20 or 100 and $0.1 \leq c [\eta] \leq 10$.   Figure \ref{figureonyen1993aDp1}b  shows $D_{p}$ and fits to stretched exponentials in $c$.  At larger scattering angles, a second, much faster relaxation interpreted as polymer scattering was found. Onyenemezu, et al., assert that within experimental error the Stokes-Einstein equation is always closely obeyed in their systems; cf.\ Figure \ref{figureonyen1993aDp2} for the modest non-Stokes-Einsteinian behaviors that their data reveal.

Phillies\cite{phillies1985aDp} reports on the diffusion of bovine serum albumin through solutions of 100 kDa and 300 kDa polyethylene oxides.  $D_{p}$ depended measurably on the probe concentration.  At elevated polymer $c$ and low protein concentration, $D_{p}$ was as much as a third faster than expected from the $c$-dependent solution fluidity $\eta^{-1}$.  With increasing protein concentration, $D_{p}$ fell toward values expected from the macroscopic $\eta^{-1}$.  This study pushed the technical limits of then-current light-scattering instrumentation. 

\begin{figure}[t] 
\includegraphics{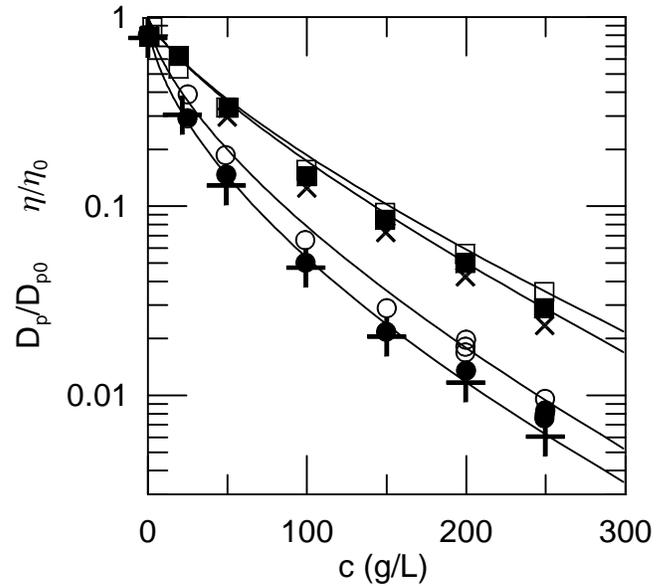}
\caption{\label{figurephillies1989bDp1} $D_{p}/D_{p0}$ of 20 nm (open points) and 230 nm (filled points) radius probes, and $\eta/\eta_{0}$ ($+$, $\times$), with solutions of 70 (squares) and 500 (circles) kDa aqueous dextran, and fits to stretched exponentials (lines), after Phillies, et al.\cite{phillies1989bDp}.}
\end{figure}

Phillies, et al.\cite{phillies1989bDp} observed probe diffusion in aqueous dextran.  Dextran concentrations covered $0 \leq c \leq 250$ g/L, using 9 different dextran samples. Probes were polystyrene spheres with radii 21 and 230 nm; solution viscosities were obtained using thermostatted capillary viscometers.  Multiple tests confirmed that probe scattering completely dominated matrix polymer scattering.  Representative measurements of $D_{p}$ and $\eta$ for two polymer molecular weights appear in Fig.\ \ref{figurephillies1989bDp1}, together with fits to stretched exponentials in $c$.  Except for the smaller spheres in solutions of the very largest dextran, Stokes-Einsteinian behavior was uniformly observed, $D_{p}$ and $\eta^{-1}$ having very nearly the same concentration dependences $\exp(- \alpha c^{\nu})$.  This paper represented the first examination of the dependence of $\alpha$ and $\nu$ on $M_{w}$ for probe diffusion through an extensive series of homologous polymers. 

Phillies, et al.\cite{phillies1987aDp,phillies1989aDp,phillies1997bDp} report three studies of probe diffusion in polyelectrolyte solutions.  Two examine probes in dilute and concentrated solutions of high\cite{phillies1987aDp} and low\cite{phillies1989aDp} partially-neutralized poly-acrylic acid.  One\cite{phillies1997bDp} examines probe diffusion in solutions of not-quite-dilute polystyrene sulfonate.

In Ref.\ \onlinecite{phillies1989aDp}, Phillies, et al.\ examined probes ($19 \leq R \leq 380$ nm) diffusing through aqueous non-neutralized and neutralized low-molecular-weight ($5 \leq M \leq 470$ kDa) poly-acrylic acid.  The parameter space is quite large, with possible effects of concentration, molecular weight, and degree of neutralization, to to mention solution ionic strength.  Phillies, et al.\cite{phillies1989aDp} were only able to sample the behaviors encountered in these systems. 

In 2/3 neutralized 5 kDa PAA, $D/D_{0}$ fell exponentially with increasing polymer concentration, but was independent of salt concentration ($0 \leq I \leq 0.1$ M NaCl) and very nearly independent of probe radius.  Probes in 150 kDa PAA showed a more complex dependence on these parameters: For polymer concentrations 0.1-20 g/L, $D_{p}$ was relatively independent from $I$ for at higher salt concentrations ($I \geq 0.01$ M), but fell by a third to two-thirds as $I$ was reduced from 0.01 M toward zero added salt, the decline being larger at elevated polymer concentration. Probes in largely-neutralized 150 kDa have stretched-exponential dependences on polymer $c$ to good accuracy, for concentrations out to 10 g/L and $D/D_{o}$ as small as 0.25, except that at very low ($c \leq 0.2 g/L$) polymer concentration $D_{p}$ is larger than expected from a fit of a stretched exponential $\exp(-\alpha c^{\nu})$ to the higher-concentration measurements. Careful analysis revealed that at very low concentrations $D_{p} = D_{p0} (1 - a c^{1/2})$ and similarly for $\eta$.  Also, the non-Stokes-Einsteinian behaviors found by Lin and Phillies\cite{lin1982aDp,lin1984aDp,lin1984bDp} in non-neutralized PAA are also seen in partially-neutralized PAAs.

\begin{figure}[t] 
\includegraphics{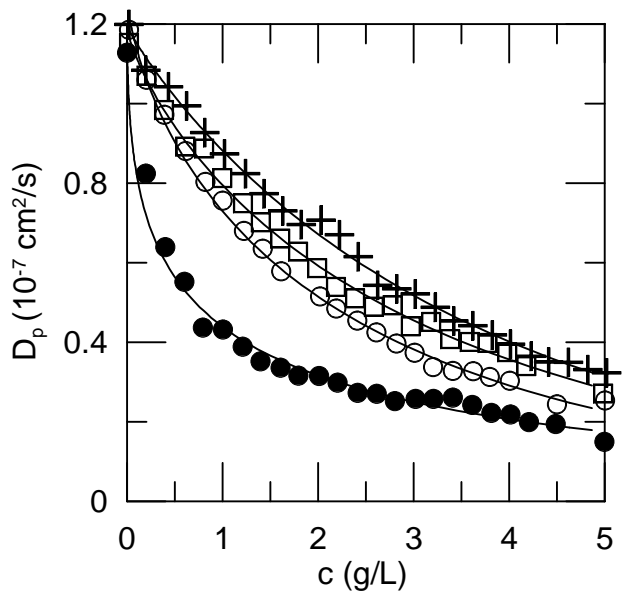}
\caption{\label{figurephillies1987aDp1} $D_{p}$ of 20 nm radius probes in 596 kDa PAA, 60 \% neutralized, at ionic strengths 0 ($\bullet$), 0.01 ($\bigcirc$), 0.02 ($\square$), and 0.1 ($+$) M, after Phillies, et al.\cite{phillies1987aDp}, and fits to stretched exponentials (lines).}
\end{figure} 

\begin{figure}[t] 
\includegraphics{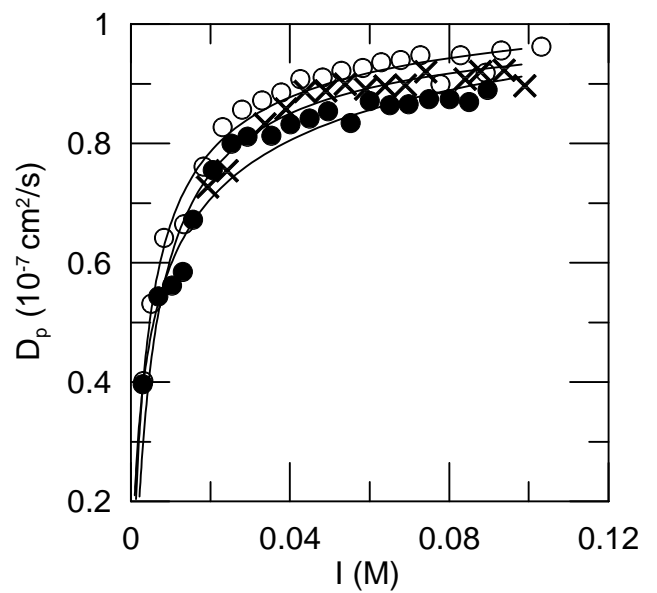}
\caption{\label{figurephillies1987aDp2} $D_{p}$ of 20 nm probes in 1 g/L 596 kDa PAA, at neutralizations 60\% ($\circ$), 85\% ($\times$), and 100\% ($\bullet$), as functions of ionic strength $I$, and fits to stretched exponentials in $I$, after Phillies, et al.\cite{phillies1987aDp}.}
\end{figure}

In a separate study, Phillies, et al.\cite{phillies1987aDp} examined the diffusion of 21, 52, 322, and 655 nm radius carboxylate-modified polystyrene spheres through 596 kDa $M_{w}$ partially-neutralized poly-acrylic acid, using light scattering to determine $D_{p}$.  The dependences of $D_{p}$ and the solution viscosity on $c$, solution ionic strength $I$ ($0 \leq I \leq 0.1$ M), and fractional neutralization $A$ of the polymer ($0.6 \leq A \leq 1$) were determined.  $D_{p}$ has a stretched-exponential dependence on $c$ (Figure \ref{figurephillies1987aDp1}) and a stretched-exponential dependence $\exp(- a I^{\beta})$ on $I$ for $\beta < 0$.  Probes generally diffuse faster than predicted by the Stokes-Einstein equation, the prediction of the Stokes-Einstein equation being approached more nearly with lower polymer neutralization, larger solvent ionic strength, and larger probes.  The dependence of the apparent hydrodynamic radius of the probes on solution pH (and, implicitly, polymer neutralization) is much more pronounced for the smaller 21 and 52 nm spheres than for the larger 322 and 655 nm spheres. If one believed that the diffusion of probe particles, whose sizes exceed all of the length scales of the polymer solution, were governed by the macroscopic viscosity, then for this large-$M$ polymer the longest length scale is at least 52 nm and perhaps larger than 322 nm.

\begin{figure}[tbh] 
\includegraphics{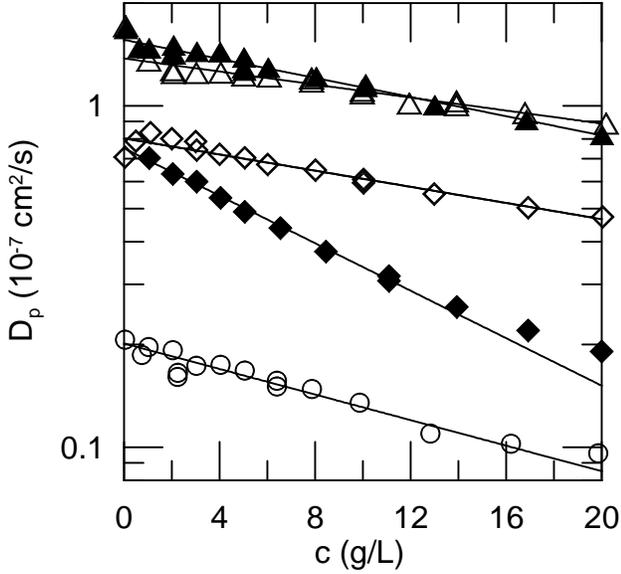}
\caption{\label{figurephillies1997bDp2} $D_{p}$ of 7 ($\vartriangle$), 34 ($\lozenge$), and 95 ($\bigcirc$) nm radius polystyrene spheres in solutions of 178 (open points) and 1188 (filled points) kDa polystyrene sulfonate, and fits to simple exponentials in $c$, after Phillies, et al.\cite{phillies1997bDp}.}
\end{figure} 

Phillies, et al.\cite{phillies1997bDp} examined polystyrene spheres, radii 7, 34, and 95 nm, diffusing through aqueous polystyrene sulfonate. The purpose of the experiments was to determine the initial slope $\alpha$ of $D_{p}$ against polymer $c$ for various polymer $M$.  Polymers had 7 molecular weights with $1.5 \leq M_{w} \leq 1188$ kDa. To minimize low-salt polyelectrolyte anomalies, the solvent included 0.2 M NaCl.  Carboxylate-modified polystyrene spheres are charge-stabilized.  To prevent aggregation on the time scale of the experiments, the solvent included 1 mM NaOh and 0.1 or 0.32 g/L sodium dodecyl sulfate. Figure \ref{figurephillies1997bDp2} shows representative data, namely $D_{p}$ for each sphere size in 178 and 1188 kDa polystyrene sulfonate, and fits to simple exponentials in $c$.  These results are interpreted below.

\begin{figure}[tbh] 
\includegraphics{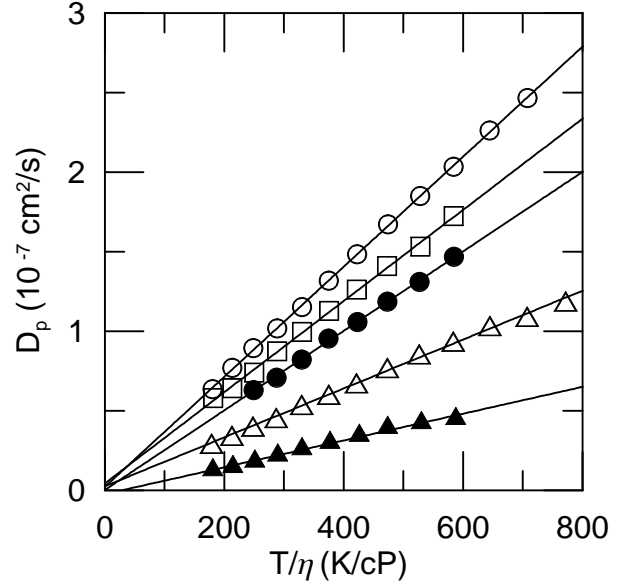}
\caption{\label{figurephillies1991aDp4} $D_{p}$ of 20 nm radius probes as a function of $T/\eta$ in 2/3 neutralized 5kDa polyacrylic acid at polymer concentrations 0 ($\circ$), 5 ($\square$), 25 ($\bullet$), 50 ($\vartriangle$), and 100 ($\blacktriangle$) g/L, and linear fits at each $c$ showing simple Walden's rule behavior, after Phillies, et al.\cite{phillies1991aDp}.}
\end{figure} 

\begin{figure}[hbt] 
\includegraphics{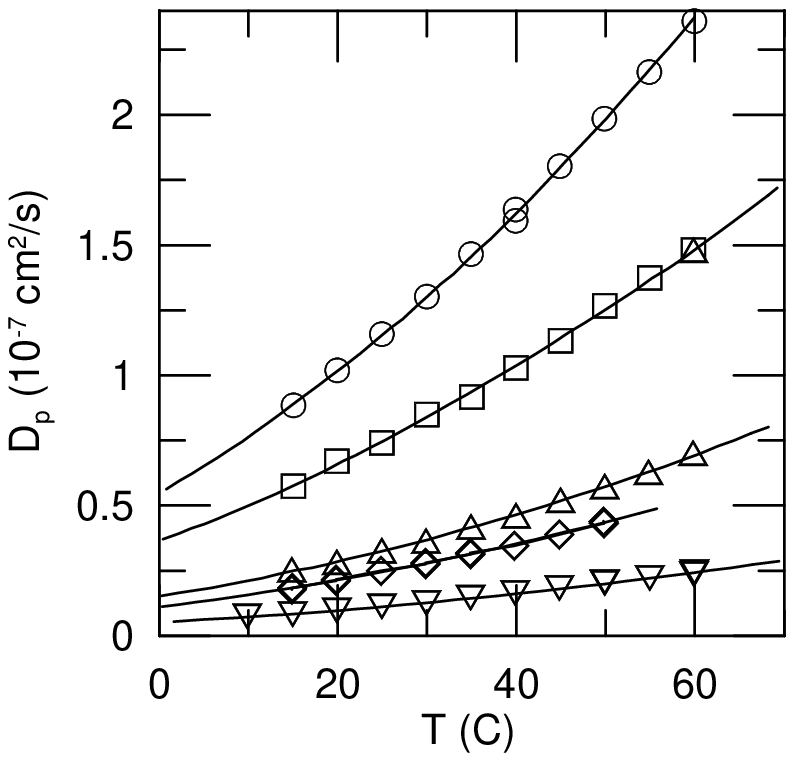}
\caption{\label{figurephillies1992cDp1} $D_{p}$ of 20 nm radius probes as a function of $T$ in 2/3 neutralized 596 kDa polyacrylic acid:0.1 M NaCl: 0.1 wt\% Triton X-100 at concentrations 0 ($\circ$), 3 ($\square$), 6 ($\vartriangle$), 8 ($\lozenge$), and 15 ($\triangledown$) g/L and fits at each $c$ to the Vogel-Fulcher-Tamman equation using the same $T_{g}$ at every $c$, after Phillies, et al.\cite{phillies1992cDp}.}
\end{figure} 

Phillies, et al.\cite{phillies1991aDp,phillies1992cDp,phillies1992eDp} report studies of the temperature dependence of probe diffusion through various polymer solutions. These studies constitute a response to a critique, made to the author at several conferences, of using the stretched-exponential concentration dependence.  Critics complained that data had not been "reduced relative to the glass temperature". To examine this issue, Phillies, et al.\cite{phillies1991aDp,phillies1992cDp,phillies1992eDp} measured $\eta$ and $D_{p}$ for probes in various solutions at a series of temperatures.  The systems studied included (i) 20.4 nm radius carboxylate-modified polystyrene sphere probes in 2/3 neutralized low-molecular-weight 5 and 150 kDa poly-acrylic acid with and without added 0.1 M NaCl\cite{phillies1991aDp}, (ii) the same probes in 2/3 neutralized intermediate-molecular-weight 596 kDa poly-acrylic acid\cite{phillies1992cDp} with or without added 0.1 M NaCL or added surfactant, and (iii) 34 nm nominal radius polystyrene spheres in solutions of dextrans having $11.8 \leq M_{w} \leq 542$ kDa\cite{phillies1992eDp}.  Polymer concentrations reached 100, 45, or 20 g/L, respectively, for the three poly-acrylic acids and 300 g/L for the dextrans.  

For probes\cite{phillies1991aDp} in solutions of low-molecular weight polyacrylic acid, the temperature dependence of of $D_{p}$ seen in Figure \ref{figurephillies1991aDp4} is entirely explained by the temperature dependence of the solvent viscosity $\eta_{s}$ and Walden's rule $D_{p} \sim T/\eta_{s}$. Furthermore, at each polymer concentration $\eta/\eta_{s}$ is very nearly independent from $T$.  This result includes data for 2-65$^{o}$ C and fourteen polymer concentrations at two polymer molecular weights. There was no residual temperature dependence of $D_{p}$ to be explained by glass temperature issues.  The notion that $D_{p}$ is sensitive to $T-T_{g}$ other than through $\eta_{s}$ is rejected by this data. On the other hand, if hydrodynamic interactions between the polymer chains and the probes were primarily responsible for controlling $D_{p}$, it would be reasonable for $D_{p}(T)$ to depend linearly on the solvent viscosity, as is observed.  Note that the chain monomer mobility is also reasonable expected to scale linearly with $\eta_{s}$, so the observations of Reference \onlinecite{phillies1991aDp} are equally consistent with reptation-type dynamic models.

\begin{figure}[bht] 
\includegraphics{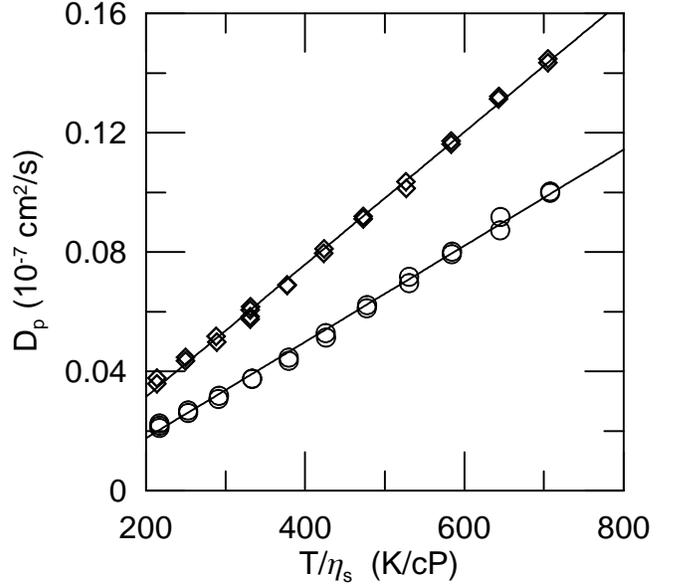}
\caption{\label{figurephillies1992eDp1} $D_{p}$ of 34 nm radius probes against $T/\eta_{s}$ in aqueous 200 g/L 83.5 kDa dextran ($\bigcirc$) and in 100 g/L 542 kDa dextran ($\square$), showing the linear dependence of $D_{p}$ on $T/\eta_{s}$, after Phillies, et al.\cite{phillies1992eDp}.}
\end{figure} 

In ref.\ \onlinecite{phillies1992cDp}, Phillies, et al.\ examined probe diffusion through solutions of intermediate-molecular-weight poly-acrylic acid, comparing the temperature dependence of $\eta$ and $D_{p}$ with Walden's rule and with the Vogel-Fulcher-Tamman equation.  For the solution viscosity, this equation is
\begin{equation}
      \eta = \eta_{0} \exp(-\frac{B}{T-T_{g}}).
      \label{eq:Dpvogel} 
\end{equation}
Here $\eta_{0}$, $B$, and $T_{g}$ are material-dependent parameters; $B$ is sometimes replaced with $\bar{B} T_{g}$.
Walden's rule was followed accurately by all of these data.  The VFT equation uniformly describes the temperature dependence of the measurements accurately.  Fits of measurements covering all $c$ studied to eq.\ \ref{eq:Dpvogel}, using the same $T_{g}$ but different $B$ at different $c$, gave excellent results as seen in Figure \ref{figurephillies1992cDp1}.  These data thus reject the possibility of adjusting $D_{p}$ at different $c$ for a hypothesized variation in $T-T_{g}$ with $c$.

Separately, Phillies, et al.\cite{phillies1992cDp} also fit their data at each polymer concentration to 
\begin{equation}
     D_{p}  = D_{p0} ( 1 + A_{1} \frac{T}{\eta} + A_{2} (\frac{T}{\eta})^{2}+ \ldots ),
     \label{eq:DpTetapoly}
\end{equation}
both with $A_{2} = 0$ forced and with $A_{2}$ as an additional free parameter.  Adding $A_{2}$ as a further free parameter had no effect on the RMS error in the fits, indicating that there was no non-linear dependence of $D_{p}$ on $T/\eta$.  Phillies, et al.\ conclude that their data were consistent with hydrodynamic models in which solvent-mediated forces are the dominant forces between polymer chains, and were equally consistent with reptation-type models in which the solvent viscosity determines the monomer friction constant.

Phillies and Quinlan\cite{phillies1992eDp} used light scattering spectroscopy to obtain the $T$-dependence of $D_{p}$ of 34 nm radius carboxylate-modified polystyrene spheres in solutions of various dextran fractions, $11.8 \leq M_{w} \leq 542$ kDa at concentrations up to 300 g/L.  Solution viscosities were also measured. Representative measurements appear in Figure \ref{figurephillies1992eDp1}.  Measurements of $D_{p}$ were fit to a modified Vogel-Fulcher-Tamman equation 
\begin{equation}
      D_{p} = D_{p0} T \exp(-\frac{B}{T-T_{g}}).
      \label{eq:Dpvogelmod} 
\end{equation}
and to eq \ref{eq:DpTetapoly}.  A very weak deviation from Walden's rule was observed, $D_{p}$ increasing  with increasing $T$ faster than expected from the viscosity. The deviations, which are at most twice the random scatter in the measurements, increase with increasing $c$ but are independent of $M_{w}$.  Phillies and Quinlan note that the deviations are consistent with a slight change in solvent quality with increasing dextran monomer concentration.  Equation \ref{eq:Dpvogelmod}, with $T_{g}$ the same for all solutions, fits almost all measurements to within 2\% rms error.  The apparent glass temperature of water:dextran solutions, as inferred from the modified VFT equation, is independent of dextran molecular weight and dextran concentration.  The measurements here serve to exclude any hypothesis that the strong concentration dependence of $D_{p}$ arises from a strong concentration dependence of $T_{g}$, namely (i) $T_{g}$ is independent of $c$, and, alternatively, (ii) after removing the temperature dependence of $\eta_{s}$, there is next-to-no remnant $T$-dependence of $D_{p}$ available to be interpreted as a variation in $T-T_{g}$.

Roberts, et al.\cite{roberts2001aDp} used PFGNMR to examined a model liquid filled-polymer system, formed from silica nanoparticles suspended in monodisperse poly(dimethylsiloxane).  Silica particles had diameters 0.35 and 2.2nm; polymers had molecular weights 5.2 and 12.2 kDa, with $M_{w}/M_{n}$ of 1.07 and 1.03, respectively.  These are actually not probe measurements; the volume fractions of probes and matrix polymers were both always substantial. $D_{s}$ of the small silica particles and the 5.2 kDa polymer in a mixture both fall linearly with increasing polymer volume fraction for polymer volume fraction $\phi$ in the range 0.2 to 0.95.  In contrast, in a mixture of the larger polymer and larger spheres, $D_{s}$ of the polymer rises and $\eta$ of the mixture decreases with increasing sphere concentration.

Shibayama, et al.\cite{shibayama1999aDp} examine the diffusion of polystyrene probe particles in poly(N-isopropylacrylamide) gels and solutions using QELSS.  The solution spectra are bimodal, with both relaxation rates scaling linearly in $q^{2}$.  A faster, relatively concentration-independent mode corresponds to motion of the polymer chains, while a slower mode corresponds to probe diffusion.  As a result of the preparation method, solutions at each concentration involved polymers having a different average molecular weight. Over the observed range of polymer concentrations, a reduced probe diffusion coefficient $D_{p}/M^{0.6}$ changed by several orders of magnitude. Shibayama, et al.\ propose that the probe diffusion coefficient should scale with concentration and polymer molecular weight as $D_{p}/D_{po} \sim (c/c*)^{-1.75}$, based on the corresponding prediction in Shiwa\cite{shiwa1987aDp} for polymer self-diffusion.  Spectra of probes in true crosslinked gels were also bimodal, but with increasing crosslinker concentration the amplitude of the probe mode is suppressed to zero.

\begin{figure}[tbh] 
\includegraphics{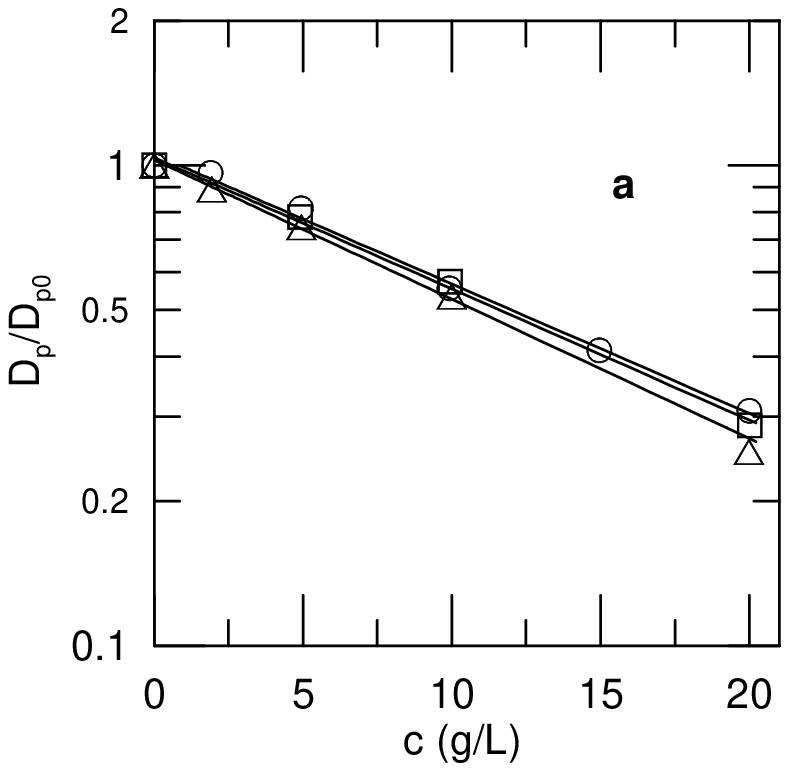}
\includegraphics{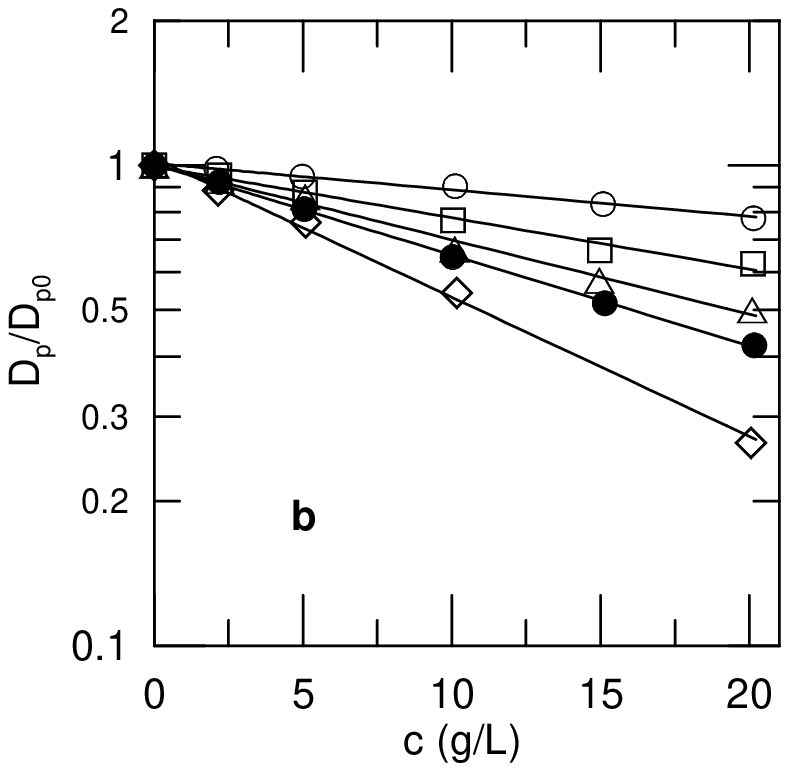}
\caption{\label{figureturner1976aDp} $D_{p}/D_{p0}$ of (a) 93 ($\bigcirc$), 183($\square$), and 246 ($\bigtriangleup$) nm butadiene spheres in aqueous 2 MDa dextran, and (b) 246 nm spheres in solutions of    20 ($\bigcirc$), 70 ($\square$), 150 ($\bigtriangleup$), 500 ($\bullet$), and 2000 ($\lozenge$) kDa dextran, all as functions of dextran concentration, following Turner and Hallett\cite{turner1976aDp}.}
\end{figure} 

Turner and Hallett\cite{turner1976aDp} used QELSS to measure the diffusion of carboxylated styrene butadiene spheres, diameters 93, 183, 246 nm, in solutions of dextrans having nominal molecular weights 20, 70, 150, 500, and 2000 kDa at polymer concentrations up to 20 g/l.  As seen in Figure \ref{figureturner1976aDp}, the normalized diffusion coefficient $D_{p}/D_{po}$ was very nearly insensitive to probe diameter, but changed more than three-fold as polymer concentration $c$ was increased.  To good approximation, $D_{p}/D_{po} = \exp(-ac)$.  $D_{p}$ at fixed $c$ depends appreciably on dextran molecular weight, $a$ increasing with increasing $M$.  The microviscosity inferred from $D_{p}$ agreed well with the viscosity measured with a rotating drum viscometer.

Ullmann and Phillies\cite{ullmann1983aDp} made a study of the diffusion of polystyrene latex spheres through polyethylene oxide: water. These data are discussed with the following paper.  They used the macroscopic $\eta$ to compute effective hydrodynamic radii for their PSL sphere probes (radii 21-655 nm), finding that $r_{h}$ had a complex dependence on polymer concentration. Addition of 0.01\% of the nonionic surfactant Triton X-100, which suppresses polymer binding, eliminated the complexity.  Probe spheres in surfactant-containing mixtures showed a much simpler behavior, namely the apparent hydrodynamic radii of the probes fell smoothly with increasing polymer $c$. The degree of failure of the Stokes-Einstein equation increased markedly with increasing probe radius.  By using measurements on probes in PEO:Triton X-100 to supply calibrating factors, Ullmann and Phillies quantitated the substantial degree of polymer adsorption by probes in surfactant-free solutions.
  
\begin{figure}[tbh] 
\includegraphics{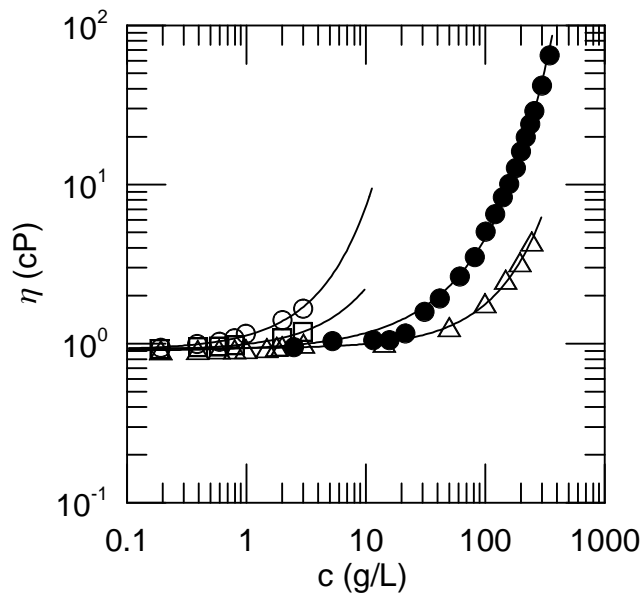}
\caption{\label{figureullmann1985aDp1} $\eta$ of aqueous 7.5 ($\vartriangle$), 18.5 ($\bullet$), 100 ($\square$), and 300 ($\bigcirc$) kDa PEO and stretched-exponential fits, after Ullmann, et al.\cite{ullmann1985aDp}.}
\end{figure} 

\begin{figure}[tbh] 
\includegraphics{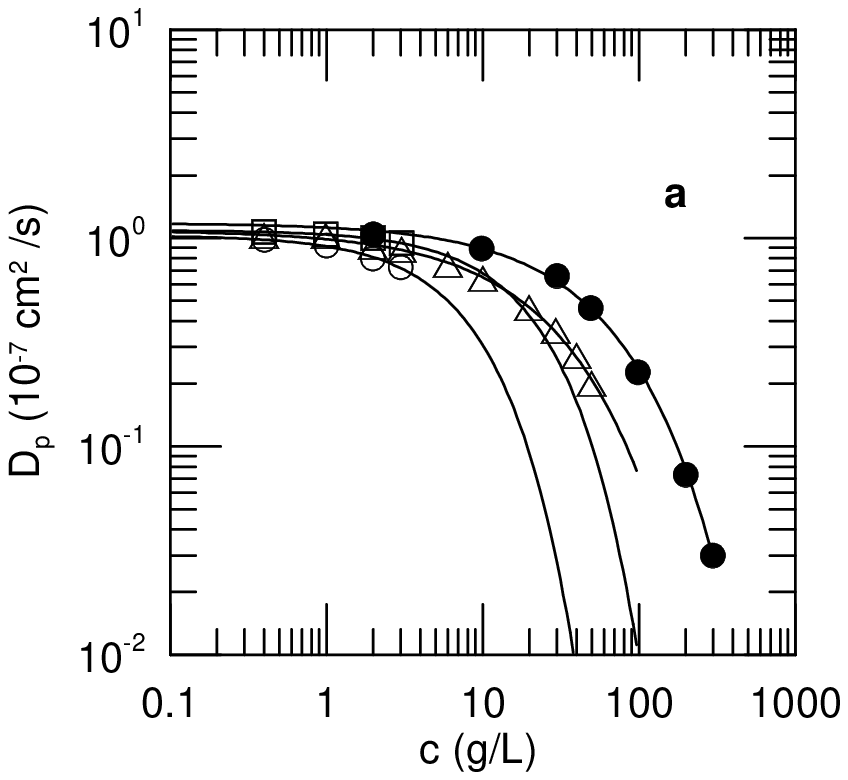}
\includegraphics{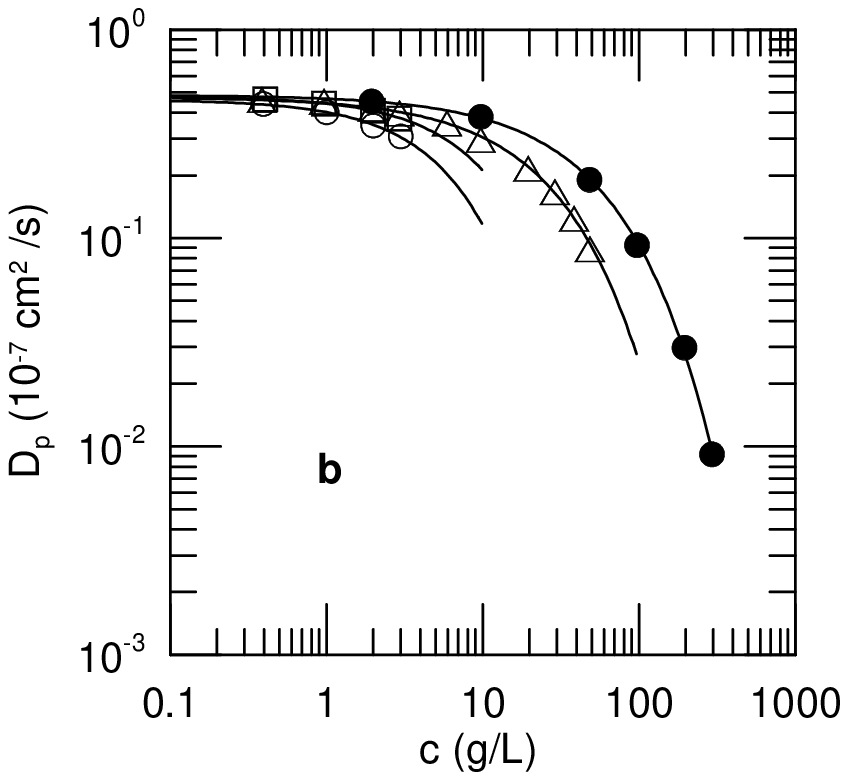}
\caption{\label{figureullmann1985aDp2} $D_{p}$ of (a) 20.8, (b) 51.7, (c) 322, and (d) 655 nm PSL in  aqueous 7.5 ($\bullet$), 8 ($\vartriangle$), 100 ($\square$), and 300 ($\bigcirc$) kDa PEO and stretched-exponential fits, after Ullmann, et al.\cite{ullmann1985aDp}.}
\end{figure} 

\begin{figure}[tbh] 
\includegraphics{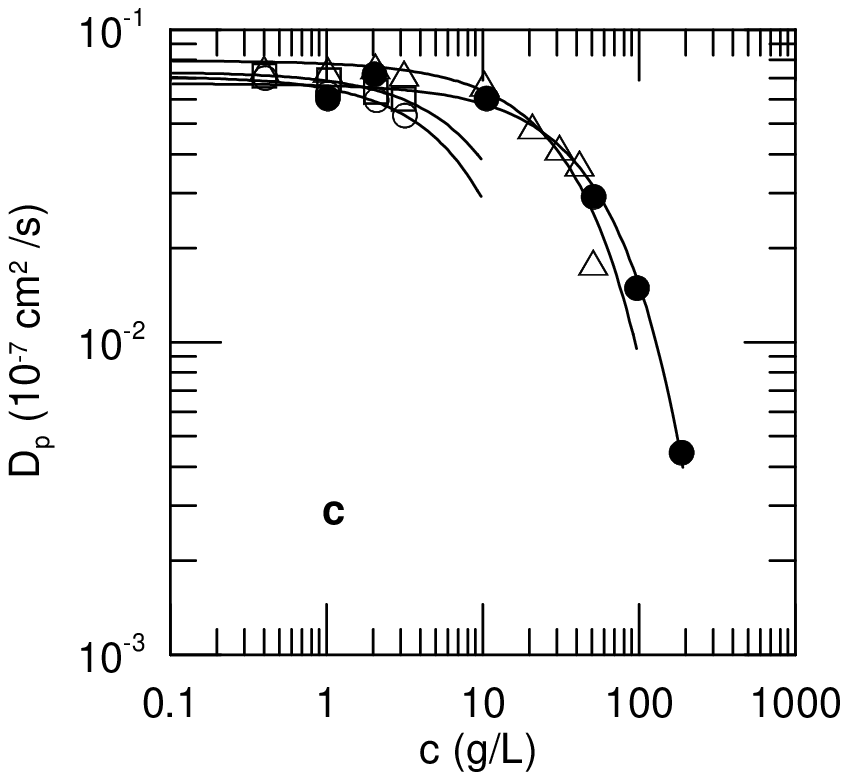}
\includegraphics{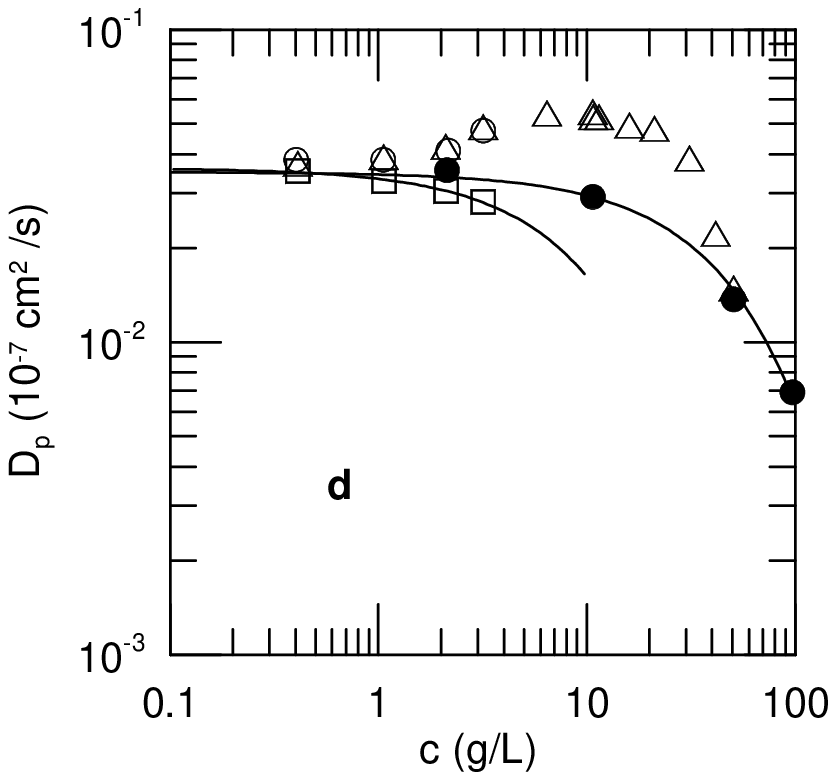}
\end{figure} 

Ullmann, et al.\cite{ullmann1985aDp} extended  Ullmann and Phillies\cite{ullmann1983aDp} to study viscosity (Figure \ref{figureullmann1985aDp1}) and  diffusion (Figure \ref{figureullmann1985aDp2}) of 20.8, 51.7, 322, and 655 nm diameter carboxylate-modified PSL spheres in solutions of 7.5, 18.5, 100, and 300 kDa polyethylene oxides.  $D_{p}$ is substantially modified by the addition of Triton X-100, which is believed to suppress polymer binding by the spheres.  $D_{p}$ follows a stretched exponential in polymer concentration, except that 655 nm diameter spheres in the 18.5 kDa polymer show reentrant behavior, with $D_{p}$ first climbing 50\% above $D_{o}$ and then declining markedly. The relationship between $D_{p}$ and $\eta$ is complex in the lower-$M$ polymers; at large $M$, $D_{p}$ follows a stretched exponential in polymer concentration.  $D_{p}/D_{o}$ depends relatively weakly on probe radius.

\begin{figure}[tbh] 
\includegraphics{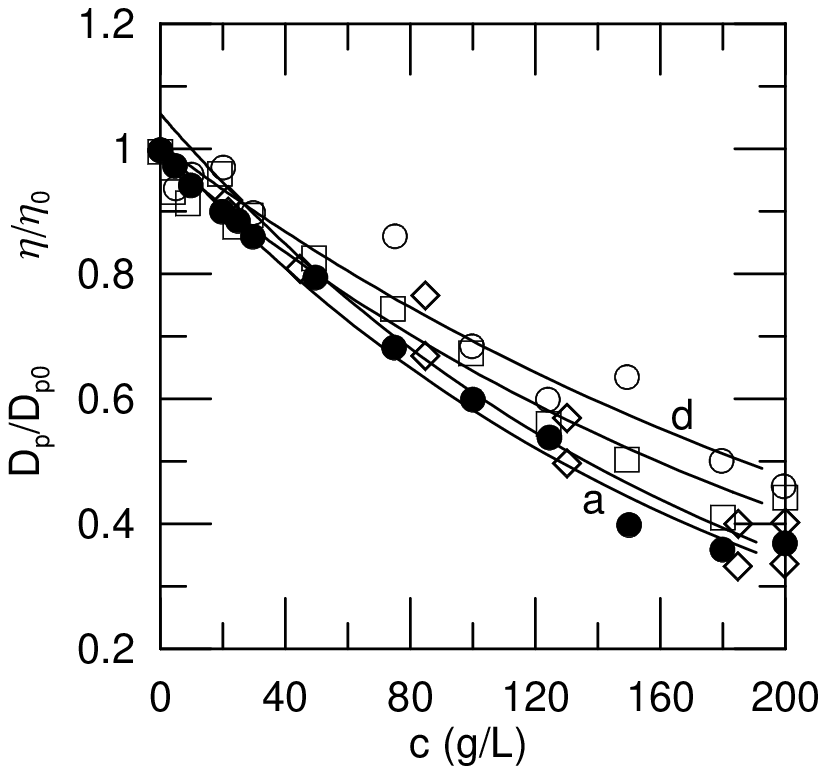}
\caption{\label{figureullmann1985bDp} Solutions of bovine serum albumin, showing their viscosity ($\bullet$) and self-diffusion coefficient  ($\lozenge$) (from refs.\ \onlinecite{keller1971aDp,kitchen1976aDp}), and $D_{p}$ of 322 ($\square$) and 655 ($\circ$) nm polystyrene spheres, and fits (lines a-d, respectively) to stretched exponentials in protein $c$, from data of Ullmann, et al.\cite{ullmann1985aDp}.}
\end{figure} 

Ullmann, et al.\cite{ullmann1985bDp} studied with QELSS the diffusion of 52, 322, and 655 nm radius polystyrene spheres in solutions of bovine serum albumin (BSA) in 0.15M NaCl, pH 7.0.  They compared $D_{p}$ with $\eta$ as determined using capillary viscometers, as seen in Figure \ref{figureullmann1985bDp}.  $D_{p}$ of the two larger spheres had a stretched-exponential dependence on $c$.  The Stokes-Einstein relation failed, $D_{p} \eta$ of the 322 and 655 nm spheres increasing with increasing $c$.  $D_{p}$ of the 52 nm spheres showed re-entrant behavior, $D_{p}$ at first increasing above its value in pure solvent and then returning to the values expected from $\eta$.  At very small $c$, $D_{p}$ of the 52 nm spheres showed a local minimum.  Dilution experiments showed that the minimum arose from aggregation of partially-protein-coated spheres due to the BSA in solution.  When spheres were diluted from concentrated protein solution to more dilute protein solution, $D_{p}$ of the 52 nm spheres returned linearly to its zero-$c$ value, with no sign of aggregation effects.   

\begin{figure}[tbh] 
\includegraphics{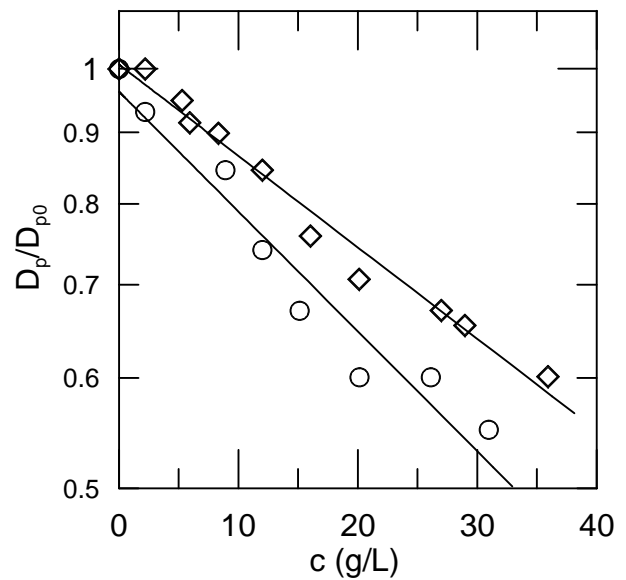}
\caption{\label{figurewattenbarger1992aDp1} Probe diffusion coefficient\cite{wattenbarger1992aDp} of bovine serum albumin in DNA solutions containing 0.01 ($\circ$) or 0.1 ($\lozenge$) M NaCl, and fits to stretched exponentials, after Wattenbarger, et al.\cite{wattenbarger1992aDp}.}
\end{figure} 

Wattenbarger, et al.\cite{wattenbarger1992aDp} used FRAP to examine the diffusion of bovine serum albumin (BSA) through solutions of a 160 base-pair DNA at DNA concentrations 2-63 g/L and NaCl concentrations 0.01 and 0.1M.  DNA molecules had lengths ca.\ 56 nm, identified as being approximately 1 persistence length. As seen in Figure \ref{figurewattenbarger1992aDp1}, $D_{p}$ was approximately exponential in $c$.  Increasing the solution ionic strength increases $D_{p}/D_{po}$, especially at larger DNA concentrations.

\begin{figure}[tbh] 
\includegraphics{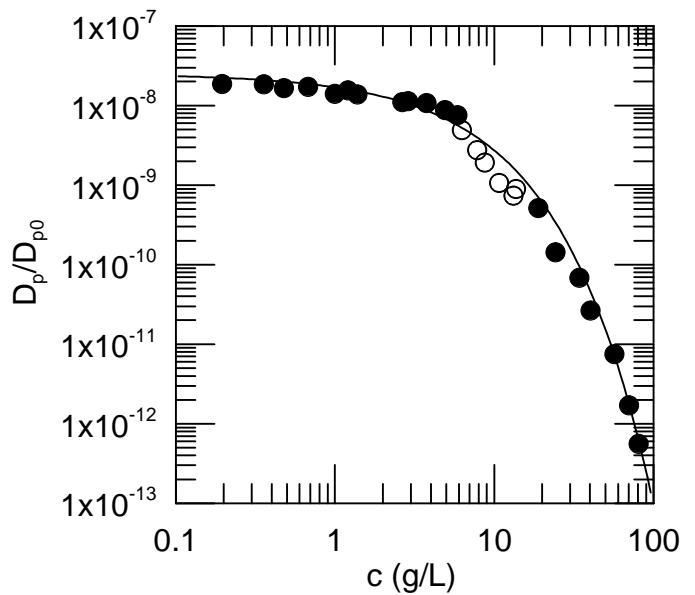}
\caption{\label{figurewon1994aDp1} Diffusion coefficient\cite{won1994aDp} of 200 nm polystyrene spheres in 1.3 MDa PVME: toluene and fit of filled points to a stretched exponential, after Won, et al.\cite{won1994aDp}.}
\end{figure} 

Won, et al.\cite{won1994aDp} report on the diffusion of 200 nm radius PSL spheres in solutions of 1.3 MDa poly(vinylmethylether): toluene using QELSS, as shown in Figure \ref{figurewon1994aDp1}. Additional measurements of $D_{p}$ made with forced Rayleigh scattering were in very good agreement with the QELSS data.  PVME concentrations reached up to 100 g/l, i.e., $c[\eta]$ up to 36.  QELSS spectra were generally unimodal; a small amplitude slow mode was sometimes seen.  The decline in $D_{p}$ with increasing $c$ was followed over more than four orders of magnitude.  The product $D_{p} \eta/D_{po} \eta_{o}$ shows re-entrant behavior, first climbing to nearly 3.5, and then falling back to 1.0 by 10 g/L PVME.  A comparison of PSL spheres in this system with spheres in polystyrene:dimethylformamide found that $D/D_{o}$ is to first approximation a universal function of $c [\eta]$, except for the concentrations at which re-entrant departure from Stokes-Einstein behavior is found.  Tracer diffusion of linear and star polystyrene molecules through PVME solutions has also been studied extensively\cite{lodge1986aDp,lodge1987aDp,lodge1989aDp,wheeler1987aDp,wheeler1989aDp,won1993aDp}.

\begin{figure}[tbh] 
\includegraphics{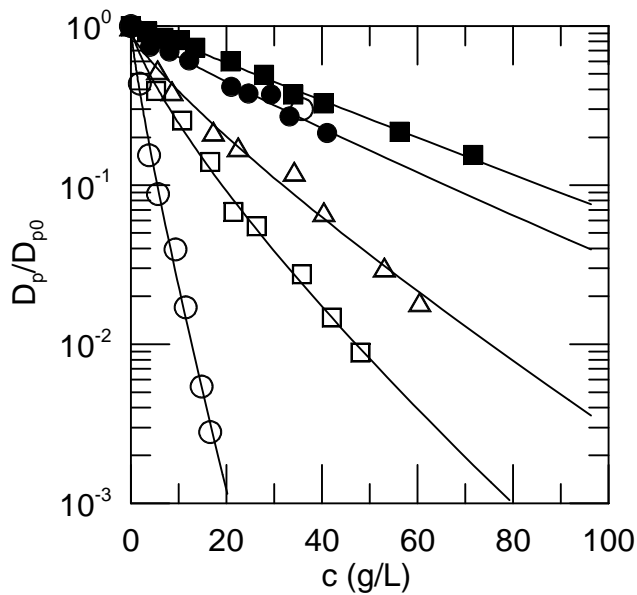}
\caption{\label{figurezhou1989aDp1} Diffusion coefficient of 160 nm silica spheres in solutions of 57 ($\bigcirc$), 95 ($\square$), 182 ($\bigtriangleup$), 610 ($\bullet$), and 1900 ($\blacksquare$) kDa polyisobutylene, and fits to stretched exponentials, after Zhou and Brown\cite{zhou1989aDp}.}
\end{figure} 

Zhou and Brown\cite{zhou1989aDp} measured $D_{p}$ of stearic-acid-coated silica spheres in polyisobutylene (PIB): chloroform using QELSS, as seen in Figure \ref{figurezhou1989aDp1}.  Comparison was made with PFGNMR determinations of polymer self-diffusion and QELSS measurements of polymer mutual diffusion.  Polymer molecular weights ranged from 57 kDa to 4.9 MDa.  Sphere motion was diffusive, with a $q^{2}$-dependent linewidth, leading to diffusion coefficients seen in Figure \ref{figurezhou1989aDp1}.  Inverse Laplace transform of the spectra clearly resolved a weak peak due to polymer mutual diffusion from an intense peak taken to reflect probe motion.  Probe diffusion coefficients very nearly tracked the solution viscosity, $D_{p}\eta$ increasing very slightly over the range of concentrations studied.


Many of the above studies compare $D_{p}$ with the macroscopically measured viscosity $\eta$.  A series of studies confirm that in viscous simple liquids the $D_{p}$ measured with light scattering tracks $\eta$, no matter whether $\eta$ is changed by varying the temperature or the composition of the liquid.  Phillies and Fernandez\cite{phillies1983aDp} confirmed $D_{p}$ of polystyrene spheres in water depends linearly on $T/\eta$, showing that PSL spheres do not swell or contract with changing $T$.  Phillies\cite{phillies1981aDp} measured $D_{p}$ of bovine serum albumin and 45.5 nm radius PSL in water:glycerol and the same PSL spheres in water:sorbitol for temperatures 5-50 C and concentrations as high as 86 wt\% glycerol and 65 wt\% sorbitol. Phillies found probe diffusion in these mixtures followed accurately the Stokes-Einstein equation, even with $\eta$ as large as 1000 cP. Wiltzius and van Saarloos\cite{wiltzius1991aDp}, using 19, 45.5, and 107.5 nm polystyrene spheres in 99.5\% glycerol for $280 \leq T \leq 320$ K, found to high precision that the temperature dependence of $D_{p}$ was independent of probe size, contrary to an earlier literature report\cite{kiyachenko1985a}.  Phillies and Clomenil\cite{phillies1992dDp} examined the possibility that this disagreement in the literature was related to temperature-dependent changes in the shape of the spectrum $S(q,t)$.  They made high precision measurements of the spectral first and second cumulants $K_{1}$ and $K_{2}$ of optical probes in water:erythritol, water:glycerol, and neat glycerol, their new result being that $K_{2}$ in these systems was independent from $T$.  Their measurements serve to reject any hypothesis that the earlier literature disagreement arose because different spectral fitting processes accommodated differently to temperature-dependent variations in a not-quite-exponential line shape of $S(q,t)$.  From these simple-liquid studies, failures of the Stokes-Einstein equation in viscoelastic polymer solutions cannot to be ascribed simply to the large zero-shear viscosity of those solutions.  

\subsubsection{Hydroxypropylcellulose}

The translational diffusion of probes in hydroxypropylcellulose has been studied extremely extensively.  Earlier work was assisted by the practical matters that HPC samples are available with a wide range of molecular weights, dissolve well in water, and have interesting thermodynamic properties, including a transition from good to theta solvent behavior with increasing temperature as well as a liquid-to-liquid-crystal transition at extremely high polymer concentration.  Much of the later work was motivated by the viscosity measurements of Phillies and Quinlan\cite{phillies1995aDp}, who observed $\eta(c)$ for solutions of 300 kDa, 1 MDa, and 1.3 MDa HPC over wide concentration ranges.  Phillies and Quinlan\cite{phillies1995aDp} found an unusual viscometric transition not observed in most polymers studied with optical probe diffusion. At first, work focused on finding evidence corroborating the reality of the transition. Later work focused on the search for a  physical interpretation of this transition. 

Phillies and Quinlan\cite{phillies1995aDp} established that up to a transition concentration $c^{+}$ the concentration dependence of $\eta$ of HPC solutions follows accurately a stretched exponential 
\begin{equation}
     \eta = \eta_{o} \exp(-\alpha c^{\nu}).
    \label{eq:etastretch}
\end{equation}
At concentrations $c > c^{+}$, the concentration dependence of $\eta$ follows equally accurately a  power law
\begin{equation}
     \eta = \bar{\eta} c^{x}.
     \label{eq:etapower}
\end{equation}
Here $\alpha$ is a scaling prefactor, $\nu$ and $x$ are scaling exponents, and $\eta_{o}$ and $\bar{\eta}$ are dimensional prefactors. To avoid model-dependent phrasings, Phillies and Quinlan termed the $c<c^{+}$ and  $c>c^{+}$ domains the 'solutionlike' and 'meltlike' regimes. At the transition, $\eta$ is continuous.  Furthermore, the transition is analytic: Both the functions and their first derivatives are continuous.  There is no crossover regime: one form or the other describes $\eta(c)$ at every concentration.  Systematic reviews of the literature\cite{phillies1988aDp,phillies1992rDp} showed that such transitions happen in some but not all polymer solutions.  In natural units the transition concentration $c^{+} [\eta]$ for different systems has a wide range of different values. HPC:water is distinguished by the very low polymer concentration at which the transition occurs, namely $c^{+} [\eta] \approx 4$ rather than the more typical $c^{+} [\eta] \approx 35$ or larger.  The nonuniversality of the transition concentration, the lack of a matching transition in $D_{s}(c)$,  indeed, the complete lack of a transition in many polymer systems, all show that the viscosity transition here should not be confused with a hypothesized transition to reptation dynamics.  

For clarity, the remaining literature in this Section is discussed in chronological order. Work by Phillies and Quinlan had been preceded by extensive optical probe diffusion studies of HPC:water solutions:
Brown and Rymden\cite{brown1986aDp} used QELSS to examine 72 nm radius PSL spheres diffusing in solutions of hydroxyethylcellulose (HEC), hydroxypropylcellulose (HPC), carboxymethylcellulose (CMC), and poly(acrylic acid)(PAA).  The focus was polymer-induced cluster formation, evidenced by the form of substantial decreases in $D_{p}$ and increases in the second spectral cumulant as seen at very low (0.001 g/g) concentrations of HEC and HPC. These changes were substantially reversed by the addition of 0.15\% Triton X-100.  $D_{p}$ of spheres was reduced by the addition of small amounts of fully-charged pH 9 CMC, but addition of TX-100 had no effect in CMC solutions. Brown and Rymden also examined sphere diffusion in nondilute polymer solutions. Relatively complex dependences of $D_{p}$ on concentration were suppressed by the addition of TX-100.  In the presence of TX-100, simple stretched-exponential concentration dependences were observed, but the second spectral cumulant still increased with increasing polymer concentration.

\begin{figure}[tbh] 
\includegraphics{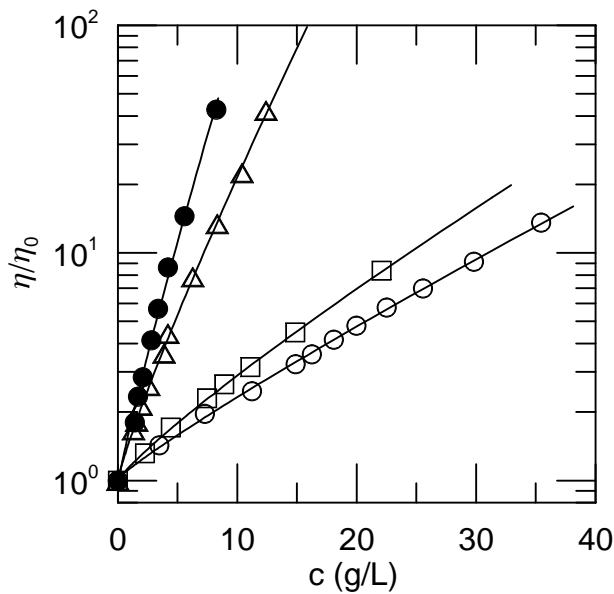}
\caption{\label{figureyang1988aDp1} Viscosity\cite{yang1988aDp} of $M_{w}$ 110 ($\bigcirc$), 140 ($\square$),  450 ($\vartriangle$), and 850 ($\bullet$) kDa HPC, and fits of $\eta(c)$ to exponential concentration dependences.}
\end{figure} 

\begin{figure}[tbh] 
\includegraphics{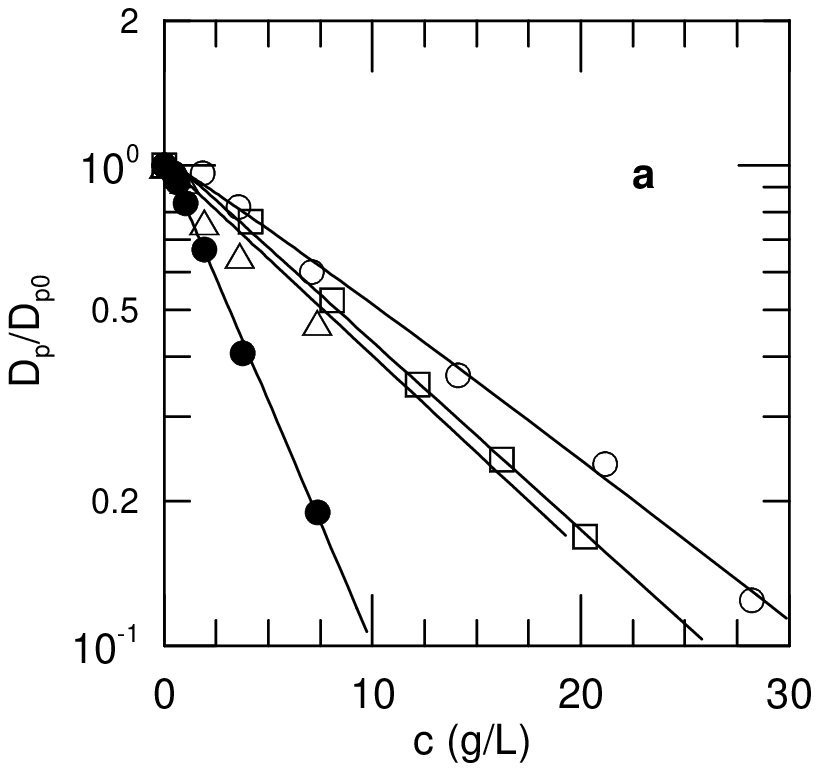}
\includegraphics{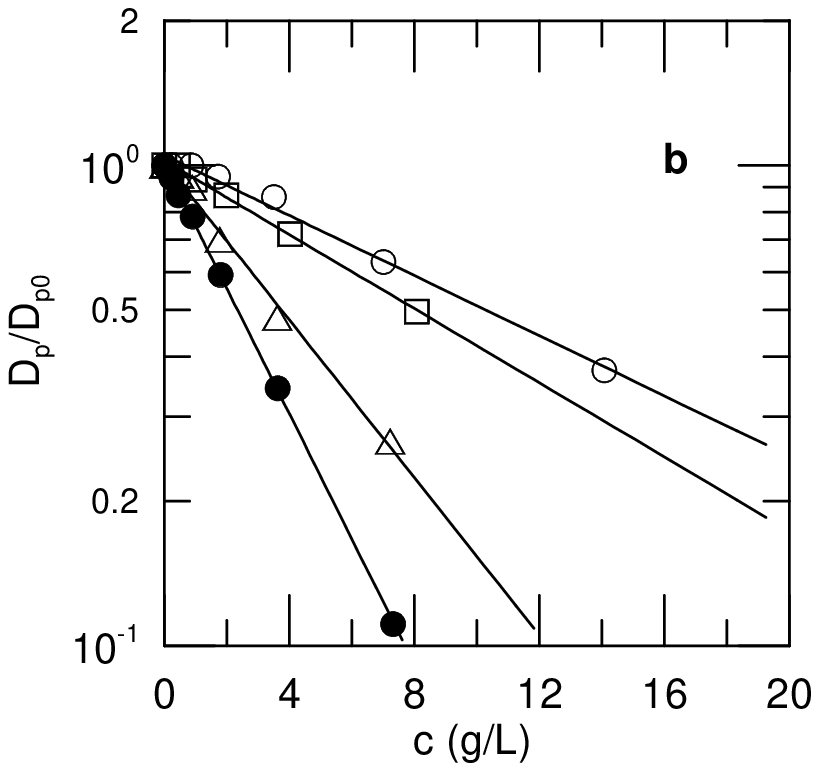}
\end{figure} 

\begin{figure}[tbh] 
\includegraphics{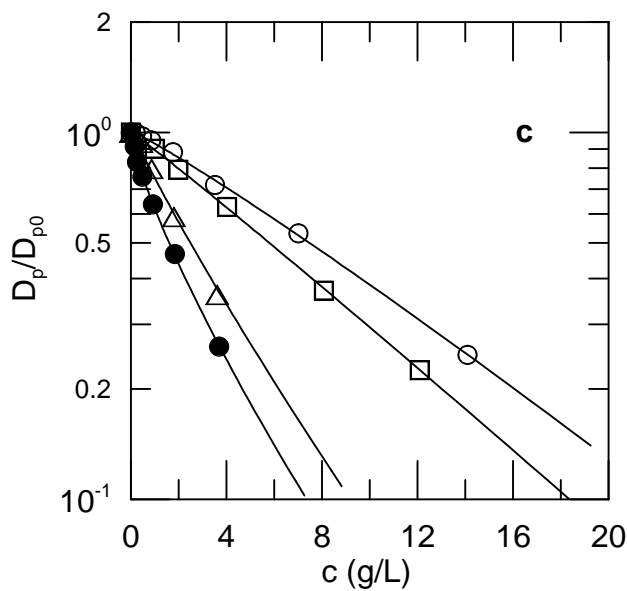}
\caption{\label{figureyang1988aDp2} Diffusion coefficient of (a) 120, (b) 210, and (c) 350 nm diameter PSL in solutions of $M_{w}$ 110 ($\bigcirc$), 140 ($\square$), 450 ($\vartriangle$), and 850 ($\bullet$) kDa HPC, and fits of $D_{p}(c)$ to stretched-exponential concentration dependences, after Yang and Jamieson\cite{yang1988aDp}.}
\end{figure} 

Yang and Jamieson\cite{yang1988aDp} examined the diffusion of polystyrene spheres in HPC, measuring the shear viscosity $\eta$ (Figure \ref{figureyang1988aDp1}), and using QELSS to obtain $D_{p}$  (Figure \ref{figureyang1988aDp2}).  Polymer molecular weights based on static light scattering measurements were 110, 140, 450, and 850 kDa; sphere diameters were 120, 210, and 350 nm.  Polymer concentrations extended up to 40 g/l.  Yang and Jamieson found $D_{p}$ to be independent of polymer concentration over a range of not-quite-zero concentrations.  At larger concentrations, $D_{p}$ fell as a stretched exponential in $c$. With Triton X-100 added to suppress polymer binding, $D_{p}(c)$ was larger than expected from $\eta$.  The ratio of the macroscopic and microscopic viscosities $\eta/\eta_{\mu}$ increases with increasing $c$ and $M$. A modest dependence of $\eta/\eta_{\mu}$ on probe $R$ is seen, small probes having the smaller $\eta_{\mu}$. 

Russo, et al.\cite{russo1988bDp} report on HPC solutions with PSL spheres as optical probes.  Polymer $M$ was 292 kDa; sphere radii were 40 and 90 nm.  Polymer concentrations were  $\leq 4$ wt\%. Light scattering spectra and zero-shear viscosities were determined at a series of polymer concentrations.  Addition of Triton X-100 dramatically markedly reduced the hydrodynamic radii determined using QELSS.  $D_{p}$ of nonaggregated spheres very nearly tracks $\eta^{-1}$, with a modest 12\% deviation for the larger spheres; however, only two measurements with $\eta/\eta_{0} > 10$ were reported.

Mustafa and Russo\cite{mustafa1989aDp} applied QELSS to measure the diffusion of 90.7 nm radius polystyrene latex through solutions of high-molecular-weight (1 MDa) hydroxypropylcellulose at concentrations up to 1.5 g/L. Comparison was made between solutions to which Triton X-100 had been added in order to eliminate polymer binding to the probes, and solutions to which Triton X-100 had not been added.  QELSS spectra were profoundly bimodal, the fast mode being half or a quarter as intense as the slow mode.  $q^{2}$ measurements confirmed that the slow mode is diffusive, and indicated that the fast mode is probably diffusive.  For both modes, the Stokes-Einstein equation fails, probe motion being up to two-fold faster than expected from the macroscopic solution viscosity.

\begin{figure}[tbh]
\includegraphics{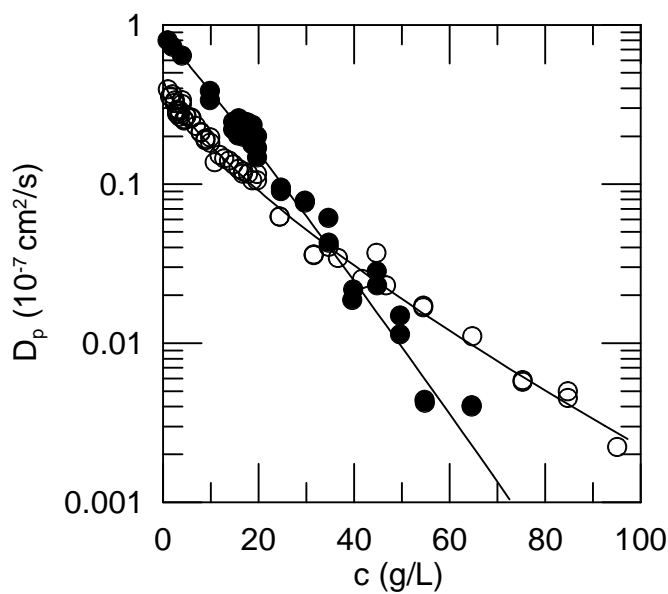}
\caption{\label{figurephillies1993bDp1} $D_{p}$ of 67 nm polystyrene spheres in aqueous 139 kDa HPC at 10 ($\circ$) and 39 ($\bullet$) C, after Phillies and Clomenil\cite{phillies1993bDp}.}
\end{figure}

Phillies and Clomenil\cite{phillies1993bDp} made a detailed study of the diffusion of 67 nm PSL spheres in 139 kDa HPC at 10$^{o}$C (good-solvent conditions) and 41$^{o}$C (near-theta-solvent conditions), as seen in Figure \ref{figurephillies1993bDp1}.  At each temperature, $D_{p}$ depended on polymer concentration via a stretched exponential $\exp(-\alpha c^{\nu})$.  The exponent $\nu$ was 1 under theta conditions but 0.74 under good-solvent conditions, showing probe diffusion can be sensitive to solvent quality.  The dependence of $\nu$ on solvent quality is not entirely useful as a test of models for polymer dynamics: The observed dependence is equally predicted by the Langevin-Rondelez\cite{langevin1978aDp}, Altenberger\cite{altenberger1986aDp}, and hydrodynamic scaling\cite{phillies1987dDp,phillies1988cDp} treatments.

\begin{figure}[tbh] 
\includegraphics{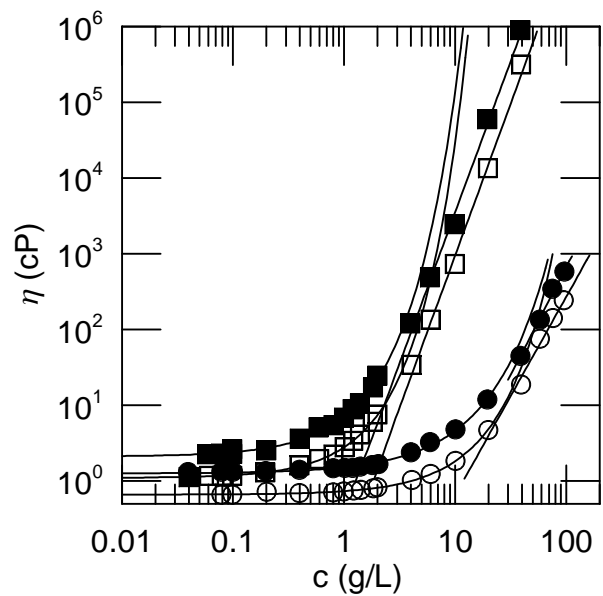}
\caption{\label{figurephillies1993aDp1} $\eta$  of 139 ($\circ$) and 1280 kDa ($\square$, displaced two-fold vertically for clarity) hydroxypropylcellulose at temperatures 10 C (filled points) and 39 C (open points) and fits to stretched exponentials (smooth curves) and power laws (straight lines), after Phillies, et al.\cite{phillies1993aDp}.}
\end{figure} 

\begin{figure}[tbh] 
\includegraphics{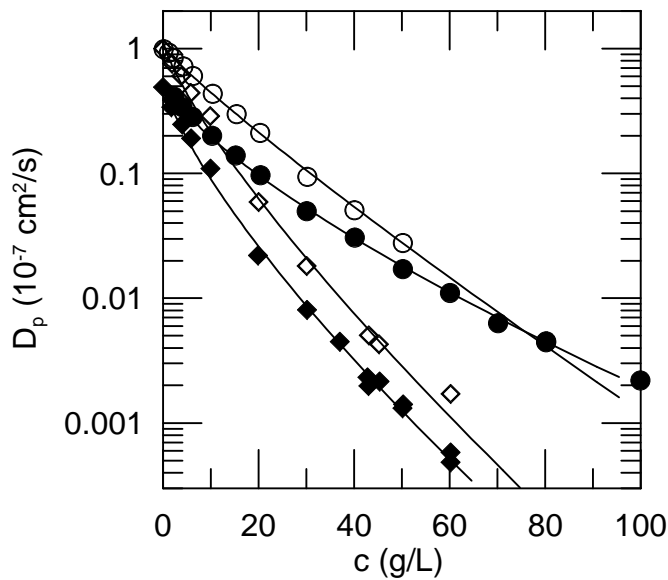}
\caption{\label{figurephillies1993aDp2} $D_{p}$ of 67 nm polystyrene spheres in aqueous 139 ($\circ$) and 415 kDa ($\square$) hydroxypropylcellulose at 10 C (filled points) and 39 C (open points) and fits to stretched exponentials (smooth curves), after Phillies, et al.\cite{phillies1993aDp}.}
\end{figure} 

\begin{figure}[thb] 
\includegraphics{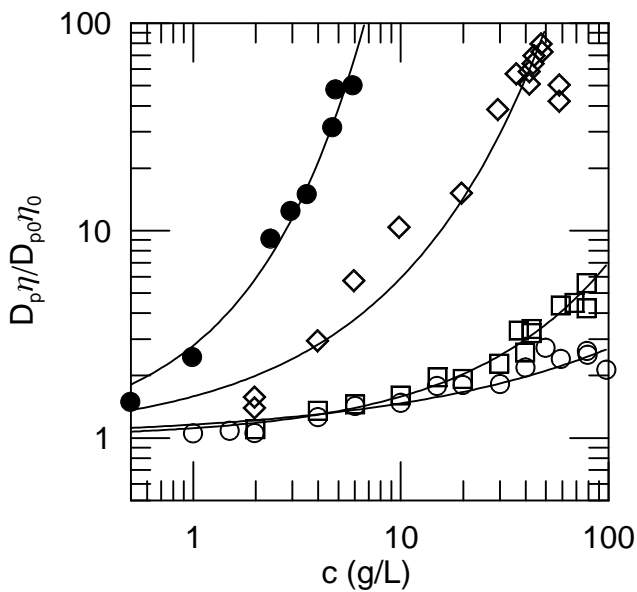}
\caption{\label{figurephillies1993aDp9} $D_{p} \eta/ D_{p0} \eta_{0}$ of 67 nm polystyrene spheres in aqueous 139 ($\circ$), 146 ($\square$), 415 ($\lozenge$), and 1280 ($\blacksquare$) kDa  HPC at 10 C, after Phillies, et al.\cite{phillies1993aDp}.}
\end{figure} 

Phillies, et al.\cite{phillies1993aDp} measure diffusion of 67 nm PSL spheres via QELSS in 139, 146, 415, and 1280 kDa hydroxypropylcellulose (HPC) at temperatures 10, 25, and 39$^{o}$ C, for polymer concentrations up to 100 g/l.  Shear viscosities were measured for the same solutions.  Aqueous HPC is a good-solvent system at 10C and approaches a pseudotheta transition at 41 C, so varying the temperature allows a variation in solvent quality.
$\eta$ (Figure \ref{figurephillies1993aDp1}) was found to have a stretched-exponential concentration dependence up to an apparent transition concentration $c^{+}$, and an apparent power-law concentration dependence at higher concentrations.  These observations led to the much more systematic study of Phillies and Quinlan\cite{phillies1995aDp}. Spectra were analysed with Koppel's method of cumulants.  Individual spectra also fit well to stretched exponentials in time.  Over the observed concentration regime, $D_{p}$ fell by 1.5 to 2.5 orders of magnitude, as seen in Figure \ref{figurephillies1993aDp2}. Over the same range of $c$, $D_{p} \eta/D_{p0} \eta_{0}$ increased by factors of several to several hundred, the increase being largest for the largest-$M$ polymer (as seen in Figure \ref{figurephillies1993aDp9}), and larger at 10$^{o}$C than at 39$^{o}$C. In fits to Williams-Watts spectral forms, $\theta$ and $\beta$ fall as stretched exponentials in $c$, the decline being more dramatic at larger polymer $M$.  $S(q,t)$ was obtained with an experimental S/N of 300-1000.  A careful search was made for a very fast spectral mode that might have perturbed cumulant fits; none was found. Later work from the same group found that $S(q,t)$ for probes in HPC:water is generally bimodal.  Ref.\ \onlinecite{phillies1993aDp} was found to have examined the more prominent shorter-lived relaxational mode.

\begin{figure}[t] 
\includegraphics{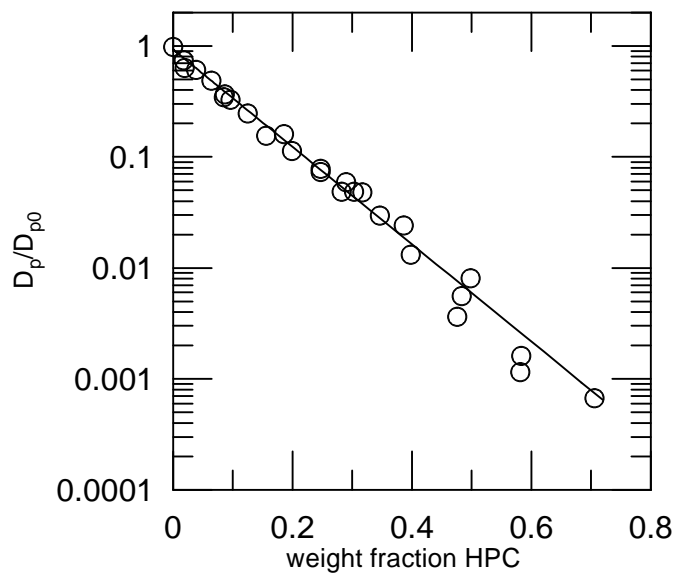}
\caption{\label{figuremust1993aDp5} $D_{p}$ of fluorescein dye in hydroxypropylcellulose solutions, showing a simple exponential dependence of $D_{p}$ on $c$, with no discontinuity at the liquid crystal phase transition near weight fraction 0.4, after Mustafa, et al.\cite{mustafa1993aDp}}.
\end{figure} 

Mustafa, et al.\cite{mustafa1993aDp} used fluorescence recovery after photobleaching to examine diffusion of fluorescein dye through 130, 292, and 855 kDa HPC solutions.  $D_{p}$ was measured at various grating spacings and temperatures at HPC concentrations from dilute solution up through the HPC liquid:liquid crystal phase transition.  As seen in Figure \ref{figuremust1993aDp5}, $D_{p}$ decreases approximately exponentially in polymer concentration but is independent of polymer molecular weight.  There is no discontinuity in $D_{p}(c)$ on crossing the HPC liquid:liquid crystal phase transition.

Ngai and Phillies\cite{ngai1996aDp} extend the Ngai-Rendell coupling model\cite{ngai1994aDp} to treat probe diffusion and polymer dynamics in polymer solutions.  This is not an experimental paper; it forces extant experimental data to confront a particular theoretical model, which in the paper was extensively reconstructed to treat the particular experimental methods under consideration.  Ngai and Phillies consider zero-shear viscosities and optical probe diffusion spectra for HPC solutions, extracting from them the Ngai-Rendell model relaxation time $\tau_{o}$ and coupling exponent $n$.  Optical probe spectra can be used to obtain $n$ in four independent ways, namely from $\eta(c)$ and from the concentration, time, and wave-vector dependences of $g^{(1)}(q,t)$.  The four paths from $\eta$ and $g^{(1)}(q,t)$ lead to consistent values for $n$, especially for larger probes and higher polymer concentrations at which the scaling-coupling arguments invoked in the paper are most likely to be valid.

Phillies and LaCroix\cite{phillies1997aDp} used QELSS to study diffusion of four PSL sphere species in aqueous 300 kDa HPC at concentrations up to 15 g/L.  Spheres had diameters 21, 102, 189, and 760 nm.  Field correlation functions were fit to the sum of a weak, fast initial exponential and a stretched exponential $\exp(-\theta t^{\beta})$.  The fast initial exponential was never more than 2.5\% of $g^{(1)}(q,t)$.  $\beta$ and the relaxation pseudorate $\theta$ both had stretched-exponential concentration dependences, the dependences being the stronger for the larger probes.   For the 21 nm spheres, $\theta \eta$ increased with increasing $c$; for the large spheres, $\theta \eta$ was very nearly independent from $c$.  The relaxation time $\tau = \theta^{-1/\beta}$ depends on $c$ less strongly than $\eta$ does, so that for smaller spheres $\tau/\eta$ falls with increasing $c$.  For larger spheres, $\tau /\eta$ is approximately independent of $c$.  However, because these spectra follow $\exp( - (t/\tau)^{\beta})$ with $\beta \neq 1$, it would be erroneous to write $\tau = D q^{2}$ for some $D$.  In consequence, the constancy of $\tau/\eta$ represents pseudo-Stokes-Einsteinian behavior, not Stokes-Einsteinian behavior, for $\tau$.  These results are subsumed in the later studies of Streletzky and collaborators, discussed below.

\begin{figure}[tbh]
\includegraphics{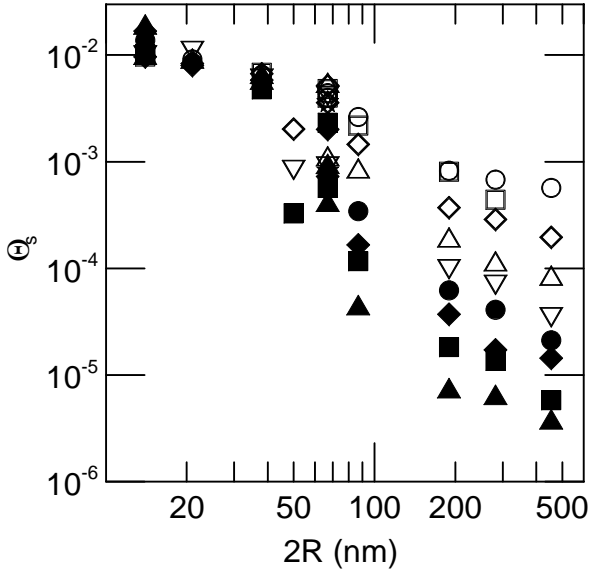}
\caption{\label{figurestreletzky1998aDp1} Sharp mode pseudorate constant versus probe size at HPC concentrations 0 ($\bigcirc$), 0.5($\square$), 1 ($\lozenge$), 2 ($\vartriangle$), 3 ($\triangledown$), 3.75 ($\bigstar$), 4($\bullet$), 5 ($\blacklozenge$), 6 ($\blacksquare$), and 7 ($\blacktriangle$) g/L, showing a demarkation between small and large probe behavior for $R$ between 20 and 30 nm, after Streletzky and Phillies\cite{streletzky1998aDp}.}
\end{figure}

Streletzky and Phillies\cite{streletzky1998aDp} use QELSS to study the diffusion of 14-455 nm diameter PSL spheres in 1MDa hydroxypropylcellulose-water for concentrations $1 \leq c \leq 7$ g/L.  Probe spectra fit well to a sum of two stretched exponential modes $\exp(- \theta t^{\beta})$, characterized in this paper as 'slow' and 'fast'.  The dependences of spectral fitting parameters on probe radius, polymer concentration, wave vector, and solution viscosity were examined.  Probe behaviors fell into two classes, distinguished by probe size.  Figure \ref{figurestreletzky1998aDp1} shows sample results, namely $\theta_{s}$ against probe radius $R$ at different concentrations: for small probes, $\theta_{s}$ is very nearly independent of $c$, while for large probes $\theta_{s}$ decreases sharply with increasing $c$.  The characteristic length separating large and small probes, which appears to be the same at all polymer concentrations, is approximately the size of a complete polymer chain, probes having a diameter less than the polymer hydrodynamic radius being 'small' and probes having a diameter larger than the polymer radius of gyration being 'large'. A similar distinction between small and large probes is seen for $\beta$ of the slow mode, with $\beta \approx 1$ for large probes and $\beta$ decreasing with increasing $c$ for small ($R \leq 35$ nm) probes.  On the other hand, $\theta$ of the fast mode is nearly independent of concentration regardless of probe size. 

Streletzky and Phillies\cite{streletzky1998bDp} present an approach to interpreting the QELSS spectral mode structure for optical probes diffusing in HPC:water, based on their prior work.  They consider only probes in the HPC solutionlike regime, whose phenomenology is sharply divided between 'small' and 'large' probes, small and large being defined relative to the size of the polymer chains.  The crossover from small- to large- probe behavior is found at the same probe radius at all polymer concentrations. They interpret three time-scale-separated modes to be: (i) a large-probe 'slow' mode that appears to reveal particle motion in a viscous medium, (ii) large-probe 'fast' and small-probe 'slow' modes that occur on the same time scale, and appear to be related to chain internal dynamics, and (iii) a small-probe 'fast' mode.  The large-probe 'slow' mode is approximately governed by the solution viscosity, and goes linearly to zero as $q^{2} \rightarrow 0$.  Later work by these authors established that the 'fast' and 'slow' modes were better described as 'sharp' and 'broad', depending on their time stretching exponent $\beta$, because 'fast' and 'slow' do not consistently describe the mode lifetimes. The Ngai scaling-coupling model only works for modes in category (ii), which for both small and large probes is the shorter-lived of the two modes.

\begin{figure}[tbh]
\includegraphics{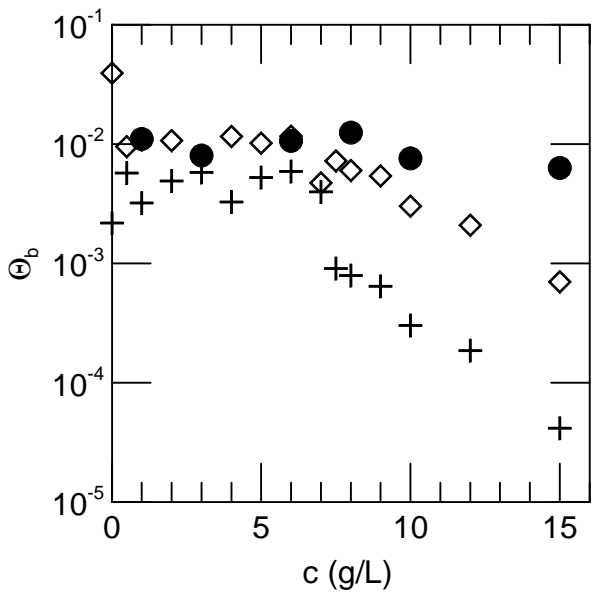}
\caption{\label{figurestreletzky1999aDp1} Broad mode pseudorate constant $\theta_{f}$ against polymer concentration for 50 ($\bullet$), 87 ($\lozenge$), and 189 ($+$) nm probes, showing the change in concentration dependence near $c^{+} \approx 6$ g/L, after Streletzky and Phillies\cite{streletzky1999aDp}.}
\end{figure}

\begin{figure}[tbh]
\includegraphics{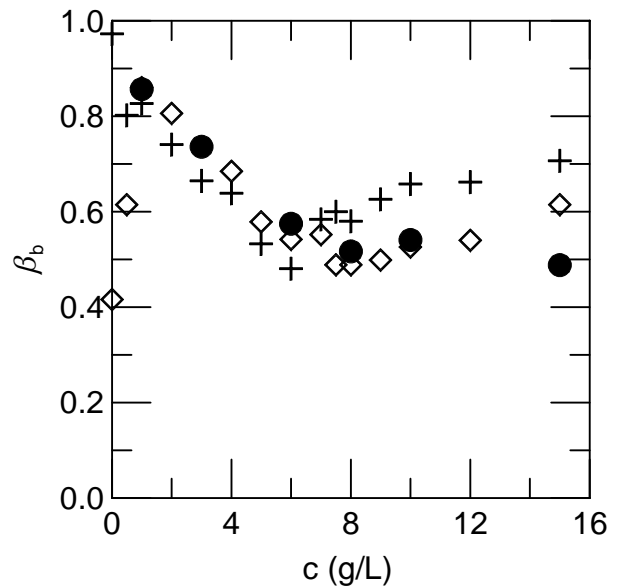}
\caption{\label{figurestreletzky1999aDp2} Broad mode stretching exponent $\beta_{f}$ against polymer concentration for  50 ($\bullet$), 87 ($\lozenge$), and 189 ($+$) nm probes, showing the change in concentration dependence near $c^{+} \approx 6$ g/L, after Streletzky and Phillies\cite{streletzky1999aDp}.}
\end{figure}

Streletzky and Phillies\cite{streletzky1999aDp} compare optical probe diffusion and viscosity measurements on HPC:water.The immediate question treated by  Streletzky and Phillies\cite{streletzky1999aDp} was whether the transition is real, whether the apparent transition at $c^{+}$ is simply an artifact of ref.\ \onlinecite{phillies1995aDp}'s numerical fitting procedures, or as a third alternative whether the stretched exponential behavior is merely a numerical approximant lacking deep physical meaning.  Streletzky and Phillies\cite{streletzky1999aDp}proceeded by using QELSS to measure the diffusion of large (50, 87, and 189 nm diameter) PSL probes in aqueous 1MDa HPC, including concentrations above and below $c^{+}$.  After testing alternatives, spectra were fit to the sum of a fast stretched exponential and a slow pure exponential
\begin{equation}
     g^{(1)}(q,\tau) =  A_{o} (A_{f} \exp(-\theta_{f} c^{\beta_{f}}) + (1- A_{f}) \exp(-\theta c).
     \label{eq:g1qtstretched}
\end{equation}
At $c^{+}$, every single spectral parameter changes its concentration dependence. As seen in Figure \ref{figurestreletzky1999aDp1}, below $c^{+}$, $\theta_{f}$ is independent of $c$; at larger $c$,  $\theta_{f}$ decreases as a stretched exponential in $c$.  As seen in Figure \ref{figurestreletzky1999aDp2}, $\beta_{f}$ declines with increasing $c$ until $c^{+}$ is reached; at higher concentrations $\beta_{f}$  is nearly independent of $c$.  The fast-mode fractional amplitude $A_{f}$ is nearly zero at $c=0$; it climbs with increasing $c$ until it reaches $A_{f} \approx 0.4-0.6$, which it does at a concentration near $c^{+}$.  $\theta$ follows stretched exponentials in $c$; near $c^{+}$, for the 87 and 189 nm probes there is a sudden change in the slopes of the $\theta(c)$ curves. At $c^{+}$, the correlations of $\theta$ and $\theta_{f}$ with $\eta$ both change.  $\theta_{f} \eta$ climbs near-exponentially with increasing $c$ until $c^{+}$ is reached; at larger $c$, $\theta_{f} \eta$ increases weakly with increasing $c$.  The slope $d \theta \eta/dc$ also changes near $c^{+}$, though less markedly than the slope of $\theta \eta$ changes.  $\theta$ and $\theta_{f}$ are both linear in $q^{2}$.  Streletzky and Phillies conclude that the marked changes in the concentration dependences of the optical probe diffusion parameters near $c^{+}$ prove the physical reality of the solutionlike-meltlike transition.

Streletzky and Phillies\cite{streletzky1999cDp} attempt to apply the Ngai-Phillies\cite{ngai1996aDp} coupling/scaling model for probe diffusion to their QELSS spectra\cite{streletzky1998aDp} of mesoscopic probes diffusing in 1MDa HPC.  The model consistently works for one of the two spectral modes but not the other. For large probes (probe $R > R_{g}$ of the polymer), the model works for the shorter-lived stretched-exponential mode, but not for the longer-lived pure exponential mode.  For small spheres, both modes relax as stretched exponentials; the coupling-scaling description works for the shorter-lived mode but not for the longer-lived mode.  In the original paper, modes are called 'fast' or 'slow', not longer- or shorter- lived.  Because 'fast' or 'slow' were identified based on early parts of the slope of $g^{(1)}(q,t)$ rather than the integrated lifetime, and 'fast' and 'slow' do not map one-to-one to short- and long-lived, the original paper identified the modes for which coupling-scaling works as the 'fast' mode of small probes and the 'slow' mode of large particles, leaving it less clear which physics was applicable when.

Streletzky and Phillies eventually demonstrated that these results are consistent with the earlier papers of Phillies and Lacroix\cite{phillies1997aDp} and Ngai and Phillies\cite{ngai1996aDp}.  The demonstration consists of showing that Phillies and Lacroix had only been able to study the shorter-lived mode of probes in 300 kDa HPC, for which the Ngai-Phillies model\cite{ngai1996aDp} is correct; in both systems, it is the shorter-lived mode that follows coupling-scaling.

Streletzky and Phillies\cite{streletzky2000aDp} make a systematic presentation of their findings on probe diffusion in aqueous HPC.  Spectra were found to be sums of two stretched exponentials. Their principle results are: (i) probes in the solutionlike $c <c^{+}$ regime, in their scattering behavior, fall into two classes, small and large.  The classes are separated by a well-defined {\em concentration-independent} boundary.  The length scale separating the probe classes is approximately the hydrodynamic radius of a polymer chain, not a hypothesized mesh size $\xi$.  (ii) There are two distinct concentration regimes, solutionlike and meltlike, in the sense of Phillies and Quinlan\cite{phillies1995aDp}.  At the solutionlike-meltlike concentration boundary, parameters describing the spectra of large probes show dramatic changes in their concentration dependences.  (iii) In the solutionlike regime, for large probes $\theta \eta$ depends relatively weakly on polymer concentration.  In the same regime, $\theta \eta$ of small probes and $\theta_{f} \eta$ of all probes increases greatly (typically more than 100-fold over the observed concentration range) with increasing $c$.  Streletzky and Phillies offer an extensive series of physical interpretations that rationalize the summarized data.

Phillies, et al.\cite{phillies2003aDp} extend prior work on HPC:water with a QELSS study of the diffusion of small 35 nm probes in nondilute 1MDa HPC.  They introduce a novel alternative process for characterizing light scattering spectra.  Mindful of the exponential integral
\begin{equation}
    \int_{0}^{\infty} dt \exp(-\Gamma t) = \Gamma^{-1},
    \label{eq:expint}
\end{equation}
they took spectra that could be accurately represented as a sum of exponentials or stretched exponentials, and characterized a spectrum 
\begin{equation}
    g^{(1)}(t) = \int_{0}^{\infty} \ d\Gamma \ A(\Gamma) \exp(- \Gamma t) 
    \label{eq:g1int}
\end{equation}
via its time moments
\begin{equation}
     M_{n} \equiv \int_{0}^{\infty} dt \ t^{n} g^{(1)}(t) = \frac{\gamma(1+n)}{\langle \Gamma^{1+n} \rangle}.
     \label{eq:integraltime}
\end{equation}
Here $\gamma(1+n) = n!$ is the conventional Gamma function, the $M_{n}$ being the time moments.  In general, $M_{n}/n! = \langle \Gamma^{-1-n} \rangle$.  If $A(\Gamma)$ is normalized, $M_{0} = \langle \Gamma^{-1} \rangle$ is the mean relaxation time. For a stretched-exponential mode, the mean relaxation time is
\begin{equation}
     M_{0} \equiv \int_{0}^{\infty} dt \ \exp(-\theta t^{\beta}) = \gamma(1+1/\beta)/\theta^{1/\beta}.
     \label{eq:thetaaverage}
\end{equation}

A direct computation of the time moments is sensitive to noise at large $t$.  Phillies, et al.\ instead obtained analytic forms that accurately fit their spectra, and obtained the moments as integrals of the analytic forms.  Phillies, et al.\ conclude that time moments provide a substantially less complex characterization of relaxational modes than do alternative parameterizations of the same modes of the same system.

\begin{figure}[tbh]
\includegraphics{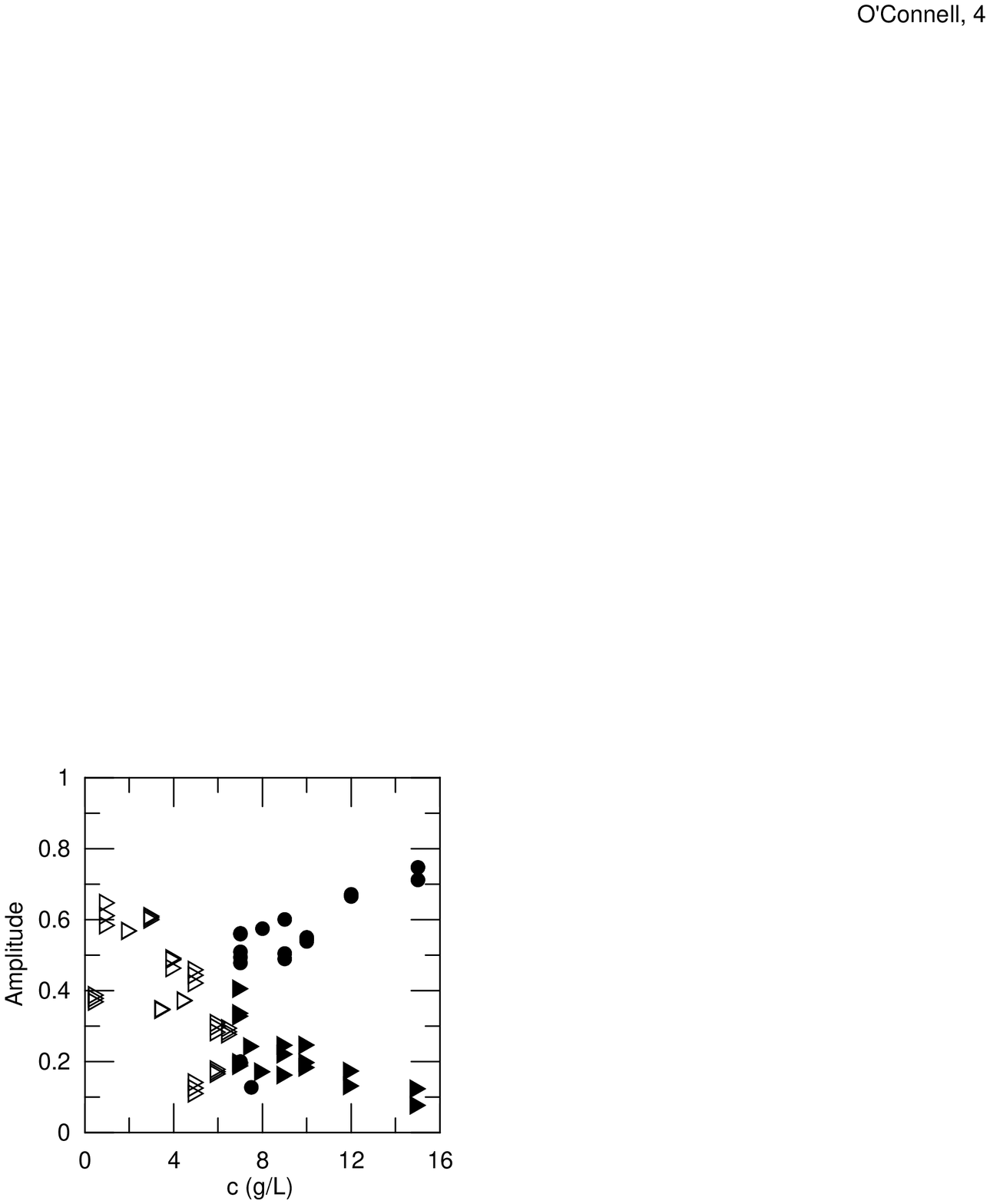}
\caption{\label{figureoconnell2005aDp4}Fractional amplitudes of the fast (triangles) and slow (circles)
spectral modes of HPC: water as functions of polymer concentration.  Open points:
two-stretched-exponential fits; filled points:
three-stretched-exponential fits.  The intermediate-mode amplitude
has very nearly the same concentration dependence as the fast-mode
amplitude, after O'Connell, et al.\cite{oconnell2005aDp}.}
\end{figure}

\begin{figure}[tbh]
\includegraphics{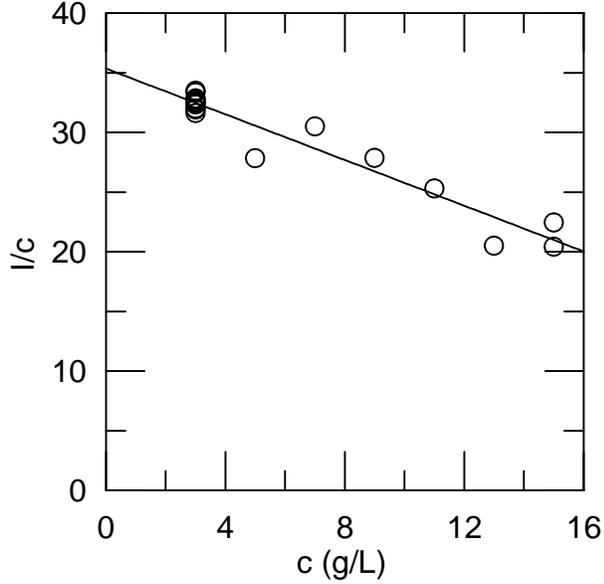}
\caption{\label{figureoconnell2005aDp1}   Normalized intensity $I/c$ of aqueous
HPC solutions, after O'Connell, et al.\cite{oconnell2005aDp}.}
\end{figure}

\begin{figure}[tbh]
\includegraphics{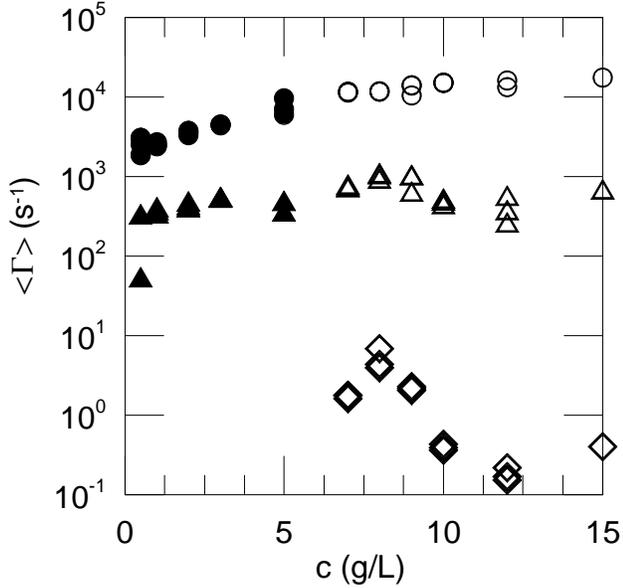}
\caption{\label{figureoconnell2005aDp3}Mean lifetimes of the fast (circles), intermediate (triangles), and slow (diamonds) modes of HPC in water, from two-stretched-exponential fits (filled points), and three-stretched-exponential fits (open points), after O'Connell, et al.\cite{oconnell2005aDp}.}
\end{figure}

O'Connell, et al.\cite{oconnell2005aDp} report using QELSS to examine the mode structure of probe-free aqueous 1 MDa HPC solutions.   Below approximately 6 g/L HPC, QELSS spectra of HPC:water are well characterized as a sum of two stretched-exponential spectral modes having approximately equal amplitudes. At 6 g/l, which is the crossover concentration for the viscometric solutionlike-meltlike transition, an emergent slow mode takes up half the total scattering intensity (Figure \ref{figureoconnell2005aDp4}).  The slow mode does not arise from aggregates: With increasing $c$, $I/c$ shows (Figure \ref{figureoconnell2005aDp1}) a slow decrease  continuous through the transition. This study demonstrates both the utility of measuring simultaneously optical probe spectra and matching spectra of probe-free systems, and of measuring absolute scattering intensities. 

Phillies, et al.\cite{phillies2003aDp} found that  small probe spectra shift at $c^{+}$ from a bimodal to a trimodal form.  The new very slow mode, which appears at the viscometric transition concentration $c^{+}$, has relaxation times of several seconds.  The two modes found below $c^{+}$, described by the authors as the sharp and broad modes, partially merge at $c^{+}$.  Above $c^{+}$, the original sharp and broad modes have very nearly the same mean relaxation times, but very different widths $\beta$.  Phillies, et al.\ emphasize that their data show that while some probe and polymer relaxations happen on similar time scales, they are not the same: the probe and polymer modes have concentration dependences of opposite sign, so the probe and polymer modes on the same time scale do not simply reflect probes and polymers  moving in unison.

Ref.\ \onlinecite{phillies2003aDp} presents results confirming Streletzky and Phillies's\cite{streletzky1998aDp} prior interpretation that HPC solutions have a dominant, concentration-independent characteristic dynamic length scale, namely the radius of a polymer chain, which for this species is $R \sim 50$ nm.  In particular: (i) There are distinct small-probe and large probe phenomenologies, with the division between small and large probes being about 50 nm, the same at all polymer concentrations. (ii) For small probes, the relative amplitude of the sharp and broad modes depends markedly on scattering vector $q$ with a crossover near $q^{-1} \approx 70$ nm. (iii) The mean relaxation rate of the small-probe broad mode increases markedly near $q^{-1} \approx 50$ nm.  (iv) The probe intermediate mode becomes much more strongly concentration dependent when it becomes longer-lived than the polymer intermediate mode, which occurs at the rheological transition concentration only if the probe is the correct size, namely 50 nm.

Phillies, et al.\cite{phillies2003aDp} propose that the dynamic changes at $c^{+}$ reflect the formation of a Kivelson glass\cite{kivelson1994aDp} in these systems. In a Kivelson glass, one encounters long-lived equilibrium structures with frustrated growth that prevents them from growing into a crystalline solid.  The concentration of these structures increases at low temperature, leading to glass formation.  In polymer solutions studied by Phillies, et al., solution properties are very nearly athermal at accessible temperatures. Increasing concentration replaces decreasing temperature as the factor driving structure formation.

\subsection{Probe Rotational Diffusion}

By examining the polarized and depolarized light scattering spectra at a series of scattering angles, or by measuring the depolarized spectrum at zero scattering angle, it is possible to infer the rotational diffusion coefficient $D_{r}$ of probe particles that depolarize scattered light.

Camins and Russo\cite{camins1994aDp} used zero-angle depolarized light scattering to study the rotational diffusion of poly(tetrafluoroethylene) colloidal probes in gelling polyacrylamide.  The probes were optically-anisotropic prolate ellipsoids of modest (3:1) axial ratio, the effective hydrodynamic radii being $140 \pm 40$ nm, dispersed in polyacrylamide systems undergoing gelation.  During gelation, the depolarized relaxation slows and broadens markedly, changing from a near-exponential relaxation to a stretched exponential with exponent $\beta \approx 0.5$.  Interpretation of spectra in terms of mobile and immobile probe fractions suggests that most probes become immobile during the gelation process, but at least some are at least somewhat free to move.

Cheng and Mason\cite{cheng2003aDp} discuss the extension of optical probe diffusion methods to study rotational diffusion in 900 kDa poly(ethylene oxide).  The optical probe was a micron-size $\alpha$-eicosane microdisk whose angular orientation could be tracked with time resolution of $0.3 \mu$s using microscopy.  Because the time-dependent orientation and angular displacement $\theta(t)$ of the disk polar axis were tracked directly, Cheng and Mason were able to determine $\langle (\theta(t))^{2} \rangle$ and estimate $\mid G^{*} (\omega) \mid$ for a single polyethyleneoxide solution.

\begin{figure}[tb] 
\includegraphics{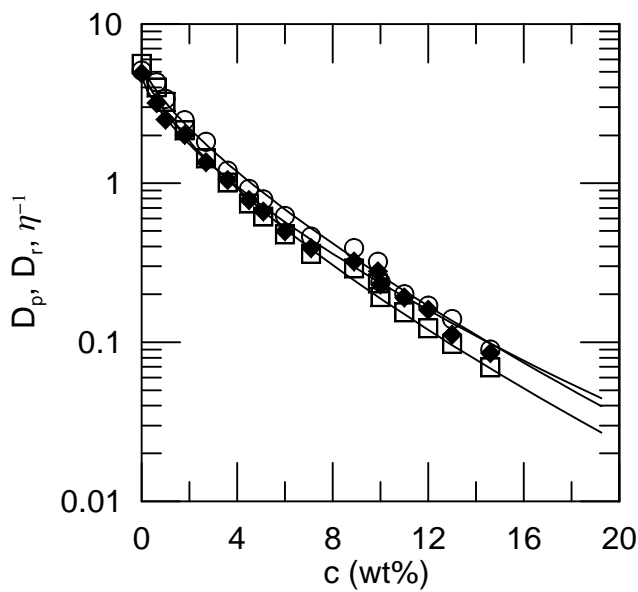}
\caption{\label{figurecush2004aDpX} 
$D_{p}$ ($\bigcirc$) and $D_{r}$ ($\square$) of tobacco mosaic virus in 647 kDa dextran solutions, and $\eta^{-1}$ ($\blacklozenge$) of those solutions, after Cush, et al.\cite{cush2004aDp}, and (lines) fits to stretched exponentials in $c$.}
\end{figure}

\begin{figure}[tb] 
\includegraphics{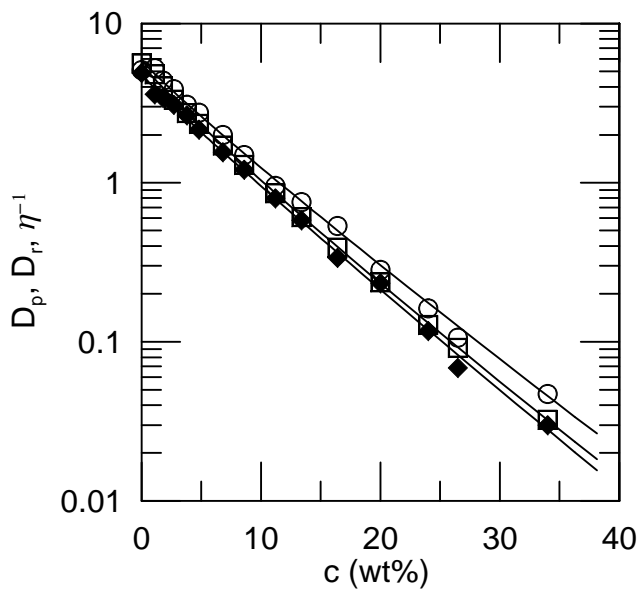}
\caption{\label{figurecush2004aDpY} 
$D_{p}$ ($\bigcirc$) and $D_{r}$ ($\square$) of tobacco mosaic virus in 428 kDa ficoll solutions, and $\eta^{-1}$ ($\blacklozenge$) of those solutions, after Cush, et al.\cite{cush2004aDp}, and (lines) fits to stretched exponentials in $c$.}
\end{figure}

Cush, et al.\cite{cush2004aDp} used depolarized light scattering to study the rotational diffusion of tobacco mosaic virus (TMV) in dextran and ficoll solutions.  TMV is a rod of length $L \approx 300$ nm and diameter $d \approx 18$ nm \cite{cush2004aDp}.  This study replaces the prior results from the same research group\cite{cush1997aDp} using essentially the same method, on precisely the same system, by applying corrections due to the optical activity of dextran and ficoll and more subtle instrumental issues.  The concentration dependences of $D_{r}$ and $D_{p}$ of TMV in 670 kDa dextran and 428 kDa ficoll were examined for concentrations up to 15 or 34 wt\%, respectively, as seen in Figures \ref{figurecush2004aDpX} and \ref{figurecush2004aDpY}.  Polydispersities of the dextran and ficoll were 1.94 and 4.3, with $R_{g}$ measured from light scattering of 22 and 20 nm, respectively.  Over this concentration range, $\eta$ increases by nearly 100-fold, while $D_{r}$ and $D_{p}$ fall by nearly the same factor.  Within experimental scatter, the concentration dependences of $D_{p}$, $D_{r}$, and $\eta^{-1}$ in ficoll solutions were very nearly the same. $D_{p}\eta$ and $D_{r} \eta$  increased perhaps 50\% in dextran solutions.  The authors also studied the molecular weight dependence of $D_{r}$ and $D_{p}$ at fixed 14.6 wt\% dextran concentration, finding $D_{r}$ and $D_{p}$ have power-law dependences on polymer $M_{w}$ for $40 \leq M_{w} \leq 2750$ kDa. Cush, et al.\ reason from the minimal issues with Stokes-Einsteinian behavior for $D_{r}$ and $D_{p}$ that their data support a continuum picture of polymer dynamics in which 'hydrodynamic effects surpass any due to topological constraints', even in solutions having nearly a hundred times the viscosity of water.

Hill and Soane\cite{hill1989aDp} examine the rotational diffusion of collagen molecules in aqueous 4 Mda polyethyleneoxide using the electrooptic Kerr effect.  Collagen is a rigid rod, with diameter and length of ca.\ 1.5 and 170 nm, respectively.  In a Kerr effect experiment, a linearly polarized laser beam is passed through a sample on which an electric field has briefly been imposed.  Hill and Soane cite 24 kV/cm as their largest attainable electric field.  Partial induced alignment of the molecules in the cell leads to an induced birefringence and thence to a change in the transmission of the polarized light, through a quarter-wave plate and second optical polarizer, into the detector.  The relaxation time for the Kerr effect is determined by rotational diffusion of the weakly aligned objects in the Kerr cell. The collagen preparations had a bimodal mass distribution, so two relaxation times were observed, both strongly concentration dependent. 

Jamil and Russo\cite{jamil1998aDp} studied translational and rotational diffusion of poly(tetrafluoroethylene) latex in aqueous sodium polystyrenesulfonate (NaPSS).  Spectra were single exponentials. $D_{p}$ and $D_{r}$ were extracted from the $q^{2}$ dependence of the spectral linewidths.  Addition of NaCl causes the probes to aggregate; further addition of NaPSS reverses the salt-induced probe aggregation, even though it appears that NaPSS does not bind to the latex particles.   $D_{r}$ tracks the changes in viscosity of the solution attendant to addition of polymer, in the sense that $\eta D_{r}$ is independent of polymer concentration.  However, at fixed $c$, $\eta D_{r}$ does decrease as the salt concentration is increased.

\begin{figure}[bth] 
\includegraphics{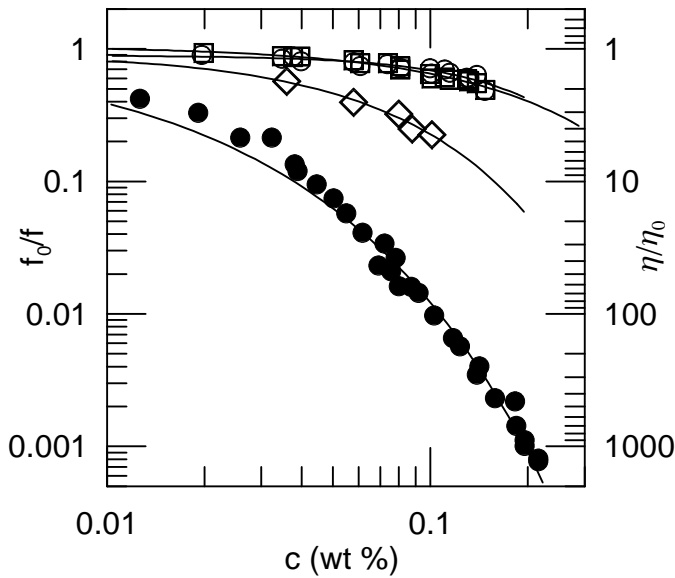}
\caption{\label{figurekoend2004aDpx} $D_{p} (\square)$, $D_{r} (\bigcirc)$, and $s (\lozenge)$ of 92.5 nm radius spheres in 4 MDa xanthan, all reported [left axis] as nominal normalized drag coefficients $f_{0}/f$, and shear viscosity $\eta/\eta_{0} (\bullet)$ [right axis], after data of Koenderink, et al.\cite{koenderink2004aDp}, and fits to stretched exponentials in $c$. }
\end{figure} 

Koenderink, et al.\cite{koenderink2004aDp} examine the motion of perfluorinated hydrocarbon spheres through xanthan solutions.  Depolarized QELSS spectra were measured at a series of angles and fit to second-order cumulant expansions.  The spheres had radii 93 nm; xanthan molecular weight was 4 MDa.  Koenderink, et al., measured solution viscosity, shear thinning, storage and loss moduli, translational and rotational diffusion coefficients $D_{p}$ and $D_{r}$ of the probes, and probe sedimentation coefficient $s$, and made an extensive and systematic comparison with the available theoretical background. As seen in Figure \ref{figurekoend2004aDpx}, xanthan is far more effective at increasing the solution viscosity than at hindering translational or rotational diffusion of the probes, and is more effective at hindering sedimentation than at hindering diffusion.  The observed variation of $D_{p} \eta$ and $D_{r} \eta$ with $c$ in xanthan solutions is much larger than the variation observed for probes in many other systems.

Le Goff, et al.\cite{legoff2002aDp} polymerized actin fragments with various degrees of labelling, using a protocol which led to ends made of heavily-labelled actin and an extremely long center segment that was sparsely labelled or unlabelled.  Fluorescence microscopy was used to track the motion of the labelled ends, determining thereby the distribution and time dependence of the fluctuating end-to-end distance and the angular reorientation of the end-to-end vector.  Proof-of-principle data was obtained.  Repeat of these experiments in solutions at a series of concentrations should provide extremely valuable information about translation, rotation, and internal motion of somewhat rigid polymer chains.

\begin{figure}[tb] 
\includegraphics{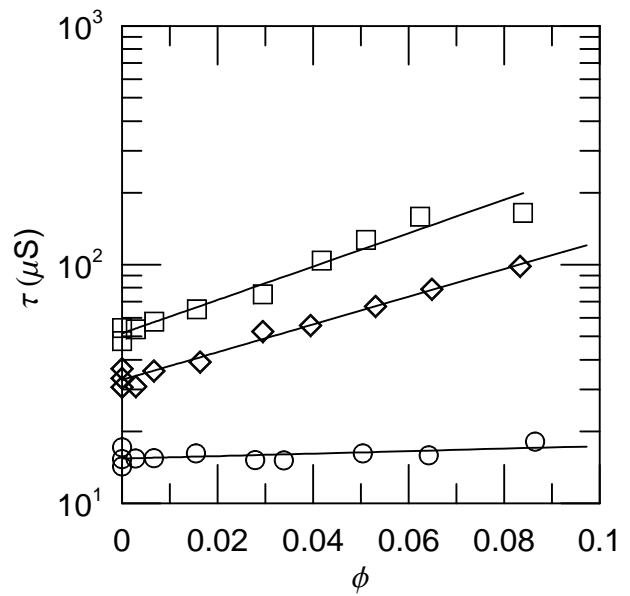}
\caption{\label{figurephalak2000aDp2} Rotational relaxation times of 68 $(\bigcirc)$, 128 $(\lozenge)$, and 170 $(\square)$ nm PBLG rods in solutions of 62 nm silica spheres as a function of sphere concentration\cite{phalak2000aDp}, and simple-exponential fits to $\tau_{r}(c)$.}
\end{figure}

Phalakornkul, et al.\cite{phalak2000aDp} used transient electric birefringence to examine rotational relaxation of poly-($\gamma$-benzyl-$\alpha$-L-glutamate) (PBLG) in solutions of silica spheres.  PBLG preparations had lengths 68, 128, or 170 nm.  Spheres had radius 31 nm. Birefringence spectra were analyzed with the Laplace inversion program CONTIN, which found consistent fast and slow modes and an occasional very weak, very fast relaxation. At very low sphere concentration, the slow mode reduces to rod rotation. The fast mode appears to correspond to the first bending mode of the PBLG.  The slow mode of the short rods had a concentration-independent relaxation time $\tau_{s}$. Increasing the rod length greatly increases the concentration dependence of $\tau_{s}$.  For the longest rods, $\tau_{s}$ increases three-fold as sphere volume fraction is increased from 0.0 to 0.08, while the relaxation time of the first internal mode increases less strongly with increasing $c$.  $\eta$ and all rod relaxation times had simple exponential concentration dependences (Figure \ref{figurephalak2000aDp2}). The concentration dependence of $\eta$ is stronger than the concentration dependence of $\tau_{s}$ of the short rods, but weaker than the concentration dependence of $\tau_{s}$ of the long rods.   

Sohn, et al.\cite{sohn1996aDp} used polarized and depolarized light scattering spectroscopy to examine the diffusion of various latex particles containing magnetite inclusions through solutions of 70 kDa polystyrene sulfonate.  Most work was done on 50 nm radius particles. Extensive preliminary work was needed to characterize the available latex-magnetite probes, which tend to be rather polydisperse unless they are fractionated.  Addition of the polymer reverses salt-induced aggregation of the probes. On increasing the polymer concentration the translational and rotational diffusion coefficients $D_{t}$ and $D_{r}$ both fell. At fixed polymer concentration, addition of NaCl did not have an obvious effect on $D_{t}$ but tended to reduce $D_{r}$.  At fixed salt or polymer concentration, the number of polymer or salt concentrations (respectively) that were examined was not large, so it is difficult to make detailed inferences about the effect of the solutes on probe diffusion.

\subsection{Particle Tracking Methods}

In an extremely important paper, Apgar and Tseng, et al.\cite{apgar2000aDp} measured the {\em distribution} of mean-square displacements for probes in a variety of solutions.  They used fluorescence microscopy to track fluorescently labelled microspheres diffusing through actin solutions.  Spheres with radii in the range 0.43-0.6 $\mu$m were used.  Tracking was used to determine the mean-square displacements as functions of time for a large number of probe particles, and to obtain statistics for mean-square deviations over each of several time intervals. In homogeneous water-glycerol solutions, the distribution of mean-square displacements is Gaussian. In solutions of dilute actin, concentrated F-actin, and F-actin/fascin, the distributions are very definitely not Gaussian.  Correspondingly, analyses of measurements on particle motion in F-actin solutions, if based on an assumption of Gaussian particle displacements, have a doubtful starting point.

Chen, et al.\cite{chen2003aDp} used video microscopy to measure the motion of single particles and pairs of particles through a viscoelastic  solvent, namely $\lambda$-DNA in water.  Measurements were made of one- and two-particle diffusion for spheres with diameters 0.46, 0.97, and 2.0 $\mu$m and interpreted using 'two-point microrheology'\cite{levine2001aDp}, which compares the diffusion coefficient of a single particle with the relative diffusion of pairs of particles.  Interpretation was made in terms of a model involving polymer depletion zones around each particle.

Crocker, et al.\cite{crocker2000aDp} describe a novel physical technique in which video microscopy is used to capture the positions of pairs of diffusing particles at an extended series of times.  Cross correlations in the particle displacements yield a range-dependent cross-diffusion coefficient $D_{ij}(r,t)$, where $r$ the distance between particles $i$ and $j$, and $t$ is the elapsed time during which the displacement occurs.  The Laplace transform of $D_{ij}$ is taken to be inversely proportional to a frequency-dependent viscosity $\tilde{G}(s)/s$.  The method was applied to study 100 and 235 nm radius spheres in glycerol, guar gum, and F-actin solutions. In glycerol solutions, the single-particle and cross-diffusion coefficients agree with each other.  In viscoelastic guar solutions, there are factor-of-two disagreements in the storage and loss moduli inferred from the single-particle and cross-diffusion coefficients and a generalized Stokes-Einstein equation; $G'$ and $G''$ from $D_{ij}(r,t)$ are much closer to macroscopic measurements. Crocker, et al.\cite{crocker2000aDp} conclude that their two-point pair diffusion method has "underlying validity", while "single particle microrheology provides qualitatively different moduli and completely fails to detect the crossover [of $G'$ and $G''$]".  Crocker, et al.'s\cite{crocker2000aDp} implicit conclusion is that single-particle microrheology does not provide valid rheological measurements. 

Dichtl and Sackmann\cite{dichtl1999aDp} studied the diffusion of individual actin chains to which substantial numbers of 35 nm fluorescent latex or 17 nm gold beads had been attached, employing a confocal laser scanning microscope to track bead motions with three-dimensional resolution.  Bead positions fluctuated; bead motions could be decomposed into components parallel and perpendicular to the visualizable polymer chain. Measurements were confined to time intervals much shorter than the terminal relaxation time of the solutions. Fluctuations perpendicular to the chain axis had a Gaussian distribution whose width varied from bead to bead by nearly an order of magnitude. Fluctuations parallel to the tube axis had a non-Gaussian bimodal distribution.   At short times, mean square displacements parallel and perpendicular to the local chain are nearly equal.  At longer times, displacements parallel to the chain continue to increase, while displacements perpendicular to the local chain tend to cease to increase.  However, the authors note that $D$ parallel to the chains, as measured by them, implies a solvent viscosity an order of magnitude larger that the actual viscosity of water.  That is, the apparent bead mobility along the chain axis is very different from the expected bead mobility in pure solvent.

Gardel, et al.\cite{gardel2003aDp} used video microscopy to track 230, 320, and 420 nm radius carboxylate-modified spheres diffusing in solutions of F-actin.  Single-particle-motions and correlated-motion of pairs of particles were computed.  F-actin solutions are considerably more effective at retarding the motion of large particles than at retarding the motions of small particles, especially at long times.  Relative motions of particles were obtained for particles separated by as much as 100$\mu$m.  Single-particle motions sometimes show longer-time plateaus. Correlated-pair displacements increase linearly in time with no plateau, but at the longest times studied the mean-square correlated pair motions are not yet larger than the observed plateaus.  $G''(\omega)$ from one-particle and two-particle measurements differ by factors of three or more, the two-particle values being in relatively close agreement with macroscopic measurements of $G''(\omega)$ made with a parallel-plate stress-controlled rheometer.

Gittes, et al.\cite{gittes1997aDp} observed the overdamped thermal displacements of single 0.9$\mu$m silica spheres 
in F-actin and polyacrylamide gels. Microscopic observation was applied to isolate single particles, whose motions were detected interferometrically.  The power spectrum of the displacements $\langle (\Delta x)^{2}_{\omega}\rangle$ was used to compute the shear modulus, which has both real and imaginary parts. Over a frequency range $0.1 \leq \nu \leq 3 \cdot10^{3}$ Hz, $G'(\omega)$ and  $G''(\omega)$ both showed power-law frequency dependences $\sim \omega^{z}$ with $z \approx 0.76$.

\begin{figure}[bth] 
\includegraphics{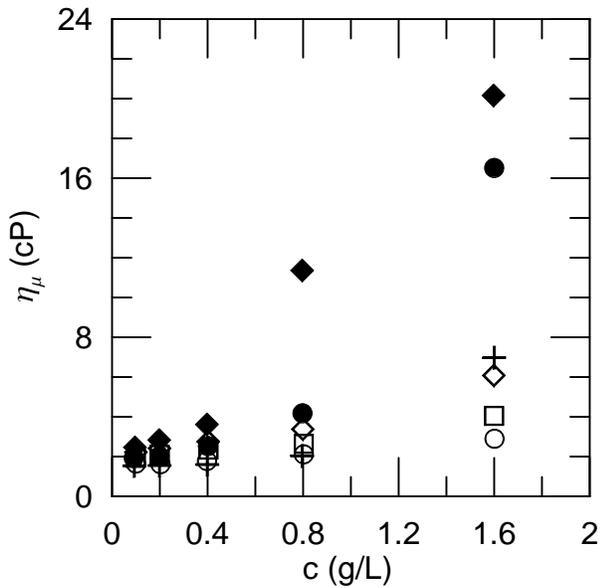}
\caption{\label{figuregoodman2002aDp1} $\eta_{\mu}$ from mean-square displacement of 500 nm radius PSL through circular DNAs ($\bigcirc, \square, \lozenge$) (masses 2.96, 5.40, and 10.31 kilobasepairs, respectively) and their single-cut linear derivatives ($\bullet, + , \blacklozenge$), after Goodman, et al.\cite{goodman2002aDp}.  }
\end{figure}

Goodman, et al.\cite{goodman2002aDp} studied the motion of 1 $\mu$m fluorescent-labelled carboxylate-modified polystyrene latex spheres through solutions of circular-supercoiled and linearized (single cut at controlled location) DNAs, size 2.96, 5.4, and 10.3 kilobasepairs (1.8-6 MDa; contour lengths 1, 1.84, 3.57 $\mu$m), using particle tracking methods.  The linear and matching circular DNAs are unambiguously identical in molecular weight.  While issues of chain configuration can arise in studies of synthetic circular polymers, the circular DNAs here are unambiguously not knotted or topologically crosslinked.  The spatial resolution on particle resolution was 5 nm.  Video monitoring was limited to a 30 Hz time resolution.  From particle-tracking, time-dependent mean-square displacements were determined as functions of DNA molecular weight, concentration and circularization.  Solution microviscosities $\eta$ were inferred from the particle motions via a generalized Stokes-Einstein equation.  At these relatively long times and small viscosities ($\eta < 20$ cP), mean-square displacements after convective corrections were in most solutions nearly linear in time.  In some linear DNA solutions, the mean-square displacement was not simply diffusive.  In these cases, $\langle \Delta x^{2} \rangle$ was replaced with a nominal creep microcompliance that is linearly proportional ($\Gamma(t) \sim \langle \Delta x^{2}(t) \rangle$) to the time-dependent mean-square displacement. 

Goodman, et al.\cite{goodman2002aDp} report evidence that particle microtracking can also provide evidence about solution microviscoelasticity that is more detailed than that provided by macroscopic measurements, for example, details of spatial variations in microviscoelastic parameters that could be ascribed to solution microheterogeneity.  The microviscosities for the linear DNAs were as much as several times larger than the microviscosity of the corresponding circularized DNA.  $c[\eta]$ was never larger than 6, but one certainly does not see the reptation prediction $\eta_{\rm ring} \gg \eta_{\rm linear}$ at concentrations and molecular weights reached here.

Lau, et al.\cite{lau2003aDp} use video tracking to measure the two-particle cross-diffusion-coefficients of naturally-occurring particles in the intracellular medium.  Comparison is made between one-particle and two-particle measurements, each giving a time course of a mean-square displacement, in the same cell at the same time. Cell interiors are characterized by extensive chemical reactions, and are not equilibrium systems.  The observed behaviors are more complex than would be obtained in a simple viscoelastic medium.  For example, in these cells the relative displacement $\Delta {\bf r}$ of two particles depends on time as $\langle \Delta {\bf r}^{2}(t) \rangle \sim t^{1.5}$.

Mason, et al.\cite{mason1997aDp} examine a single 520 nm diameter sphere diffusing through 3 wt\% 5 MDa polyethylene oxide:water and DNA solutions, using laser deflection particle tracking to determine the motion of the single probe particle on which a weak laser beam has been focused.  The technique is directly responsive to motion in a plane, variations in the intensity of the light scattered in various directions acting as an optical lever to amplify the time-dependent particle motions.  They compare mechanical and DWS measurements of $G'$ and $G''$ in the same solutions.  Direct measurement from their graph indicates the two methods gave values for viscoelastic parameters that agree to within perhaps 30\%, and sometimes better.

\begin{figure}[tb] 
\includegraphics{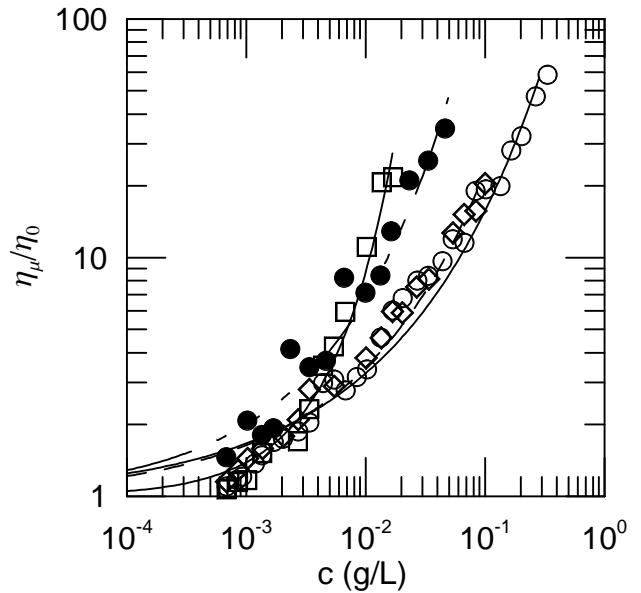}
\caption{\label{figurepapag2005bDp}  Microviscosity of polystyrene sulfonate combs via particle tracking of 274 nm radius polystyrene latex probes, and fits to stretched exponentials. Polystyrene sulfonates differ in structure, namely (a) ($\bigcirc$) 2.1 Mda chain with 24 side branches, (b) ($\lozenge$) same backbone as (a) but 3/8 as many side chains, (c)  ($\bullet$) double the backbone length of (a) but 5/8 as many side chains, and (d) ($\square$) double the backbone length of (a) and $\gg$6 times as many side chains, together with forced fits to stretched exponentials, after Papagiannapolis, et al.\cite{papag2005aDp}.}
\end{figure}

Papagiannapolis, et al.\cite{papag2005aDp} used diffusing wave spectroscopy and video particle tracking to observe the diffusion of polystyrene latex spheres through solutions of fully-neutralized polystyrene sulfonate comb polymers.  Probes had 548 nm diameter for particle tracking measurements and 112nm diameter for the DWS measurements.  The combs differed roughly two-fold in their main-chain length, in the number of branches, or in the length of the branches. Video tracking observed motion during times $\geq 3 \cdot 10^{-2}$s, during which the mean square displacement increased linearly in time, permitting determination of $D_{p}$ and thencefrom the solution viscosity $\eta_{\mu}/\eta_{o}$ at different polymer concentrations.  From Figure \ref{figurepapag2005bDp}, changing the number of sidechains at fixed concentration has a very limited effect on $\eta_{\mu}$, while doubling the length of the backbone at fixed monomer concentration very markedly increases $\eta_{\mu}$. Fits of $\eta_{\mu}$ to stretched exponentials in $c$ are quite unsatisfactory at low concentration, the measured $\eta_{\mu}(c)$ being indistinguishable from the solvent viscosity up to polymer concentrations at which the best-fit stretched exponential predicts $\eta_{\mu}(c)$ should be readily distinguishable from unity. 

Schnurr, et al.\cite{schnurr1997aDp} demonstrate a novel interference microscopy method for studying probe motion.  In their technique, mesoscopic beads are suspended in a solution or gel, their images are identified through a microscopic, and optical interferometry through the microscope stage taking each bead to be one arm of the interferometer was used to track bead motion.   The power spectral density of the displacements and the Kramers-Kronig relation are applied to infer the imaginary and real part of the complex shear modulus of the gels. Here silica beads with radii 0.25-2.5 $\mu$m were studied in F-actin and polyacrylamide gels.  

Tseng and Wirtz\cite{tseng2001aDp} used particle tracking  based on video light microscopy to measure the motion of 485 nm radius polystyrene spheres in solutions of F-actin and $\alpha$-actinin.  Mean-square displacements were  determined as a function of diffusion time and characterized statistically.  Storage and loss moduli and phase angle were computed from the displacements.  The distribution of displacements generally did not have a Gaussian form; its mean square displacement at long times increased less rapidly than linearly in time.  The observed displacement distributions depend in a complex way on the F-actin and $\alpha$-actinin concentrations.

Valentine, et al.\cite{valentine2004aDp} used single particle and two-particle tracking methods to examine the diffusion of carboxylate-modified polystyrene spheres in fibrin-F-actin-scruin networks.  The objective was to study the effect of modifying the probe surface chemistry.  Different spheres were uncoated, heavily coated with bovine serum albumin, or surface modified by systematic coupling of large numbers of short methoxy-poly(ethylene glycol) groups.  Particle motion was examined using video-tracking microscopy.  Valentine, et al., found that the dynamic behavior of the particles changed very substantially as surface treatments were changed.  However, particles whose single-particle displacement distributions differ considerably due to different surface coatings can have very similar two-particle displacement correlations.

Xu, et al.\cite{xu1998aDp,xu1998bDp} used mechanical rheology, particle tracking microrheology, and DWS to characterize the rheological properties of actin filaments. Experimental interests included the concentration dependence of the elastic modulus and the frequency dependence at high frequency of the magnitude of the complex viscoelastic modulus.  DWS spectra were interpreted by invoking the Gaussian assumption.

Xu, et al.\cite{xu2002aDp} used video tracking microscopy to observe the diffusion of 970 nm diameter fluorescent polystyrene spheres through water:glycerol and aqueous wheat gliadin solutions.  Increasing the gliadin concentration slows probe diffusion.  In glycerol, the distribution of particle displacements was a well-behaved gaussian.  For probes in gliadin suspensions: At low gliadin concentrations, measurements of the distribution $P(\Delta R)$ of mean-square displacements against time found that different particles all had the same $P(\Delta R)$.  In contrast, in concentrated gliadin solutions $P(\Delta R)$ for different particles showed a wide range of different time dependences.  Correspondingly, at low (250 g/L) gliadin concentrations $P(\Delta R)$ at fixed time is a gaussian, but at large (400 g/L) gliadin concentrations $P(\Delta R)$ is extremely non-gaussian.

\subsection{True Microrheological Measurements}

This section considers true microrheological measurements.  In a true rheological measurement, external forces or displacements are imposed, and consequent displacements and/or forces are measured.  True microrheological measurements, not to be confused with 'microrheology' studies of Brownian motion, differ from classical macroscopic rheological studies in that the probes or apparatus parts function on a mesoscopic length scale.  If one believes that probe diffusion measurements could be inverted to obtain $\eta$, $G'$, or $G''$, it becomes interesting to compare those viscosities and moduli both with the corresponding quantities measured in classical instruments and also with the same quantities measured with true microrheological instruments build on the same size scale as the diffusing probes. 

Amblard, et al\cite{amblard1996aDp} used video microscopy and magnetic tweezers to study probe motion in viscoelastic F-actin systems.  The probes were polystyrene spheres having radii of 75, 760, and 750 nm.  Video microscopy determines particle positions; the tweezers can apply a constant force to a particle.  The F-actin filaments at concentration 0.1 g/l had an estimated length of 20 $\mu$m, persistence length ca 14 $\mu$m, and a mesh size $\xi \sim 1 \mu$m.  Video microscopy determined the mean-square displacement as a function of time.  Magnetic tweezers allowed application of a fixed force, permitting determination of displacement under the influence of an applied force, also as a function of time.  With small beads ($d/\xi < 1$), the mean-square distance travelled during diffusion followed $\langle x^{2}(t) \rangle \sim t^{1}$.  For large beads ($d/\xi > 3$), over a range of times 0.03-2.0 s, Amblard, et al., found $\langle x^{2}(t) \rangle \sim t^{q}$, with $q=0.76 \pm 0.03$ for motion with an external driving force, and with $q=0.73 \pm 0.01$ for free diffusion.  For driven motion, the apparent drag  on the probe increases after the probe has travelled 10-20 $\mu$m. Use of probe surface coatings including streptavidin, surfactant, or bovine serum albumin had no effect on other results.  Specific surface interactions were concluded not to affect these findings greatly.  Here we see a direct demonstration that the drag processes for driven motion and for thermal motion can be the same:  The mean-square displacements for both processes scale approximately as $t^{3/4}$.The time dependence of the thermal motion is not $\sim t^{1}$, so the underlying thermal particle motion is not simple Brownian diffusion. 

Bishop, et al.\cite{bishop2004aDp} studied the driven rotational motion of small spheres inside a model for the intracellular medium.  The spheres were 1-10$\mu$m birefringent vaterite (CaCO$_{3}$) crystals.  The driving force is provided by illuminating the probes with circularly polarized light, and measuring the degree of polarization of the light after it has passed through the sample.  The rotation rate is obtained by examining the transmission of one linear polarization of the incident light.  Bishop, et al., examined rotational motion of their probes within a drop of hexane and in bulk water, finding that their measured microviscosities were in good agreement with the viscosities measured macroscopically.

Hough and Ou-Yang\cite{hough1999aDp} report using optical tweezers to drive the motion of a single 1.58 $\mu$m silica microsphere through solutions of 85kDa end-capped (C$_{12}$H$_{25}$-) polyethylene oxide in water.  The tweezer position was driven with piezoelectrically controlled mirrors at frequencies as large as 40,000 rad/s. Measurements of the magnitude $A$ and phase $\delta$ of the sphere oscillations as functions of frequency were inverted, using the model that the sphere is a forced damped harmonic oscillator, to obtained $G'(\omega)$ and $G''(\omega)$.  The dynamic moduli were "...quite different from those obtained by a macroscopic rheometer, and are sensitive to surface treatment of the bead."

Keller, et al.\cite{keller2001aDp} introduce an oscillatory magnetic bead rheometer, in which 2.25$\mu$m latex beads containing imbedded Fe$_{2}$O$_{3}$ are placed in an oscillating magnetic field, and the bead positions are tracked with video microscopy.  The amplitude and phase shift of the bead motion are determined, allowing calculation of $G'(\omega)$ and $G''(\omega)$.  The paper was a proof of principle demonstrating a method for making measurements at higher frequencies (up to 40 Hz), so only a single solution of nondilute F-actin was examined.

Schmidt, et al.\cite{schmidt2000aDp} compare microscopic and macroscopic measurements of the storage and loss moduli of F-actin and gelsolin solutions.  Schmidt, et al.\ examined F-actin solutions with $0.5 \leq c \leq 2$ g/L, and controlled average lengths $1.5 \leq \ell \leq 10 \mu$m, all for frequencies 0.004-4 Hz.  The microscopic particles were 4.5$\mu$m paramagnetic beads.  These results are from true microrheological measurements: the microspheres were subject to a known external force and the amplitude and phase shift of their motions relative to the driving force were obtained. The macroscopic probe was a rotating disk rheometer.  With a viscous small-molecule liquid, microrheology and macrorheology agree as to the measured viscosity.  Moduli of F-actin solutions measured microscopically were substantially smaller than moduli measured with a rotating disk rheometer. The network relaxation time estimated from microscopic data is the same as or substantially larger than the time measured with the macroscopic instrument.  The frequency dependences from microrheological and from macrorheological measurements are somewhat similar, but are clearly not the same: The microrheological measurements typically show stronger dependences of $G'$ and $G''$ on $\omega$.  

\begin{figure}[thb] 
\includegraphics{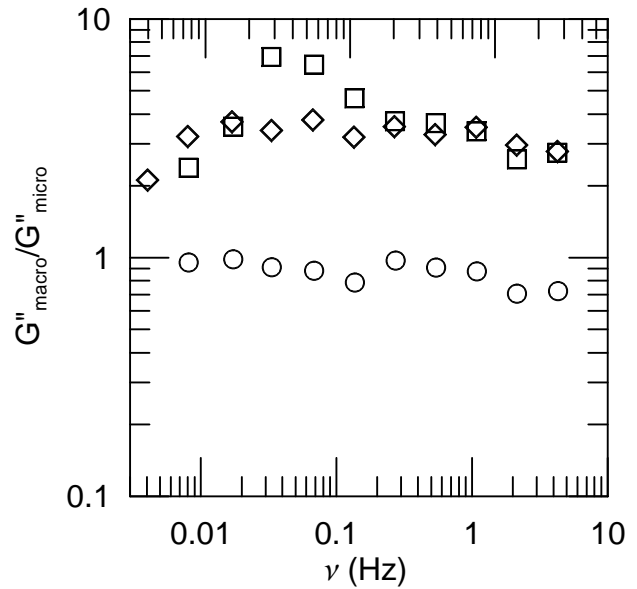}
\caption{\label{figureschmidt2000aDp1} $G''_{\rm macro}/G''_{\rm micro}$ of F-actin solutions, including pure water ($\bigcirc$), 2 g/L F-actin solutions with random lengths ($\square$), and 2 g/L F-actin solutions with $\ell = 10 \mu$m ($\lozenge$), using data from Schmidt, et al.\cite{schmidt2000aDp}, showing that true moduli measured macroscopically and microscopically are unequal.}
\end{figure}

Figure \ref{figureschmidt2000aDp1} shows representative parts of Schmidt, et al.'s measurements of $G''_{\rm macro}/G''_{\rm micro}$.  Over a wide range of frequencies, $G''_{\rm macro}/G''_{\rm micro}$ for F-actin solutions  is in the range 2-8. Schmidt, et al.\cite{schmidt2000aDp} cite Maggs\cite{maggs1998aDp} as predicting similar results.  Maggs examined spherical nanoparticles diffusing through intertwined actin filaments in solution.  Maggs' model treats probes bending filaments and distorting the local mesh, using scaling arguments and the presence of two independent length scales, namely the persistence length and the mesh size.  The model does not include hydrodynamic interactions between probes and the mesh. The apparent storage and shear moduli are seen to be sensitive to the apparatus length scale. 

Schmidt, et al.\cite{schmidt2000bDp} studied the rheological properties of solutions of fd-virus, using classical mechanical and magnetic tweezers rheometry to determine $G'$ and $G''$.  Macroscopic and microscopic measurements are reported to be in reasonable agreement.  Comparison was made with some modern theoretical calculations\cite{morse1998aDp}; the observed frequency dependence of  $G'$ at low frequency was much weaker than predicted. Comparison was also made with studies of actin solutions, permitting separation of fundamental physical properties from peculiarities of particular chemical systems. Schmidt, et al., observe that actin aggregation is extremely sensitive to a wide variety of proteins and other environmental factors.  They propose that a practical way to avoid these challenges to the use of F-actin solutions as a model for testing theories of polymer solution dynamics is to use alternative physical systems, such as the fd-virus that they studied, as sound models for testing theories of polymer solution dynamics.

From the published literature, in small-molecule liquids true microrheology and classical  measurements give the same results.  In many though not all polymer solutions, true microrheology and classical rheometry do not agree, the viscosity measured with true microrheology generally being smaller than the viscosity determined by a classical, macroscopic rheometer.  

Particle tracking gives direct access to correlation functions $\langle \Delta \vec{R}(t_{1}, t_{2}) \Delta \vec{R}(t_{3}, t_{4}) \rangle$.  These functions in turn give access to the memory functions for the Langevin-equation 'random' force in ways that light scattering spectroscopy does not.  Bandwidth for video tracking presently limits that technique, but bandwidth improvements are a matter of time, and alternative paths to determining displacements $\Delta \vec{R}(t_{1}, t_{2})$ exist\cite{schnurr1997aDp}.

\subsection{Probes in Gels and Biological Systems}

True gels are not solutions.  However, some models of polymer solution dynamics invoke analogies with polymer motion through gels, at least on favorable time scales, so it is clearly worthwhile to compare probe motion through polymer solutions with polymer motion through true gels.  Much work has focused on (1) cross-linked polyacrylamide gels and (2) cross-linked actin and other protein gels, but there are also results on (3) probes in the complex interior of living cells and other gelling systems.

\subsubsection{Probes in Polyacrylamide Gels}

Allain, et al.\cite{allain1986aDp} studied the diffusion of 0.176 $\mu$m diameter polystyrene spheres through solutions of acrylamide/N,N'-methylene bisacrylamide during irreversible gel formation.  QELSS was used to measure the relaxation spectra of the probes, which dominated the scattering by their host solutions.  The probe diffusion time $\tau \sim D_{p}^{-1}$ was obtained from a second order cumulant fit to the spectra. During the gelation process, $\tau$ of the probes increased 120-fold with increasing duration of the gelation process.  Over the same time interval, the macroscopic solution viscosity increased only 12-fold, so the probe diffusion time was not simply proportional to the macroscopic viscosity.

Matsukawa and Ando\cite{matsukawa1996aDp} used PFG NMR to study polyethylene glycol and water diffusing through fully swollen cross-linked polyacrylamide gels.  PEG molecular weights were 4.25, 10.89, and 20 kDa.  $D_{p}$ was taken to be proportional to $D_{po} \exp(-\kappa R)$, where $\kappa$ is a gel length scale that could be varied by changing the degree of swelling.  

Nishio, et al.\cite{nishio1987aDp} used QELSS to monitor the diffusion of polystyrene latex spheres in polyacrylamide: water.  Probe radii were 25 and 50 nm.  Probes were added to the solutions prior to adding the ammonium persulphonate polymerization initiator.  The fraction concentration of bisacrylamide crosslinker was varied from zero (leading to linear polyacrylamide solutions) to 5\% (leading to a strong gel).  Under fixed optical conditions, the extent to which the correlation function decayed at long times decreased with increasing crosslinker concentration and with decreasing scattering angle, indicating that above 1.6\% bisacrylamide more and more particles are confined, especially over longer distances. Nishio, et al.\ make an inversion of their data to determine the distribution of effective pore lengths for different probe radii (and, implicitly, pore diameters) showing that the distribution is quite wide. 

Park, et al.\cite{park1990aDp} used holographic relaxation spectroscopy to measure $D_{p}$ of a dye and a labelled protein through polyacrylamide gels as a function of polyacrylamide concentration $c$. The holographic method measures diffusion over distances orders of magnitude larger than any structure in the gel. QELSS was used to infer a nominal correlation length $\xi$ for the gels from their apparent diffusion coefficient $D$ via 
\begin{equation}
 D = \frac{k_{B} T}{6 \pi \eta \xi},
 \label{eq:gellength}
\end{equation}
where $k_{B}$ is Boltzmann's constant, $T$ is the absolute temperature, and $\eta$ is the solvent viscosity. On comparison with literature data on diffusion by D$_{2}$O, sucrose, and urea through the same medium, Park, et al find from their measurements that the probe diffusion coefficient $D_{p}$ depends on $c$ and probe radius $R$ via
\begin{equation}
     D_{p} = D_{po} \exp(- a c^{\nu} R^{\delta})
   \label{eq:Dppark}
\end{equation}
for $\nu \approx 0.94$ and $\delta \approx 0.59$.  That is, Park, et al.\ found that real polyacrylamide gels are size filters that selectively retard the diffusion of larger probe particles. Park, et al.\ also find from QELSS measurements on gels that $\xi \sim c^{-0.64}$, leading them to note that $D_{p}/D_{po}$ is to good approximation a function of the single variable $R C/\xi$.

Reina, et al.\cite{reina1990aDp} used QELSS to measure the diffusion of 25 and 50 nm radius PSL probe particles in polyacrylamide solutions and cross-linked polyacrylamide gels over a range of polymer concentrations and scattering angles.  In simple solutions, the probe spectrum is close to a single exponential, whose decay rate falls with increasing polymer $c$ until the gel threshold is reached. Above the threshold, a second slow mode appears; the relaxation rates of both modes then increase with increasing polymer concentration. At the threshold, particle trapping becomes evident from the lack of complete relaxation of the scattering spectrum. Above the threshold, a complex $q$-dependence of the spectrum determined by the probe diameter and the gel mesh spacing is observed.

Suzuki and Nishio\cite{suzuki1992aDp} examined 60 nm radius PSL spheres in polyacrylamide gels, determining the extent to which the spectrum decays toward zero as $t \rightarrow \infty$, and the dependence of the spectrum on monomer concentration and scattering angle.  Spectra were polymodal.  Qualitative properties of spectra were interpreted from physical models.

\subsubsection{Probes in Protein Gels}

Fadda, et al.\cite{fadda2001aDp} used QELSS and static light scattering to monitor the diffusion of 225 nm radius polystyrene latex probes through gelatin solutions during the gelation process.  Static light scattering determined the particle radius.  The presence of a deep first minimum in $I(q)$ showed that the particles were highly monodisperse. The spectrum $\langle I(0)I(\tau)\rangle$ of the probes was monitored as a function of time $t$ after quenching from high to low temperature.  The spectrum was approximated as a pure exponential at small $\tau$ and a stretched exponential in $\tau$ at large $\tau$.  When gelation set in, the short-time decay became faster, the long-time decay became very long, and a normalized measure of spectral intensity fell.  Comparison was made with particular models of gel formation.

\begin{figure}[t] 
\includegraphics{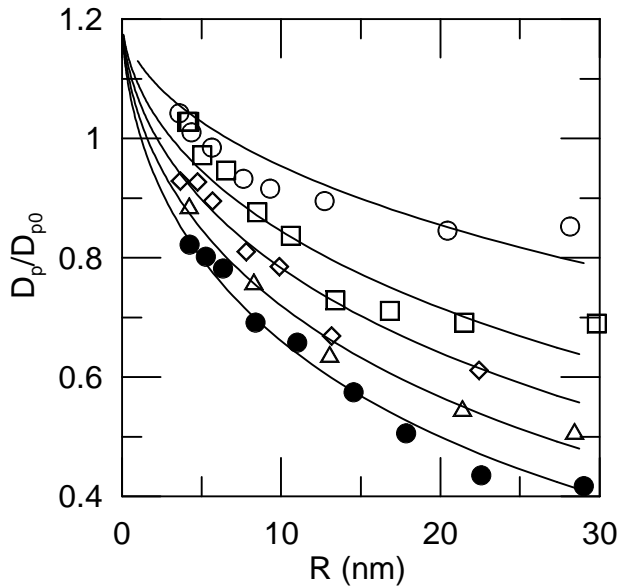}
\caption{\label{figurehou1990aDp2} Diffusion\cite{hou1990aDp} of ficoll probes through solutions of overlapped F-actin chains at actin concentrations 1 ($\bigcirc$), 3($\square$), 5 ($\lozenge$), 8 ($\bigtriangleup$), and 12 ($\blacksquare$) g/L, and a fit to a single joint stretched exponential in $c$ and $R$.}
\end{figure} 

Hou, et al.\cite{hou1990aDp} used probe diffusion measured with FRAP to examine several physical models of the intracellular medium.  While simpler model systems containing either globular particles or long chains did not reproduce the behavior of probes moving in vitro, probes diffusing through a mixture of globular and long-chain proteins did show most physical properties seen with \emph{in vivo} probe diffusion studies.  The probe particles were a series of fluorescently-tagged size fractionated ficolls.  The background matrix included concentrated globular particles (ficolls or bovine serum albumin(BSA) ) and/or heavily overlapped F-actin filaments.  The experimental focus was studying nine probes in each of a few solutions, so concentration dependences for $D_{p}$ were examined imprecisely. Probes diffusing through solutions of globular particles showed a $D/D_{o}$ that was only weakly dependent on probe radius $R_{h}$.  For probes in these solutions, $D/D_{o}$ was always larger than  $\eta_{o}/\eta$, $D \eta/D_{o} \eta_{o}$ increasing with increasing solution viscosity. In contrast, $D/D_{o}$ of probes diffusing through F-actin solutions falls markedly with increasing probe radius, but extrapolates to $D/D_{o} =1$ as $R_{h} \rightarrow 0$. Figure \ref{figurehou1990aDp2} shows the nine probes diffusing through heavily overlapped actin matrix solutions.  A joint fit of all measurements gave $D/D_{0} = 1.21 \exp (-0.067 R^{0.55} c^{0.38})$, as shown.

Hou, et al.\cite{hou1990aDp} compared their results with findings of Luby-Phelps, et al.\cite{lubyphelps1986aDp,lubyphelps1987aDp} on probes in the cytoplasm of living cells.  In living cells, $D/D_{o}$ depends strongly on probe radius $R_{h}$ but extrapolates to $D/D_{o} \approx 0.3$ as $R_{h} \rightarrow 0$.  In contrast to results in {\em in vivo} probes, probes in simple ficoll and bovine serum albumin solutions have a $D/D_{o}$ that is substantially independent of $R_{h}$.  Probes in F-actin solutions have $D/D_{o} \approx 1$ in the limit of small $R_{h}$.  However, probes in a mixture of F-actin and concentrated (7-10\%) ficoll or concentrated (7\%) BSA show the same properties as probes in cytoplasm, showing that probe diffusion in cytoplasm is plausibly governed by the simultaneous presence of a network phase and concentrated globular macromolecules.

Madonia, et al.\cite{madonia1983aDp} prepared nongelling and gelling hemoglobin solutions and used QELSS to watch the diffusion through them of 43 and 250 nm radius probe particles. At the onset of gelation, $D_{p}$ of the 250 nm probes falls rapidly.  Under identical experimental conditions, as gelation proceeds $D_{p}$ of the 43 nm spheres increases slightly. The gel structure functions as a size filter, trapping the larger particles but not obstructing significantly the motions of the smaller particles.

Newman, et al.\cite{newman1989aDp} used QELSS to monitor the diffusion of polystyrene latex spheres through F-actin solutions.  Four probe species with radii of 55, 110, 270, and 470 nm were employed.  Actins were polymerized with MgCl$_{2}$, KCl, or CaCl$_{2}$; however, all samples studied were liquids, not gels.  Prior to polymerization, $D_{p}$ in the actin solutions was not significantly different from $D_{p}$ of the same probe in pure water. On addition of salt to start the polymerization process, $D_{p}$ of the latex spheres began to fall, and the second  cumulant of the probe spectra began to increase, showing that the distribution of probe displacements is no longer a Gaussian. Changes in spectral parameters stop after several hours.  $D_{p}$ was determined for all probes, actin concentrations $1 \leq c \leq 20\mu$M, and several scattering angles.  $D_{p}$ in solutions polymerized with Mg or Ca followed a stretched exponential $D_{p}/D_{p0} = \exp(- \alpha c^{\nu} R^{\delta})$ for $\nu =1.08 \pm 0.09$ and $\delta = 0.73 \pm 0.05$. In contrast, for probes in solutions polymerized with K, one finds $\delta \approx 0$.

Newman, et al.\cite{newman1991aDp} used QELSS to examine the diffusion of PSL spheres (radii 55, 105, and 265 nm) through solutions of polymerized actin.  The actin concentration was fixed at 0.65 g/L throughout the experiments. Addition of gelsolin, a protein that interacts with actin to shorten the actin filaments, was used to determine the effect of actin filament length on probe diffusion at constant total actin concentration. Analysis of QELSS spectra indicated that  if little gelsolin was present the actin network trapped many probe particles, at least over long distance scales. Trapping was more effective for the larger probe particles.  On shortening the actin filament length by adding gelsolin, the probe diffusion coefficient increased and the trapping vanished.  The inferred microviscosity fell from 5-20 times that of water when actin filaments were very long (no added gelsolin) down to twice that of water at the largest gelsolin concentration examined.  After addition of gelsolin, changing the probe radius five-fold had little effect on the inferred microviscosity.  Addition of gelsolin changed modestly the probe spectrum line shape: the second spectral cumulant was large at very low gelsolin concentration, when some particles were trapped, but dropped a great deal when all probes were free to diffuse.

Schmidt, et al.\cite{schmidt1989aDp}, as part of an extremely systematic study of actin dynamics, examined the diffusion of 35, 50, and 125 nm radius latex spheres through fully polymerized actin networks.  The actin networks were also characterized using QELSS, FRAP, and electron microscopy.  Sphere diffusion through the networks followed $D_{p} = D_{o} \exp(- \alpha c^{\nu} R^{\delta})$ with prefactor $\alpha \approx 1$, and exponents $\nu \approx 1$ and $\delta \approx 2$. For actin concentrations $0 < c \leq 2$ mg/ml, 125 nm radius spheres were retarded by up to sixfold, while the concentration dependence of $D_{p}$ of the smaller spheres was weak.  Over the range of radii studied, namely $R$ comparable with the mesh spacing seen in an electron micrograph, heavily interlaced networks of actin molecules act as size filters, selectively retarding the motion of the larger probe particles. 

Stewart, et al.\cite{stewart1988aDp} used holographic relaxation spectroscopy (HRS) to monitor the diffusion of dye and tagged proteins through fibrin gels.  In an HRS experiment, a pair of crossed laser beams are used to bleach a holographic pattern in solution, creating an index of refraction grating formed from unbleached molecules, and perhaps separate gratings formed from bleached, photochromically modified, or photochemically bound molecules.  Stewart, et al.\ found that small dye molecules diffused through fibrin gels as if they were in pure water. The diffusion of labelled BSA molecules was retarded by the gel, though less so if the gel was formed in the presence of Ca$^{+2}$, while in contrast the photobleaching reaction caused some labelled immunoglobulin G molecules to bind to the fibrin gel and become immobile.

Wong, et al.\cite{wong2004aDp} study diffusion of probe particles in F-actin networks, using video tracking microscopy to observe particle motion.  Mean-square displacements were measured directly.  They were not diffusive: $\langle (\Delta x)^{2}\rangle$ was not $\sim t^{1}$.  Examination of the motions of individual particles showed that particle dynamics were heterogenous.  While the motion of some particles showed complete trapping, the particle never moving far from its starting point, the motions of other particles showed only partial trapping: particles alternately were trapped within restricted regions and made rare saltatory jumps to other trapping regions. Diffusion by 250, 320, and 500 nm radius polystyrene spheres was compared with the diffusive exponent $\gamma$ of $\langle (\Delta x)^{2}\rangle \sim t^{\gamma}$ for networks at a series of F-actin concentrations, showing that $\gamma$ is a universal function of $R/\xi$, with probe  radius $R$ and network mesh size $\xi$.

\subsubsection{Probes in Living Cells and Other Systems}

Arrio-Dupont\cite{arriodupont2000aDp} used a fluorescence recovery technique to measure the diffusion of a series of fluorescently-labeled proteins through the interior of cultured muscle cells.  Probe radii extended from 1.3 to 7.2 nm.  Cellular interiors are size filters, selective obstructing the diffusion of the larger probes.  The retardation of the diffusion of larger globular proteins by the cell interiors is substantially more extensive than the retardation of similarly large dextrans by the same interiors\cite{arriodupont1996aDp}, perhaps because dextrans are flexible and globular proteins are inflexible.

Kao, et al.\cite{kao1993aDp} used fluorescence recovery after photobleaching and picosecond resolution fluorescence polarization measurements to determine translational and rotational diffusion coefficients and fractional fluorescence recovery levels of fluorescein derivatives in Swiss 3T3 fibroblasts and viscous solutions.  Careful analysis permitted separate determination of the fluid-phase viscosity, the level of probe binding, and the inhibition of probe diffusion due to the cell volume occupied by the relatively immobile cytomatrix.  The combination of these effects accounts for the retardation of small-molecule solutes by cellular interiors.

Luby-Phelps\cite{lubyphelps1987aDp} and collaborators used FRAP to study the diffusion of a series of fractionated ficolls, both in the cytoplasm of 3T3 cells and in concentrated (10-26\%) protein solutions. Ficoll fractions had hydrodynamic diameters 60-250 $\AA$. $D_{p}/D_{o}$ of ficolls in protein solutions was independent of the size of the ficoll molecules.  That is, protein solutions are not size filters; they retard to approximately the same fractional extent the diffusion of small and large probe particles.  In contrast, cell cytoplasm is a size filter.  $D_{p}/D_{o}$ of ficolls in cytoplasm falls sixfold as the ficoll diameter is increased from 60 to 500 $\AA$.  For the largest ficoll, FRAP finds that a third of the ficoll particles are trapped by cytoplasm; only 2/3 of the particles are able to diffuse through substantial distances.  Luby-Phelps, et al. propose that objects larger than 500 $\AA$ or so are mechanically prevented from diffusing through cell cytoplasms.

Luby-Phelps, et al.\cite{lubyphelps1993bDp} describe a fluorometric method using two homologous dyes for determination of the local solvent viscosity, and apply the technique to determine the viscosity of water in the cytoplasm of a cell.  The fluorometric behavior of one dye (but not the other) is sensitive to the local solvent mobility, while both dyes are approximately equally responsive to other environmental influences.  The dyes were bound to ficoll for microinjection purposes.  It was found that the solvent viscosity is not greatly different from the viscosity of bulk water, resolving a long and well-known literature dispute.

Seksek, et al.\cite{seksek1997aDp} studied the diffusion of labelled dextrans and ficolls within fibroblasts and epithelial cells using FRAP.  Dextrans had molecular weights 4, 10, 20, 40, 70, 150, 580, and 200 kDa.  Ficolls were size-fractionated, four fractions being used for further studies.  Probes were microinjected into cells. With large probes, the cytoplasm and nucleus can be studied separately.  Small probes pass through the nuclear membrane, so nucleus and cytoplasm had to be studied simultaneously.  In MDCK cells, $D_{p}/D_{po}$ did not depend on the size of the probes.  Seksek found that under some conditions processes other than translational diffusion could lead to photobleaching recovery.  

Probe diffusion measurements are often said to be related to the phenomena that control the release of medical drugs from semirigid gels. Shenoy and Rosenblatt\cite{shenoy1995aDp}  provide an example in which release times are measured directly.  They examined succinylated collagen and hyaluronic acid matrices, separately and in mixtures, using bovine serum albumin and dextran as the probes whose diffusion was to be measured.

\subsection{Probe Spectra Interpreted with the Gaussian Assumption}

This section notes a series of papers, sometimes linked under the cognomen {\em microrheology}, that rely on the assertion that the incoherent structure factor for the diffusion of dilute probes through a viscoelastic matrix may in general be rewritten via $\langle \exp(i {\bf q} \cdot {\bf r}_{i} \rangle = \exp( - q^{2} \langle r^{2} \rangle/ 2)$, the so-called 'gaussian assumption'.  As noted above, this assumption is incorrect except in a certain very special case.  The special case in which this assertion is correct is the case in which the incoherent structure factor is characterized by a single pure exponential, so that $- \log(\langle \exp(i {\bf q} \cdot {\bf r}_{i} \rangle )$ is linear in $t$.  Interpretations based on the Gaussian assumption, when applied to non-exponential spectra, are therefore highly suspect.


Bellour, et al.\cite{bellour2002aDp} used DWS to measure probe diffusion of 0.27, 0.5, 1, and 1.5 $\mu$m radius polystyrene spheres in solutions of cetyltrimethylammonium (CTA) bromide and sodium hexane sulphonate, in some cases after ion exchange removal of the Br$^{-}$.  CTA salts in aqueous solutions form giant extended flexible micelles.  At elevated surfactant concentrations, the micelles are nondilute and potentially entwine.  However, unlike regular polymers, the micelles have a lifetime for a mid-length scission process that effectively permits micelles to pass through each other.  At elevated surfactant concentrations, the DWS spectrum had a bimodal relaxation form.

Dasgupta, et al.\cite{dasgupta2002aDp} used DWS and QELSS to study polystyrene spheres diffusing through 200 and 900 kDa polyethylene oxide solutions. A wide range of polymer concentrations, $15 c^{*}-45 c^{*}$ was examined, taking $c^{*}$ of the two polymers to be 0.48 and 0.16 wt\%, respectively.  Polystyrene probes had radii 230, 320, 325, 485, and 1000 nm, the PSL being carboxylate-modified except for the 325 nm probes, which were sulphate modified. QELSS spectra were interpreted by invoking the 'Gaussian' assumption, even though the nominal $\overline{(\Delta x(t))^{2}}$ was clearly not linear in $t$.  QELSS spectra were truncated and only reported for times longer than 10 mS.

Gisler and Weitz\cite{gisler1999aDp} used diffusing wave spectroscopy to examine the motion of polystyrene spheres through F-actin solutions.  The Gaussian assumption was invoked to convert DWS spectra to nominal mean-square particle displacements.  The inferred displacements were then used to infer viscoelastic properties of F-actin gels.   

Heinemann, et al.\cite{heinemann2004aDp} used diffusing wave spectroscopy to examine 720 nm polystyrene latex spheres moving through aqueous solutions of potato starch, using $\gamma$-dodecalactone to induced aggregation of the starch molecules. DWS spectra were inverted by invoking the gaussian approximation. Inverted spectra were used to generate nominal values for the storage and loss moduli as functions of frequency.

Kao, et al.\cite{kao1993bDp} used diffusing wave spectroscopy, in a specialized instrument in which the digital correlator is replaced by a Michelson interferometer, to examine the diffusion of 103-230 nm radius colloidal spheres over the first 20 nS of their displacements.  The Gaussian approximation for particle displacements in quasielastic scattering was invoked.

Kaplan, et al.\cite{kaplan1993aDp} use DWS to study structure formation in alkyl ketene dimer emulsions.  Spectra were interpreted by invoking the Gaussian assumption.  These systems are used in paper manufacture as sizing agents, but lead to technical difficulties if they gel.  The scattering particles were intrinsically present in the system. The time evolution of spectra over a period of weeks was observed.  Within a few days, systems that were going to gel could readily be distinguished from those that would remain stable.  It is important to emphasize that even if a measurable is not simply related to underlying physical properties, the measurable may still represent a valuable practical analytical tool for industrial purposes.

Knaebel, et al.\cite{knaebel2002aDp} used DWS, static light scattering, small-angle neutron scattering, and mechanical rheometry to characterize an alkali-soluble emulsion system. These solutions exhibit shear thinning and thixotropic behavior.  The low-shear viscosity increases dramatically with increasing concentration, up to a limiting concentration beyond which $\eta$ increases more slowly with increasing $c$.  The DWS measurements were analyzed by invoking the Gaussian assumption.

Lu and Solomon\cite{lu2002aDp} use DWS to measure the diffusion of 0.2 to 2.2 $\mu$m polystyrene spheres and 0.25 $\mu$m colloidal silica beads in solutions of hydrophobically-modified ethoxylated urethane, which is an associating polymer, at polymer concentrations 0-4 wt\%.  Mechanical rheology measurements were made on the same systems.   DWS measurements were analyzed using the Gaussian displacement assumption.  From the reported nominal $\langle \Delta r^{2}(t) \rangle$, the underlying single-scattering spectrum was not a pure exponential.

Mason and Weitz\cite{mason1995aDp} applied DWS to study the diffusion of 420 nm polystyrene spheres, at large concentration in ethylene glycol ($\phi=0.56$), at lower concentration ($\phi =0.02$), and in a 15 wt \%  solution of 4 MDa polyethylene oxide.  The Gaussian assumption was used to convert $g^{(1)}(q,t)$ to a mean square particle displacement and thence to a 'time-dependent diffusion coefficient'.  A generalized Stokes-Einstein equation translated the diffusion coefficient into microscopic storage and loss moduli.  

Narita, et al.\cite{narita2001aDp} used DWS and QELSS to study probe diffusion in solutions of 95 kDa polyvinylalcohol (PVA) and crosslinked PVA gels.  The probes were 107 and 535 nm PSL spheres.  QELSS spectra were visibly bimodal at all polymer concentrations studied.  The relaxation time of the fast QELSS mode was diffusive.  The slow mode showed a stronger-than-$q^{2}$ dependence of its linewidth.  DWS spectra lacked visible long-time shoulders, and were interpreted using the Gaussian assumption.   

Nisato, et al.\cite{nisato2000aDp} use QELSS and DWS to examine 85 and 107 nm polystyrene latex probe spheres in  polyacrylic acid gels that had been chemically-crosslinked by addition of methylene bis-acrylamide.  DWS spectra were reported for forward and backward scattering geometries, nominally giving determinations of motion on two different length scales.  QELSS spectra had a long-time limit that was far above the baseline calculated from the mean scattering intensity.  The Gaussian assumption was used to infer from the long-time limit of the dynamic structure factor a long-time limit of the mean displacement $\langle \Delta r^{2}(\infty) \rangle$; the assumption was also applied to interpret DWS spectra.

Palmer, et al.\cite{palmer1998aDp} used DWS to study the motions of 480 nm radius latex spheres in actin and actin:$\alpha$-actinin solutions.  They reported spectra, nominal mean-square displacements, time-dependent diffusion coefficients, and magnitude of the complex modulus as inferred from a generalized Stokes-Einstein equation.  A Gaussian assumption was used to interpret spectra. 

Palmer, et al.\cite{palmer1999aDp} used DWS to examine the diffusion of 480 nm radius polystyrene beads through F-actin, concentrations 0.42-6.89 g/l.  The Gaussian assumption was invoked to analyze spectra in terms of mean-square displacements and inferred frequency dependent viscoelastic moduli.  The authors also measured $G'$ and $G''$ with a mechanical rheometer, and compare with the DWS data.

Pine, et al.\cite{pine1988aDp} presented the first experimental demonstration of DWS as applied to complex fluids, based on the earlier study of Maret and Wolf\cite{maret1987aDp} on dynamic light scattering in the multiple-scattering limit. The Gaussian assumption was implicit in their use of the earlier work of Maret and Wolf\cite{maret1987aDp}.  DWS was first applied to a 0.01 volume fraction solution of 497 nm diameter PSL spheres diffusing freely in water, showing excellent agreement between the measured spectrum and the predicted theoretical form for the DWS spectrum, if the photon transport mean free path $\ell^{*}$ was used as a fitting parameter.  Pine, et al., report that $\ell^{*}$ from their dynamic measurements was consistently smaller than inferred from static backscattering measurements.  Comparison was made with mixtures of 312 and 497 nm diameter spheres (volume fractions 0.01, 0.04, respectively) that had been deionized to produce a colloidal glass.  The DWS spectrum changed markedly on formation of the glass.

Popescu, et al.\cite{popescu2002aDp} present a novel QELSS apparatus, based on their earlier analysis\cite{popescu2001aDp},  that uses extremely-short-coherence-length visible light in its measurements.  In Popescu, et al.'s method, light from a superluminescent diode with a coherence length of 30$\mu$m was sent perpendicular to the window surface into a scattering volume.  Light, backscattered from particles that are located within a few coherence lengths of the surface, combines coherently with light back-reflected by the window, allowing measurement in heterodyne mode of the QELSS spectrum, even of highly scattering samples.  Popescu's analysis invokes the Gaussian assumption 'for times much shorter than the decaying time of the autocorrelation function'. 

Sohn, et al.\cite{sohn2004aDp} present a theoretical analysis of Popescu, et al.\cite{popescu2002aDp,popescu2001aDp}'s microvolume QELSS apparatus, based on invoking the Gaussian assumption.  The possibility of working directly in frequency domain, and describing the spectrum as a sum of Lorentzians rather than a sum of exponentials, is examined.

Rojas-Ochoa, et al.\cite{rojas2002aDp} make a systematic study of DWS spectra of monodisperse interacting hard sphere systems.  Small-angle neutron scattering was used to determine the sphere size and the static structure factor, which was significantly perturbed by interparticle interactions.  The photon transport mean free path was determined directly by measuring the optical transmittance of samples of various thicknesses.  The hydrodynamic radius of the spheres was obtained via QELSS applied to dilute samples. DWS was used to measure particle motions over a few to a few hundred microseconds.  The solvent was Newtonian. Under these conditions, which would not arise in a viscoelastic matrix solution, the Gaussian assumption is applicable.  Comparison of this no-free-parameter determination of $D_{p}$ with predictions of orthodox theory for the concentration dependence of $D_{p}$ found excellent agreement. Thus, under the very restrictive conditions under which the theoretical models for $D_{p}$ and diffusing wave spectroscopy are applicable, the model gives good results.

Romer, et al.\cite{romer2000aDp} used DWS to study gel formation in colloid preparations.  Gelation was induced by varying the solution ionic strength, using enzymatic degradation of a neutral organic compound to create a slow change uniform across the sample in the ionic strength of the solution.  The Gaussian assumption was invoked to convert DWS spectra to determinations of the mean-square displacement of individual particles, even though particles in gelling systems have long-time correlations in the forces on them, so that the nominal $\langle (\Delta r)^{2}\rangle$ from DWS for identical particles in a gelling system is not linear in time. Romer, et al.\cite{romer2001aDp} present additional DWS and classical mechanical rheology measurements on the same system, showing that DWS and rheometric properties show dramatic changes after similar elapsed times, the elapsed time being characteristic of the system's sol-gel transition.

Rufener, et al.\cite{rufener1999aDp} used DWS to study concentrated G-actin that had been systematically polymerized into F-actin. The Gaussian assumption was invoked to interpret spectra of polystyrene sphere probes.  Spectra of small probes decayed to the expected baseline.  Spectra of large probes decayed only part way to the expected baseline, leading to the inference that large but not small spheres were trapped by the gel network.

van der Gucht, et al.\cite{vandergucht2003aDp} used QELSS to study the diffusion of 125 and 250 nm radius modified silica particles in solutions of the self-assembling monomer bis(ethylhexylureido)toluene.  They also measured the low-concentration viscosity, dynamic shear moduli, and static light scattering intensity. Concentration-driven self-association gives $\eta(c)$ a peculiar form: $\eta(c)$ is nearly constant out to a crossover concentration $c^{*}= 0.11$g/L monomer, and then increases as a power law in $c$. QELSS spectra were relatively unimodal at low polymer concentration, but gain a slow mode at larger $c$.  The concentration at which the slow mode appears is nearly an order of magnitude larger than $c^{*}$.  Light scattering spectra were interpreted using the Gaussian assumption even when spectra were clearly not single exponentials.

van Zanten, et al.\cite{vanzanten2004aDp} used DWS to examine 195, 511, 739, 966, and 1550 nm diameter PSL spheres diffusing in 330 kDa polyethylene oxide:water solutions for PEO concentrations $0.2 \leq c \leq 15$ wt\%, the Gaussian assumption being invoked.  A cone and plate viscometer was used to study transient creep and dynamic oscillatory responses.

\section{Systematics}

\subsection{Systems}

Here we have examined the literature on the diffusion of probes through polymer solutions.  Nearly 200 probe size: polymer molecular weight combinations were examined at a range of polymer concentrations. There is a solid but not extremely extensive body of work on the temperature dependence of probe diffusion in polymer solutions. A half-dozen studies of probe rotational motion and more than a dozen reports based on particle tracking are noted, along with a half-dozen sets of true microrheological measurements, in which mesoscopic objects perform driven motion in polymer solutions.  Two dozen studies of diffusion in which spectra are interpreted using the "gaussian approximation" are identified.  

As itemized in Tables I-V, probes have included polystyrene spheres, divinylbenzene-styrene spheres, silica spheres, tobacco mosaic virus, bovine serum albumin, ovalbumin, starburst dendrimers, fluorescein, hematite particles, unilamellar vesicles, micelles, ficols, PBLG rods, and low-molecular weight dextrans.  Polystyrene spheres, ficols, and dextrans were used in the bulk of the probe measurements.  Matrix polymer chains include a wide range of water-soluble polymers.  Only limited data exist on probes in solutions of the organophilic polymers that form the staple for the remainder of the polymer literature.

Rotational motion measurements require the use of orientationally anisotropic objects.  Rodlike particles include tobacco mosaic virus, collagen, actin fragments, and PBLG.  Optically anisotropic spheres, the anisotropy arising from oriented internal domains or magnetite inclusions, were examined by Cush, et al.\cite{cush1997aDp,cush2004aDp} and Sohn, et al.\cite{sohn2004aDp}.

Results on probe diffusion fall into three phenomenological classes.  In the first two classes, diffusion is usefully characterized by a single relaxation time and hence a well-defined probe diffusion coefficient.  In the third class, light scattering spectra are more complex.  The classes are:

(i) Systems in which $D_{p}$ decreases as $c$ is increased, with $dD_{p}/dc$ monotonically growing more negative with increasing $c$.

(ii) Systems showing re-entrant concentration behavior, in which over some concentration range $D_{p}$ first increases with increasing $c$ and perhaps then decreases again.

(iii) Systems whose spectra have bimodal or trimodal relaxations, corresponding to the relaxation of probe concentration fluctuations via several competing modes. 

We begin with systems having a well-defined $D_{p}$.

\subsection{Temperature Dependence}

In systems in which the solvent quality does not change strongly with temperature, $D_{p}$ scales with temperature as $T/\eta_{s}$, $\eta_{s}$ being the (temperature-dependent) solvent viscosity.  This result applies equally to probes in solutions containing no polymer, in which probe-solvent interactions govern diffusion, and to probes in concentrated polymer solutions, in which probe-polymer interactions however mediated must dominate the diffusion process. Studies confirming this result include
Bremmell, et al.\cite{bremmell2001aDp} on probes in water:glycerol and Phillies, et al.\cite{phillies1991aDp,phillies1992cDp,phillies1992eDp} on probes in aqueous polyacrylic acids and aqueous dextrans.  Phillies and Quinlan\cite{phillies1992eDp} observed slight deviations from simple $T/\eta_{s}$ behavior for probes in water:dextran, which they interpret in terms of a temperature dependence of the solvent quality.  

Because $D_{p}$ in general simply tracks the temperature dependence of $T/\eta_{s}$, one infers that solvent-based hydrodynamic forces play a dominant role in probe diffusion.  Unfortunately, this inference says very little about the nature of polymer solution dynamics, because models for polymer solution dynamics largely agree on predicting an inverse dependence of relaxation rates on the solvent viscosity:  Changing the solvent viscosity changes the mobility of individual polymer subunits, the so-called {\em monomer mobility}, thus changing the rate at which polymer chains can form or release hypothesized chain entanglements.  Changing the solvent viscosity equally changes the strength of probe-chain hydrodynamic interactions, thus changing the forces coupling the motion of a probe and of nearby chains.  
Temperature dependence studies do rule out one entire class of approaches, namely approaches referring to reduction relative to a glass temperature $T_{g}$. Analyses of $D_{p}$ uniformly compare measurements made at fixed $T$, and not at fixed $T-T_{g}$.  $T_{g}$ is claimed to depend very strongly on $c$, so measurements at fixed $T$ and different $c$ are claimed to have very different $T-T_{g}$.  It was proposed that comparisons should be made at fixed $T-T_{g}$, not fixed $T$, and comparisons made at fixed $T$ are not valid. To determine $T_{g}$, thereby allowing comparisons at fixed $T-T_{g}$ albeit different $T$, Phillies, et al.\cite{phillies1991aDp,phillies1992cDp,phillies1992eDp} measured $D_{p}$ at fixed $c$ over a range of $T$.  After removing from $D_{p}$ the dependence of $\eta_{s}$ on $T$, the remnant $T$-dependence of $D_{p}$ could have revealed how $T_{g}$ depends on $c$.  Comparing results at fixed $T-T_{g}$ would eliminate any $c$-dependence of $D_{p}$ arising from the $c$-dependence of $T_{g}$.  Phillies, et al.\cite{phillies1991aDp,phillies1992cDp,phillies1992eDp} actually found that after removing from $D_{p}$ the dependence of $\eta_{s}$ on $T$, there is no remnant $T$-dependence of $D_{p}$ to interpret.  Correspondingly, either $T_{g}$ is independent of $c$, or the notion 'reduction relative to the glass temperature' is fundamentally inapplicable to probe diffusion in polymer solutions.  In either case, the data serve to reject the use of glass temperature corrections as a part of interpreting $D_{p}(c)$.  

Changes in temperature may also change the solvent quality and hence the radius of each polymer coil.  Clomenil and Phillies\cite{phillies1993bDp} examined probes in HPC:water at different temperatures, including temperatures at which water is a good solvent for HPC and temperatures approaching the pseudotheta transition.  Corresponding to the change in solvent quality with temperature, the dependence of $D_{p}$ on polymer concentration is markedly temperature dependent.  Unfortunately, as detailed in Ref.\ \onlinecite{phillies1993bDp}, the observed dependence is equally consistent with most models of polymer solutions, so these results do not sort between different models of polymer dynamics.

\subsection{Probe Diffusion and the Solution Viscosity}

The relationship between $D_{p}$ and the solution viscosity was a primary motivation for early studies of probe diffusion.  In some systems though not others, $D_{p}$ does not scale inversely as the solution viscosity measured macroscopically, so that $D_{p}\eta$ depends strongly on polymer concentration and polymer molecular weight.  In almost all systems showing deviations from Stokes-Einsteinian behavior, $D_{p}\eta$ increases with increasing $c$, sometimes by large factors. Probe particles generally diffuse faster than expected from the solution viscosity. From the experimental standpoint, this sign for the discrepancy between $D_{p}$ and $\eta$ leads to a marked simplification, because the obvious experimental artifacts such as probe aggregation and polymer adsorption all lead to probes that diffuse too slowly, not too swiftly as actually observed.

In simple fluids, the probe diffusion coefficient follows the Stokes-Einstein equation with $D_{p} \sim 1/\eta$.  If the Stokes-Einstein equation remained valid in polymer solutions, $D_{p}$ would accurately track the solution viscosity, so that measuring $D_{p}$ would simply be a replacement for classical rheological measurements.  Indeed, the early work of Turner and Hallet\cite{turner1976aDp} found that $\eta_{\mu}$ inferred from $D_{p}$ is very nearly equal to $\eta$ measured classically with a rotating drum viscometer.   In contrast, a substantial motivation driving Lin and Phillies\cite{lin1982aDp,lin1984aDp,lin1984bDp} to extend their probe diffusion work was that in their systems $\eta_{\mu}$ was very definitely not equal to the macroscopic $\eta$, even for very large probes. The differences between the works of Turner and Hallet, and of Lin and Phillies, apparently reflect differences between chemical systems.

Probes in small-$M$ polymer solutions generally show Stokes-Einsteinian behavior with $\eta_{\mu} \approx \eta$, including 160 nm spheres\cite{brown1988cDp} in 101-445 kDa PMMA:CHCl$_{3}$ for $\eta/\eta_{s}$ up to 10, 20 and 230 nm probes in aqueous 20 kDa dextran\cite{phillies1989bDp}, and 20-1500 nm spheres in aqueous 50 kDa polyacrylic acid\cite{lin1984aDp}.  On the other hand, 49 and 80 nm probes in aqueous 90 kDa poly-L-lysine\cite{nehme1989aDp} show small $c$-independent deviations from Stokes-Einsteinian behavior.  Stokes-Einsteinian behavior is also found in some large-$M$ systems.  Onyenemezu, et al.\cite{onyenemezu1993aDp} find Stokes-Einsteinian behavior within experimental accuracy for 1100 kDa polystyrene solutions having $\eta/\eta_{s}$ as large as 100.   Turner and Hallett\cite{turner1976aDp} and Phillies, et al.\cite{phillies1989bDp} reveal  $\eta_{\mu}/\eta \approx 1$ in dextran solutions even with $M$ as large as 2 MDa. In most non-dilute polymer solutions, non-Stokes-Einsteinian behavior is found\cite{bu1994aDp,busch2000aDp,cao1997aDp,cheng2002aDp,desmedt1994aDp,desmedt1993aDp,furukawa1991aDp,gold1996aDp,gorti1985aDp,jamieson1982aDp,lin1984bDp,lin1984aDp,nehme1989aDp,phillies1987aDp,ullmann1983aDp,ullmann1985aDp,ullmann1985bDp,yang1988aDp}, often leading at large $c$ to $\eta_{\mu} \ll \eta$ and $r_{H} \ll R$.  A particularly conspicuous example of non-Stokes-Einsteinian behavior appears in Fig.\ \ref{figurelin1984aDp1}, based on Lin and Phillies\cite{lin1984aDp}, involving 20-620 nm probes in aqueous 1MDa polyacrylic acid. 

Much early work on the relationship between $D_{p}$ and $\eta$ referenced the pioneering study of Langevin and Rondelez\cite{langevin1978aDp} on sedimentation in polymer solutions.  Langevin and Rondelez propose that small sedimenting particles experience the solvent viscosity, but adequately large sedimenting particles experience a much larger drag proportional to the macroscopic solution viscosity.  The distinction between small and large particles is determined by a ratio $R/\xi$, $R$ bring the probe radius and $\xi$ being a hypothesized characteristic correlation length "...dependent on concentration...but independent of the molecular weight $M$."  Langevin and Rondelez cite unpublished results of deGennes\cite{degennes1978xDp} as predicting for the sedimentation coefficient $s$
\begin{equation}
     \frac{s}{s_{0}} = \exp(- A R^{\delta} c^{\nu}) + \eta_{s}/\eta .
     \label{eq:slangevinDp}
\end{equation}
Here $A$ is a molecular-weight-independent constant and $\delta$ is a scaling exponent.   $\xi \sim 1/(A c^{\nu})$ is the distance between the hypothesized entanglement points of the transient statistical lattice seen in the deGennes model\cite{degennes1979aDp} for polymer solutions.  For small particles, this form predicts $s \approx s_{0}$, while for large particles the exponential is dominated by $\eta_{s}/\eta$, even for $\eta_{s}/\eta \ll 1$, so that $s$ is determined by the solution's macroscopic viscosity.  The Langevin-Rondelez equation\cite{langevin1978aDp} is a special case of general assertions that: polymer solutions have a longest distance scale $\xi$, properties measured over distances $s \gg \xi$ necessarily reflect macroscopic system properties, and therefore $D_{p}$ of large ($R \gg \xi$) probes must follow the Stokes-Einstein equation.  If this proposal were correct, the polyacrylic acid:water systems in which the Stokes-Einstein equation fails badly would be in this regime $R \ngeqslant \xi$, implying a surprising $\xi$ of hundreds of nanometers. 

Langevin and Rondelez report sedimentation experiments leading to $\nu \approx 0.62$, $\delta \approx 1$ being consistent with limited measurements, and $A \sim M^{0}$ in the limit of low probe concentration. Langevin and Rondelez do not treat diffusion explicitly.  The same drag coefficient determines $s$ and $D_{p}$, leading to $D_{p}/D_{p0} \sim s/s_{0}$, implying that eq.\ \ref{eq:slangevinDp} should also govern probe diffusion.  Bu and Russo\cite{bu1994aDp} examine the diffusion of extremely small probes (0.5-55 nm) in HPC:water. In one of the few published explicit tests of eq.\ \ref{eq:slangevinDp} as applied to diffusion, Bu and Russo find the dependence of $D_{p}/D_{p0}$ on $R$ to match Langevin and Rondelez's prediction. Small probes diffuse much more rapidly that expected from $\eta$; for larger probes $\eta_{\mu}$ approaches $\eta$.   

A result $A \sim M^{\gamma}$ for $\gamma \neq 0$ is entirely inconsistent with the physical model invoked by Langevin and Rondelez in their {\em ansatz} for their equation.  However, in large numbers of probe:polymer systems $D_{p}/D_{p0}$ depends strongly on $M$.  It is at present unclear whether the strong $M$-dependence of  $D_{p}/D_{p0}$ would disappear with sufficiently small probes, in which case eq.\ \ref{eq:slangevinDp} could be correct, or whether $D_{p}/D_{p0}$ is determined by an equation similar to \ref{eq:slangevinDp} but in which $A$ depends on $M$.  

\subsection{Concentration Dependence of $D_{p}$}

As seen above, in almost but not quite every system whose relaxations can be characterized by a single diffusion coefficient $D_{p}$, the concentration dependence of $D_{p}$ is described by a stretched exponential in polymer concentration.  The fitting parameters $\alpha$ and $\nu$ of the stretched exponentials are found in Tables I-V.  It is apparent that $\alpha$ varies over a substantial range, and that $\nu$ varies over a considerably narrower range. Whenever $D_{p}$ and $c$ were varied over a sufficient range to support a credible log-log plot of $D_{p}$ against $c$ (e.g., Figures \ref{figurefurukawa1991aDp1}b, \ref{figurelin1984bDp6}, \ref{figurelin1984aDp5}, and \ref{figurewon1994aDp1}), the apparent slope $d \log(D_{p})/d \log(c)$ decreases monotonically with increasing $c$, precisely as expected for data that follow a stretched exponential in $c$.  

If $D_{p}$ followed a power law in $c$, a plot of $\log(D_{p})$ against $\log (c)$ would reveal a straight line.  It is possible on the aforementioned figures to draw straight lines tangent to the observed $D_{p}(c)$ curves, but such straight lines manifestly are simply local tangents.  Furthermore, a log-log plot of a stretched exponential in $c$, no matter its parameters, always shows a smooth curve, while on the same plot a power-law is a straight line.  Measurements that actually follow a power law in $c$ therefore cannot be described accurately with a force-fitted stretched exponential, and vice versa.

\begin{figure}[t] 
\includegraphics{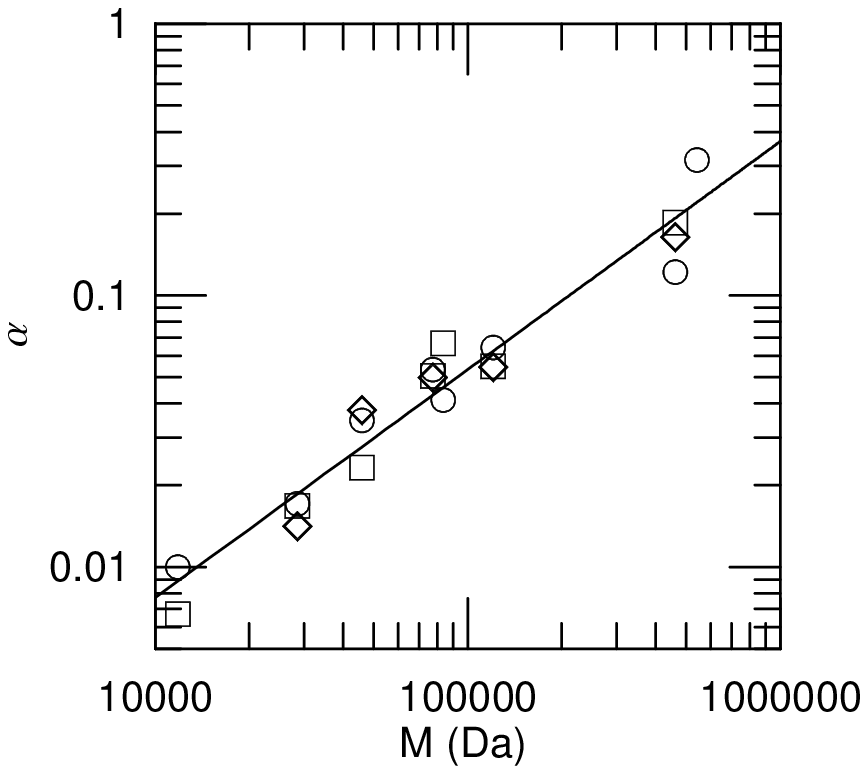}
\includegraphics{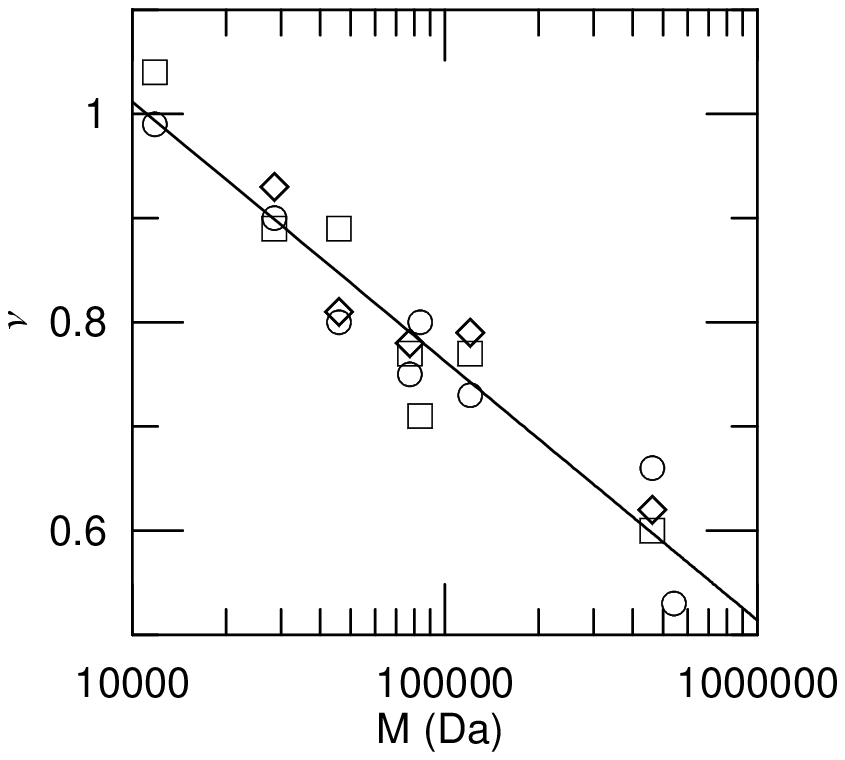}
\caption{\label{figurephillies1989bDp5} Parameters $\alpha$ and $\nu$ from fits of $D_{p}$ of 20.4 ($\circ$) and 230 ($\square$) nm spheres in dextran solutions, and $\eta$ of those solutions, to $\exp(-\alpha c^{\nu})$, against dextran $M_{w}$, showing $\alpha \sim M^{0.84}$ with $\nu$ decreasing from 1.0 toward 0.5 over the observed molecular weight range.}
\end{figure} 

\begin{figure}[tbh] 
\includegraphics{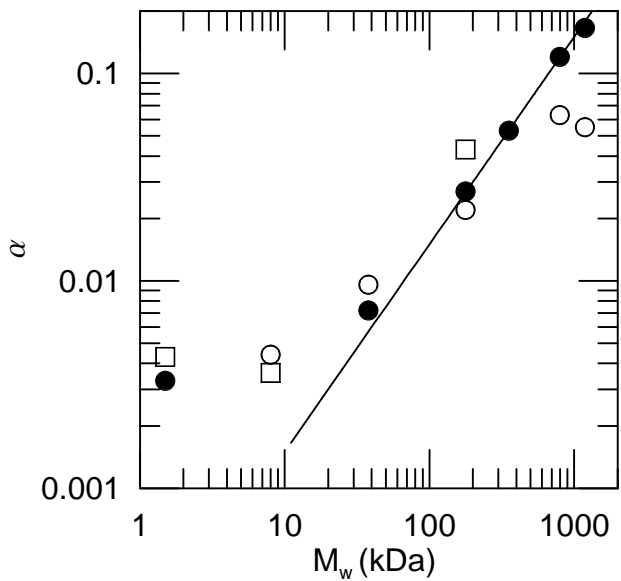}
\caption{\label{figurephillies1997bDp3} $\alpha$, the initial slope of $D_{p}(c)$ against $M$ for 7 ($\bigcirc$), 34 ($\bullet$), and 95($\square$) nm radius polystyrene spheres in aqueous polystyrene sulfonate: 0.2 M NaCl, and (line) the no-free-parameter prediction for $\alpha(M)$ in this system, after Phillies, et al.\cite{phillies1997bDp}.}
\end{figure} 

Relationships between $\alpha$, $\nu$, and $M$ for probes in an extensive series of homologous polymers were studied by Phillies, et al.\cite{phillies1989bDp} (PSL in dextran: water) and by Phillies, et al.\cite{phillies1997bDp} (PSL in polystyrenesulfonate:water.)  Representative results on these systems appear as Figures \ref{figurephillies1989bDp1} and \ref{figurephillies1997bDp2}, respectively. Figure \ref{figurephillies1989bDp5} shows $\alpha$ and $\nu$ for probes in dextran:water\cite{phillies1989bDp}. Over the observed molecular weight range, $\alpha \sim M_{w}^{0.84}$.  With increasing $M$, $\nu$ decreases from 1.0 to slightly more than 0.5.  The authors determined the polymer molecular weight distributions using aqueous size-exclusion chromatography; measured polydispersities $M_{w}/M_{n}$ ranged from 1.16 to 2.17. No effect of polydispersity on probe diffusion was identified. These results show that $\alpha$ depends strongly on $M$, and serve to reject any suggestion that polymer dynamics are strongly sensitive to the detailed shape of the polymer molecular weight distribution, as opposed to being determined by averaged molecular weights. 

Phillies, et al.\cite{phillies1997bDp} sought to determine the initial slope of $D_{p}(c)$ for probes in a series of of homologous monodisperse polymers. The intent was to test the Phillies-Kirkitelos\cite{phillies1993tDp} hydrodynamic calculation of $\alpha$.  Measurements were fit to straight lines, simple exponentials, and stretched exponentials in $c$, which describe these data out to progressively larger $c$.  The Phillies-Kirkitelos\cite{phillies1993tDp} calculation, which parallels hydrodynamic calculations of the concentration dependence of $D_{s}$ of hard spheres, has no free parameters.  The predicted $\alpha$ depends on probe radius $a_{p}$, chain thickness $a_{c}$, and $M$.  So long as $a_{p} \gg a_{c}$, the calculated $\alpha$ is very nearly insensitive to $a_{p}$ and $a_{c}$.  As seen in Figure \ref{figurephillies1997bDp3}, for $M \geq 38$ kDa good agreement  is found between the measurements and the no-free-parameter calculation.  For $M \leq 20$ kDa, the experimental $\alpha$ is too large, a deviation interpreted by Phillies, et al.\cite{phillies1997bDp} as appearing because short polystyrene sulfonate chains are not well-approximated as random coils.

\begin{figure}[t] 
\includegraphics{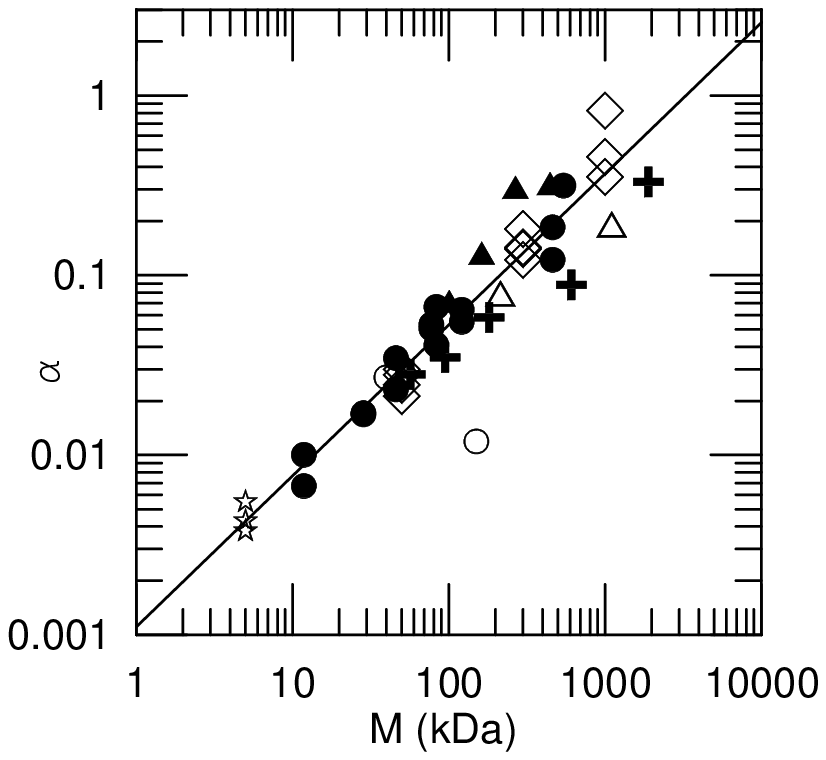}
\includegraphics{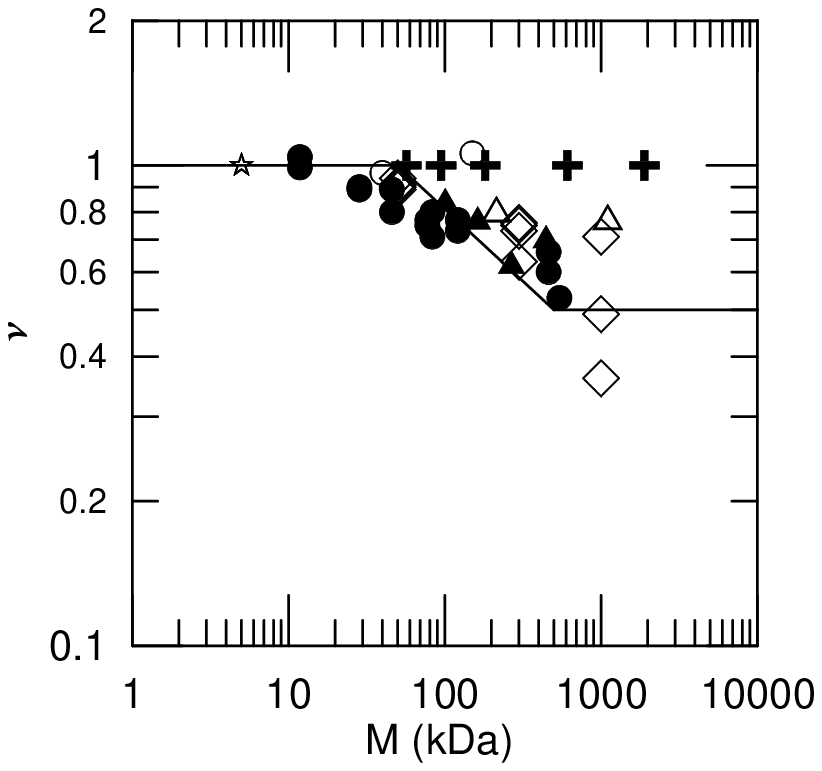}
\caption{\label{figuredataprobeDp1} $\alpha$ and $\nu$ of $\exp(-\alpha c^{\nu})$ as functions of $M$, for mesoscopic probes in solutions of PMMA ($\blacktriangle$)\cite{brown1988cDp}, dextran ($\circ$)\cite{furukawa1991aDp} ($\bullet$)\cite{phillies1989bDp}, polyacrylic acid ($\lozenge$)\cite{lin1984aDp},  polystyrene($\bigtriangleup$)\cite{onyenemezu1993aDp}, polyisobutylene ($+$), and bovine serum albumin ($\star$)\cite{ullmann1985bDp} (plotted based on its hydrodynamic radius).  The lines are $\alpha \sim M^{0.84}$, based on probes in dextran:water\cite{phillies1989bDp}, and $\nu = \nu(M)$ based on Fig.\ 65 of Ref. \onlinecite{phillies2004xDp}.}
\end{figure} 

Accurate determination of $\alpha$ and $\nu$ from non-linear least squares fits is challenging, because errors in these parameters are strongly anticorrelated.  To determine $\alpha$ and $\nu$ accurately, one needs accurate measurements of $D_{p}(c)$ at quite small, not just large, concentrations, as well as over a wide range of $D_{p}(c)/D_{p}(0)$.  Only a limited number of studies seem to attain these conditions. Furthermore, the systematic review above reveals systems with multimodal spectra, systems in which $D_{p}(c)$ exhibits re-entrant behavior, and some measurements on polyelectrolyte and rodlike polymer solutions, none of which appear exactly comparable to mesoscopic probes diffusing in solutions of near-random-coil neutral polymers.  

With these exclusions, Figures \ref{figuredataprobeDp1} exhibit $\alpha$ and $\nu$ as functions of $M$ for probes in a range of polymer solutions.  The solid line $\alpha \sim M^{0.84}$ is taken from Figure 65, Ref.\ \ref{figurephillies1989bDp5}; it represents a best fit to the measurements on dextran solutions. With few exceptions, $\alpha$ is seen to be closely correlated with the molecular weight of the matrix polymer over two orders of magnitude in its dependent and independent variables.  From Figure \ref{figuredataprobeDp1}, one sees $\nu \approx 1$ at lower polymer molecular weight, and more scattered values at larger $M$.  Probes in polyisobutylene are noteworthy for retaining $\nu \approx 1$ up to the largest $M$ examined.  The solid line in Figure \ref{figuredataprobeDp1} corresponds to the dependence of $\nu$ found for the not entirely physically dissimilar polymer self-diffusion coefficient.  As seen in Figure 65 of Ref.\ \onlinecite{phillies2004xDp}, $\nu$ from $D_{s}$ is $\approx 1.0$ at lower $M$, declining to 0.5 by $M \approx 500$ kDa. $\alpha$ and $\nu$ are thus seen to be correlated with matrix molecular weight $M$, a correlation that will be predicted by a valid physical model for these parameters.

A few systems show re-entrant behavior in which a relaxation rate increases rather than decreasing with increasing $c$. Most observations of reentrant behavior involve systems showing bimodal spectra.  For example, Bremell\cite{bremmell2001aDp,bremmell2002aDp}, and Dunstan and Stokes\cite{dunstan2000aDp} report for low-ionic-strength solutions a slow probe mode that slows with increasing $c$ and a fast probe mode whose relaxation rate first increases and then decreases with increasing $c$.  With the same polymer and elevated (0.1, 1.0 M) ionic strengths, the relaxation rates of both modes increase and perhaps plateau with increasing $c$.  Ullmann, et al. do report a probe in polyethylene oxide:water whose single relaxation rate has a true maximum with increasing $c$.  However, Ullmann, et al.'s data were from an old-style linear correlator, so their spectra might have been bimodal, the second peak passing unseen.  

In a very few systems, an apparent low-$c$ plateau is observed, in which $D_{p}$ is nearly constant over a range of low concentrations and then begins to fall above some low-concentration limiting $c$.  Good examples of this behavior are supplied by Yang and Jamieson\cite{yang1988aDp} for probes in HPC:water, and by Papagiannapolis, et al.\cite{papag2005aDp} (who find the plateau in $\eta_{\mu}$ rather than $D_{p}$).

\subsection{Particle Size Effects}

The evidence above on the effect of probe size reduces to two general statements with a qualification.  First, true gels, e.g., polyacrylamide gels, protein gels, or cellular interiors, are indeed size filters, as witness the results of, e.g., Luby-Phelps, et al.\cite{lubyphelps1986aDp,lubyphelps1987aDp}. True gels are substantially more effective at retarding the diffusion of larger probes than at retarding the diffusion of smaller probes.  Second, to first approximation polymer solutions are relatively weak size filters.  As witness the results of Hou, et al.\cite{hou1990aDp}, polymer solutions are only slightly more effective at retarding the diffusion of smaller probes that at retarding the diffusion of larger probes.  However, the approximation that polymer solutions are weak size filters breaks down for very small probes, as studied by Bu and Russo\cite{bu1994aDp}.  In a polymer solution, a very small probe may be considerably more mobile, relative to its size, than a large probe would be.

It is legitimate to ask if rheology, and probe diffusion using sufficiently large albeit mesoscopic probes, should measure the same viscosity.  It is noteworthy that the few true microrheological measurements on forced motion of micron-scale probes indicate that true microrheology and classical rheology do not always measure the same quantities. In simple liquids\cite{schmidt2000aDp} and in rigid-rod fd-virus solutions\cite{schmidt2000bDp}, macroscopic and true microscopic measurements of viscosity are consistently in good agreement.  On the other hand, in low-molecular-weight 85 kDa polyethylene oxide solutions, dynamic moduli measured with a 1.6 $\mu$m probe are\cite{hough1999aDp} 'quite different from those obtained by a macroscopic rheometer'.  In F-actin solutions, true microrheological moduli are substantially smaller than their macrorheological counterparts\cite{schmidt2000aDp}, typically by factors of 2-8.   Only very limited true microrheological measurements exist for neutral random-coil polymers in good solvents, so generalization appears hazardous.  Because microscopic and classical macroscopic rheological instruments sometimes give different values for $\eta$, there is no reason to expect that $\eta_{\mu}$ from $D_{p}$ will agree with $\eta$ from a classical macroscopic instrument, rather than agreeing with the very different $\eta$ from a microscopic instrument. 

\subsection{Comparison of Probe Diffusion and Polymer Self-Diffusion}

Phillies, Brown, and Zhou\cite{phillies1992aDp} re-examine results of Brown and collaborators\cite{brown1988cDp,zhou1989aDp} on probe diffusion by silica spheres and tracer diffusion of polyisobutylene (PIB) chains through PIB: chloroform solutions.  These comparisons are the most precise available in the literature, in the sense that all measurements were made in the same laboratory using exactly the same matrix polymer samples. Comparisons were made between silica sphere probes and polymer chains have very nearly equal $D_{p}$ and $D_{t}$ in the absence of PIB.  For each probe sphere and probe chain, the concentration dependence of the single-particle diffusion coefficient is accurately described by a stretched exponential in $c$. For large probes (160 nm silica spheres, 4.9 MDa PIB) in solutions of a small (610 kDa) PIB, $D_{p}(c)/D_{t}(c)$ remains very nearly independent of $c$ as $D_{p}(c)$ falls 100-fold from its dilute solution limit.  In contrast, with small probes (32 nm silica spheres, 1.1 MDa PIB) in solutions of a large polymer (4.9 MDa PIB), $D_{p}(c)/D_{t}(c)$ depends very strongly on $c$.  Between 0 and 6 g/L matrix PIB, $D_{t}$ falls nearly 300-fold, while over the same concentration range $D_{p}$ falls no more than 10-fold.  $D_{p}$ was followed out to the considerably larger concentration at which it, too, had fallen 300-hundred-fold from its small-matrix-concentration limit.  This result is somewhat surprising relative to some theories of polymer dynamics, because in non-dilute polymer solutions polymer chains are predicted\cite{degennes1979aDp} to have available modes of motion, such as reptation, that are predicted to be denied to rigid spheres.

That is, in solutions of a very large (4.9 MDa) polymer, at concentrations large enough that $D_{p}$ was reduced 300-hundred-fold, the matrix polymer was far more effective at obstructing the motions of rigid probes than at obstructing the motion of long flexible chains.  One could propose that at still larger $c$ the trend in $D_{p}(c)/D_{t}(c)$ reverses with increasing $c$, but there is absolutely no sign of such a phenomenon in the published experiments.  In the 4.9 MDa system, $c [\eta]$ was only taken out to 2 or 6, so it could reasonably be argued that the observed phenomenology only refers to non-dilute non-entangled systems, and that a different phenomenology might be encountered in entangled polymer solutions.

\subsection{Systems Having Multimodal Spectra}

While multimodal probe spectra have been reported in several systems, only in hydroxypropylcellulose: water is there a highly systematic examination of the full range of experimental parameters.  Because only one system showing multimodal spectra has been studied systematically, one cannot be certain to what extent one is examining specific chemical effects rather than a general phenomenology.  A few specific points seem most worth repeating: Probe and probe-free HPC solution spectra are both bimodal or trimodal, with modes on the same time scales.  However, the probe and polymer modes are not the same: The concentration dependences of a probe mode and the corresponding polymer mode can be opposite in sign.  

A wide range of phenomena indicate that these solutions have a dominant characteristic length scale $\xi$, $\xi$ being approximately the size of a polymer chain, and not some shorter distance. In particular, while small and large probes both show bimodal spectra, probes having $R < \xi$ and $R > \xi$ show very different concentration dependences for their spectral parameters.  The observed characteristic length $\xi$ is approximately $2 R_{g}$, the diameter of a random chain coil.  $\xi$ depends weakly or not at all on polymer concentration. In contrast, the hypothesized characteristic length scale between chain entanglements is a distance, smaller than a polymer chain, that depends strongly on polymer concentration.  Taken together these results serve to reject interpretations of the viscoelastic solutionlike:meltlike transition in this system as representing a transition to reptation dynamics.  However, it is not certain if chain dynamics in HPC:water are representative of generic polymer dynamics, or if special chemical effects unique to HPC are important, so it appears premature to generalize these interpretations to all polymer solutions.

\section{Summary}

The experimental literature on the motion of mesoscopic probe particles through polymer solutions is systematically reviewed.  The primary focus is the study of diffusive motion of small probe particles.  Comparison is made with literature data on solution viscosities. A coherent description was obtained, namely that the probe diffusion coefficient generally depends on polymer concentration as $D_{p} = D_{p0} \exp(-\alpha c^{\nu})$.  One finds that the scaling prefactor $\alpha$ depends on polymer molecular weights as $\alpha \sim M^{0.8}$. The scaling exponent $\nu$ appears to have large-$M$ and small-$M$ values with a crossover linking them. The probe diffusion coefficient does not simply track the solution viscosity; instead, $D_{p}\eta$ typically increases markedly with increasing polymer concentration and molecular weight.  In some systems, e.g., hydroxypropylcellulose:water, the observed probe spectra are bi- or tri-modal.  Extended analysis of the full probe phenomenology implies that hydroxypropylcellulose solutions are characterized by a single, concentration-independent, length scale that is approximately the size of a polymer coil.  In a very few systems, one sees re-entrant or low-concentration-plateau behaviors of uncertain interpretation.  From their rarity, these behaviors are reasonably interpreted as corresponding to specific chemical effects.  True microrheological studies examining the motion of mesoscopic particles under the influence of a known external force are also examined. The viscosity determined with a true microrheological measurement is in many cases substantially smaller than the viscosity measured with a macroscopic instrument.



\pagebreak

\begin{table*}[H] 
\caption{\label{tab:tablei}
Fits of the concentration dependence of the probe
coefficient $D_{p}$ to 
a stretched exponential $D_{p0} 
\exp(-\alpha c^{\nu})$, for probes having radius $R$ diffusing through polymers having molecular weight $M$.  The Table gives the fitting parameters, polymer molecular weight $M$, probe radius $R$, the polymer:solvent system, and the reference.  $D_{p0}$ is in units $10^{-7}$ cm$^{2}$/s or in units of $D_{p}(0)$.  Concentrations are converted to g/L  except g/g marks g/g as the concentration units.  Polymers include BSA--bovine serum albumin, CMC--carboxymethylcellulose, dextran, ficoll, HPC--hydroxypropylcellulose, HPSS--nonneutralized polystyrene sulfonate, PAA--polyacrylic acid, PEO--polyethylene oxide, PLL--poly-L-lysine, PMMA--polymethylmethacrylate, NaPSS--sodium polystyrene sulphonate, PVME--polyvinylmethylether.  Solvents include phosphate--aqueous phosphate buffer: $I=n$M -- background ionic strength was $n$ molar. In the $D_{p0}$ column, * indicates $D_{p}/D_{p0}$, $f$ = fast mode, $s$ = slow mode.}
\begin{ruledtabular}
\begin{tabular}{|r|r|r|r|r|l|r|} \hline
$D_{p0}$ & $\alpha$& $\nu$  & $M$(kDa) & $R$(nm)& System& Refs \\ \hline
2.35 & 0.53 & 0.36 & 3100 & 100 & polyacrylamide:H$_{2}$O & \cite{bremmell2001aDp} \\
1.061 & 0.0682 & 0.834 & 110 & 160 & PMMA: CHCl$_{3}$ & \cite{brown1988cDp} \\
1.082 & 0.1268 & 0.764 & 163 & 160 & PMMA: CHCl$_{3}$ & \cite{brown1988cDp} \\
1.359 & 0.294 & 0.618 & 268 & 160 & PMMA: CHCl$_{3}$ & \cite{brown1988cDp} \\
1.20 & 0.308 & 0.698 & 445 & 160 & PMMA: CHCl$_{3}$ & \cite{brown1988cDp} \\
42.9 & $9.43 \times 10^{-3}$ & 1 & 300 & 0.50 & HPC:H$_{2}$O & \cite{bu1994aDp} \\
19.53 & $1.33 \times 10^{-3}$ & 1 & 300 & 1.32 & HPC:H$_{2}$O & \cite{bu1994aDp} \\
13.12 & $8.89 \times 10^{-3}$ & 1 & 300 & 1.69 & HPC:H$_{2}$O & \cite{bu1994aDp} \\
7.77 & $1.77 \times 10^{-2}$ & 1 & 300 & 2.78 & HPC:H$_{2}$O & \cite{bu1994aDp} \\
5.02 & $2.65 \times 10^{-2}$ & 1 & 300 & 4.45 & HPC:H$_{2}$O & \cite{bu1994aDp} \\
3.44 & $3.44 \times 10^{-2}$ & 1 & 300 & 5.78 & HPC:H$_{2}$O & \cite{bu1994aDp} \\
2.32 & $4.51 \times 10^{-2}$ & 1 & 300 & 8.80 & HPC:H$_{2}$O & \cite{bu1994aDp} \\
1.49 & $6.31 \times 10^{-2}$ & 1 & 300 & 13.3 & HPC:H$_{2}$O & \cite{bu1994aDp} \\
0.897 & 0.107  & 1 & 300 & 17.9 & HPC:H$_{2}$O & \cite{bu1994aDp} \\
0.703 & 0.251& 1 & 300 & 55.1 & HPC:H$_{2}$O & \cite{bu1994aDp} \\
11.89 & $2.73 \times 10^{-3}$ & 1.64 & 0.1 & 1.66 & glycerol:H$_{2}$O & \cite{busch2000aDp} \\
12.82 & 0.0144 & 0.949 &  & 1.66 & ficoll 70:H$_{2}$O & \cite{busch2000aDp} \\
14.56 & 0.222& 0.49 & 160bp & 1.66 & DNA:H$_{2}$O & \cite{busch2000aDp} \\
1.007 & $4.21 \times 10^{-2}$ & 1 & 65 & 64 & polyacrylamide:H$_{2}$O & \cite{cao1997aDp} \\
0.995 & $2.27 \times 10^{-2}$ & 1 & 65 & 65 & polyacrylamide:H$_{2}$O & \cite{cao1997aDp} \\
1.007 & 0.23 & 1 & 1000 & 64 & polyacrylamide:H$_{2}$O & \cite{cao1997aDp} \\
0.98 & $7.8 \times 10^{-2}$ & 1 & 1000 & 65 & polyacrylamide:H$_{2}$O  & \cite{cao1997aDp} \\
6.48 & 0.246 g/g& 1.00 & 2000 & 3.15 & PEO:H$_{2}$O & \cite{cheng2002aDp} \\
4.26 & 0.59 g/g& 1.00 & 2000 & 5.72 & PEO:H$_{2}$O & \cite{cheng2002aDp} \\
7.72 & 0.388 g/g& 0.633 & 40 & 3.15 & PEO:H$_{2}$O & \cite{cheng2002aDp} \\
9.82 & 0.34 g/g& 0.73 & 40 & 2.71 & PEO:H$_{2}$O & \cite{cheng2002aDp} \\
4.19 & 0.661 g/g& 0.83 & 200 & 5.47 & PEO:H$_{2}$O & \cite{cheng2002aDp} \\
4.15 & 0.428 g/g& 0.80 & 200 & 5.72 & PEO:H$_{2}$O & \cite{cheng2002aDp} \\
5.73 & 0.574& 0.73 & 647 & TMV & dextran:H$_{2}$O & \cite{cush2004aDp} \\
5.91 & 0.183& 0.93 & 428 & TMV & ficoll:H$_{2}$O & \cite{cush2004aDp} \\
4.46 & 0.178 & 0.59 & 680 & 71kDa &dextran: H$_{2}$O & \cite{desmedt1994aDp}\\
3.18 & 0.318 & 0.59 & 680 & 148kDa &dextran: H$_{2}$O & \cite{desmedt1994aDp}\\
1.62 & 0.368 & 0.53 & 680 & 487kDa &dextran: H$_{2}$O & \cite{desmedt1994aDp}\\
4.20 & 0.0636 & 0.616 & 680 & 19kDa &dextran: H$_{2}$O & \cite{desmedt1997aDp}\\
2.33 & 0.0726 & 0.61 & 680 & 51kDa &dextran: H$_{2}$O & \cite{desmedt1997aDp}\\
1.67 & 0.0295 & 0.796 & 680 & 148kDa &dextran: H$_{2}$O & \cite{desmedt1997aDp}\\
1.07 & 0.076 & 0.636 & 680 & 487kDa &dextran: H$_{2}$O & \cite{desmedt1997aDp}\\
1.416*f & 0.247 & 1 & 700 & 102 & CMC: H$_{2}$O & \cite{delfino2005aDp}\\
0.449*s & 1.179 & 0.775 & 700 & 102 & CMC: H$_{2}$O & \cite{delfino2005aDp}\\
0.844*f & 0.227 & 1 & 700 & 47 & CMC: H$_{2}$O & \cite{delfino2005aDp}\\
0.539*s & 0.676 & 1 & 700 & 47 & CMC: H$_{2}$O & \cite{delfino2005aDp}\\
1.011*f & 0.762 & 1 & 700 & 14 & CMC: H$_{2}$O & \cite{delfino2005aDp}\\
0.358*s & 0.150 & 1 & 700 & 14 & CMC: H$_{2}$O & \cite{delfino2005aDp}\\
0.229 & 5.04 & 0.37 & 3100 & 100 & polyacrylamide: H$_{2}$O & \cite{dunstan2000aDp}\\
1.02* & 0.0119 & 1.06 & 150& 19 &dextran: H$_{2}$O & \cite{furukawa1991aDp}\\
1.00* & 0.027 & 0.965 & 40 & 19 &dextran: H$_{2}$O & \cite{furukawa1991aDp}\\
1.034 & 0.0927 & 0.81 & 350 & 200 & PMMA: thf & \cite{gold1996aDp} \\
1.066 & 0.0328 & 1.07 & 350 & 200 & PMMA: dmf & \cite{gold1996aDp} \\
0.964 & 0.0644 & 1.05 & 350 & 200 & PMMA: diox & \cite{gold1996aDp} \\
\hline
\end{tabular}
\end{ruledtabular}
\end{table*}
\begin{table*}
\begin{ruledtabular}
\begin{tabular}{|r|r|r|r|r|l|r|} \hline
$D_{p0}$ & $\alpha$& $\nu$  & $M$(kDa) & $R$(nm)& System& Refs \\ \hline
0.87 & 0.209 & 0.76 & 500 & 19 & NaPSS: 1  mM phosphate & \cite{gorti1985aDp} \\
0.998 & 0.251 & 0.798 & 500 & 19 & NaPSS: 5 mM phosphate & \cite{gorti1985aDp} \\
1.014 & 0.488 & 0.624 & 500 & 19 & NaPSS: 10 mM phosphate & \cite{gorti1985aDp} \\
0.87 & 0.195 & 0.887 & 500 & 19 & NaPSS: 20 mM phosphate & \cite{gorti1985aDp} \\
1.028 & 0.48 & 0.679 & 500 & 19 & NaPSS: 50 mM phosphate & \cite{gorti1985aDp} \\
1.041 & 0.454 & 0.37 & 500 & 19 & HPSS: NaPSS: 20 mM phosphate & \cite{gorti1985aDp} \\
1.007 & 0.0367& 0.546 & 70 & 19 & NaPSS: 10 mM phosphate & \cite{gorti1985aDp} \\
0.985 & $1.42 \times 10^{-3}$ & 1.68 & 500 & 0.5 & NaPSS: 10 mM phosphate & \cite{gorti1985aDp} \\
1.08 & 0.354 & 0.487 & 500 & 5 & NaPSS: 10 mM phosphate & \cite{gorti1985aDp} \\
1.015 & 0.493 & 0.614 & 500 & 19 & NaPSS: 10 mM phosphate & \cite{gorti1985aDp} \\
1.05 & 0.48 & 1 & 4000 & 193& xanthan: H$_{2}$O & \cite{koenderink2004aDp} \\
0.998 & 0.0369 & 0.93& 110 & 23 & pS:dioxane & \cite{konak1982aDp}\\
1.052 & 0.0858 & 0.803& 200 & 23 & pS:dioxane & \cite{konak1982aDp}\\
0.85* & 0.0212 & 0.94 & 50 & 20 & PAA: H$_{2}$O & \cite{lin1984aDp} \\
0.995* & 0.028 & 0.89 & 50 & 80 & PAA: H$_{2}$O & \cite{lin1984aDp} \\
0.990* & 0.0246 & 0.90 & 50 & 620 & PAA: H$_{2}$O & \cite{lin1984aDp} \\
1.26* & 0.030 & 0.91 & 50 & 1500 & PAA: H$_{2}$O & \cite{lin1984aDp} \\
0.685* & 0.14 & 0.75 & 300 & 20 & PAA: H$_{2}$O & \cite{lin1984bDp} \\
0.796* & 0.143 & 0.73 & 300 & 80 & PAA: H$_{2}$O & \cite{lin1984bDp} \\
0.823* & 0.122 & 0.76 & 300 & 620 & PAA: H$_{2}$O & \cite{lin1984bDp} \\
0.99* & 0.181 & 0.63 & 300 & 1500 & PAA: H$_{2}$O & \cite{lin1984bDp} \\
1.09* & 0.354 & 0.49 & 1000 & 20 & PAA: H$_{2}$O & \cite{lin1984aDp} \\
1.14* & 0.457 & 0.36 & 1000 & 80 & PAA: H$_{2}$O & \cite{lin1984aDp} \\
1.12* & 0.822 & 0.71 & 1000 & 620 & PAA: H$_{2}$O & \cite{lin1984aDp} \\
0.20& 0.149 & 0.73& 90 & 80 & PLL: H$_{2}$O & \cite{nehme1989aDp} \\
0.328 & 0.116 & 1   & 90   & 49  & PLL: H$_{2}$O & \cite{nehme1989aDp} \\
0.154 & 0.192 & 0.78& 1110 & 200 & PS: dmf & \cite{onyenemezu1993aDp}\\
0.154 & 0.079 & 0.81& 215 & 200 & PS : dmf & \cite{onyenemezu1993aDp}\\
1.28& 1.07  & 0.39     &560 & 20 & PAA: H$_{2}$O: $I=0M$ & \cite{phillies1987aDp} \\
1.27& 0.554  & 0.705     &560 & 20 & PAA: H$_{2}$O: $I=0.01M$ & \cite{phillies1987aDp} \\
1.19& 0.404  & 0.789     &560 & 20 & PAA: H$_{2}$O: $I=0.02M$ & \cite{phillies1987aDp} \\
1.20& 0.304  & 0.921     &560 & 20 & PAA: H$_{2}$O: $I=0.1M$ & \cite{phillies1987aDp} \\
- & .01 &  0.99& 11.8 & 20.4 & dextran: H$_{2}$O & \cite{phillies1989bDp} \\
-&  0.0171 & 0.90 & 28.5 & 20.4& dextran: H$_{2}$O & \cite{phillies1989bDp} \\
- & .0346 & 0.80 & 46 & 20.4 & dextran: H$_{2}$O & \cite{phillies1989bDp} \\
- & 5.32$\times 10^{-2}$ & 0.75 &  77.2  & 20.4 & dextran: H$_{2}$O & \cite{phillies1989bDp} \\
- &4.11$\times 10^{-2}$ & 0.80 & 83.5  & 20.4 & dextran: H$_{2}$O & \cite{phillies1989bDp} \\
- &  6.43$\times 10^{-2}$ & 0.73 & 121  & 20.4 & dextran: H$_{2}$O & \cite{phillies1989bDp} \\
- & 0.122 & 0.66& 462  & 20.4 & dextran: H$_{2}$O & \cite{phillies1989bDp} \\
- & 0.315 & 0.53 & 542  & 20.4 & dextran: H$_{2}$O & \cite{phillies1989bDp} \\
- & .0067 & 1.04 & 11.8  & 20.4 & dextran: H$_{2}$O & \cite{phillies1989bDp} \\
-& .0168 & 0.89 &28.5  & 20.4 & dextran: H$_{2}$O & \cite{phillies1989bDp} \\
- & .0232& 0.89 & 46  & 20.4 & dextran: H$_{2}$O & \cite{phillies1989bDp} \\
 - & 5.06$\times 10^{-2}$& 0.77 & 77.2  & 20.4 & dextran: H$_{2}$O & \cite{phillies1989bDp} \\
- & 0.0667 & 0.71 & 83.5 & 20.4 & dextran: H$_{2}$O & \cite{phillies1989bDp} \\
- & 0.0549& 0.77 & 121  & 20.4 & dextran: H$_{2}$O & \cite{phillies1989bDp} \\
-& .185&  0.60 & 462  & 20.4 & dextran: H$_{2}$O & \cite{phillies1989bDp} \\
0.577 & 0.206 & 0.721& 139 & 67 & HPC: H$_{2}$O 10$^{o}$ C & \cite{phillies1993aDp}\\
1.044 & 0.107 & 0.90& 139 & 67 & HPC: H$_{2}$O 39$^{o}$ C& \cite{phillies1993aDp}\\
0.556 & 0.315 & 0.633& 146 & 67 & HPC: H$_{2}$O 10$^{o}$ C & \cite{phillies1993aDp}\\*
\hline
\end{tabular}
\end{ruledtabular}
\end{table*}
\begin{table*}
\begin{ruledtabular}
\begin{tabular}{|r|r|r|r|r|l|r|} \hline
$D_{p0}$ & $\alpha$& $\nu$  & $M$(kDa) & $R$(nm)& System& Refs \\ \hline
1.05 & 0.174 & 0.793& 146 & 67 & HPC: H$_{2}$O 39$^{o}$ C& \cite{phillies1993aDp}\\
0.569 & 0.329 & 0.747& 415 & 67 & HPC: H$_{2}$O 10$^{o}$ C & \cite{phillies1993aDp}\\
0.1.159 & 0.274 & 0.789& 415 & 67 & HPC: H$_{2}$O 39$^{o}$ C& \cite{phillies1993aDp}\\
0.448 & 0.235 & 1.045& 1280 & 67 & HPC: H$_{2}$O 10$^{o}$ C & \cite{phillies1993aDp}\\
0.964 & 0.263 & 0.863& 1280 & 67 & HPC: H$_{2}$O 39$^{o}$ C& \cite{phillies1993aDp}\\
0.46 & 0.177 & 0.739& 139 & 67 & HPC: H$_{2}$O 10$^{o}$ C& \cite{phillies1993bDp}\\
0.85 & 0.0684 & 1.07& 139 & 67 & HPC: H$_{2}$O 41$^{o}$ C& \cite{phillies1993bDp}\\
1.054 & 0.062 & 1& 2000 & 93 & dextran:H$_{2}$O & \cite{turner1976aDp}\\
1.037 & 0.0629 & 1& 2000 & 183 & dextran:H$_{2}$O & \cite{turner1976aDp}\\
1.027 & 0.0688 & 1& 2000 & 246 & dextran:H$_{2}$O & \cite{turner1976aDp}\\
1.01 & 0.0127 & 1& 20 & 246 & dextran:H$_{2}$O & \cite{turner1976aDp}\\
1.027 & 0.0246 & 1& 70 & 246 & dextran:H$_{2}$O & \cite{turner1976aDp}\\
1.027 & 0.0357 & 1& 150 & 246 & dextran:H$_{2}$O & \cite{turner1976aDp}\\
1.027 & 0.0435 & 1& 500 & 246 & dextran:H$_{2}$O & \cite{turner1976aDp}\\
1.027 & 0.0667 & 1& 2000 & 246 & dextran:H$_{2}$O & \cite{turner1976aDp}\\
1.03 & 0.121 & 1 & 300 &  20.8 &PEO: H$_{2}$O & \cite{ullmann1985aDp}\\
1.09 & 0.0475 & 1 & 100 &  20.8 &PEO: H$_{2}$O & \cite{ullmann1985aDp}\\
1.09 & 0.0989 & 0.72 & 18 &  20.8 &PEO: H$_{2}$O & \cite{ullmann1985aDp}\\
1.18 & 0.050 & 0.755 & 7.5 &  20.8 &PEO: H$_{2}$O & \cite{ullmann1985aDp}\\
0.462 & 0.139 & 1 & 300 & 51.7 & PEO: H$_{2}$O & \cite{ullmann1985aDp}\\
0.48 & 0.0823 & 1 & 100 & 51.7 & PEO: H$_{2}$O & \cite{ullmann1985aDp}\\
0.474 & 0.067 & 0.82 & 18 & 51.7 & PEO: H$_{2}$O & \cite{ullmann1985aDp}\\
0.486 & 0.041 & 0.803 & 7.5 & 51.7 & PEO: H$_{2}$O & \cite{ullmann1985aDp}\\
0.071 & 0.091 & 1 & 300 & 322 &PEO: H$_{2}$O & \cite{ullmann1985aDp}\\
0.0733 & 0.0656 & 1 & 100 & 322 &PEO: H$_{2}$O & \cite{ullmann1985aDp}\\
0.0796 & 0.022 & 1 & 18 & 322 &PEO: H$_{2}$O & \cite{ullmann1985aDp}\\
0.0672 & 0.0148 & 1 & 7.5 & 322 &PEO: H$_{2}$O & \cite{ullmann1985aDp}\\
0.036 & 0.078 & 1 & 300 & 655 &PEO: H$_{2}$O & \cite{ullmann1985aDp}\\
0.0361 & 0.080 & 1 & 100 & 655 &PEO: H$_{2}$O & \cite{ullmann1985aDp}\\
0.035 & 0.072 & 1 & 18 & 7.5 &PEO: H$_{2}$O & \cite{ullmann1985aDp}\\
1.056 & $5.49 \times 10^{-3}$ & 1 & 5 nm & 5  &BSA: H$_{2}$O & \cite{ullmann1985bDp}\\
0.99 & $4.3 \times 10^{-3}$ & 1 & 5 nm & 322 &BSA: H$_{2}$O & \cite{ullmann1985bDp}\\
1.008 & $3.76 \times 10^{-3}$ & 1 & 5 nm & 655 &BSA: H$_{2}$O & \cite{ullmann1985bDp}\\
0.963 & $1.98 \times 10^{-2}$ & 1 & 115 & 5 nm &DNA:H$_{2}$O:0.01 M NaCl & \cite{wattenbarger1992aDp}\\
1.007 &  $1.51 \times 10^{-2}$ & 1 & 115 & 5 nm &DNA:H$_{2}$O:0.1 M NaCl & \cite{wattenbarger1992aDp}\\
0.251 & 0.394 & 0.75 & 1300 & 170  & PVME: toluene & \cite{won1994aDp} \\
1.05& 0.0736 & 1 &110 &60&HPC: H$_{2}$O &\cite{yang1988aDp}\\
1.06& 0.090& 1 &140 &60&HPC: H$_{2}$O&\cite{yang1988aDp}\\
0.986&0.104& 1 &450 & 60 &HPC: H$_{2}$O&\cite{yang1988aDp}\\
1.043&0.234& 1 &850 & 60&HPC: H$_{2}$O&\cite{yang1988aDp}\\
1.049& 0.072 & 1 &110 & 105 &HPC: H$_{2}$O&\cite{yang1988aDp}\\
1.024& 0.089 & 1 &140 &105&HPC: H$_{2}$O&\cite{yang1988aDp}\\
1.016& 0.189 & 1 &450 &105&HPC: H$_{2}$O&\cite{yang1988aDp}\\
1.011& 0.30 & 1 & 850 &105&HPC: H$_{2}$O&\cite{yang1988aDp}\\
1.029& 0.0995& 1 &110&175&HPC: H$_{2}$O&\cite{yang1988aDp}\\
1.016& 0.124 & 1 &140&175&HPC: H$_{2}$O&\cite{yang1988aDp}\\
0.993& 0.259 & 1 &450&175&HPC: H$_{2}$O&\cite{yang1988aDp}\\
0.935& 0.354 & 1 &850&175&HPC: H$_{2}$O&\cite{yang1988aDp}\\
0.70 & 0.33  & 1 & 1900 & 159 & PIB: CHCl$_{3}$ &\cite{zhou1989aDp} \\
0.59 & 0.0886 & 1 & 610 & 159 &PIB: CHCl$_{3}$ &\cite{zhou1989aDp} \\
0.684 & 0.0584 & 1 & 182 & 159 &PIB: CHCl$_{3}$ &\cite{zhou1989aDp} \\
0.895 & 0.035& 1 & 95 & 159 & PIB: CHCl$_{3}$ &\cite{zhou1989aDp} \\
1.036 & 0.0281 & 1 & 57 & 159 & PIB: CHCl$_{3}$ &\cite{zhou1989aDp} \\ 
\hline
\end{tabular}
\end{ruledtabular}
\end{table*}

\begin{table*}[H]
\caption{\label{tab:tableii}
Fits of the concentration dependence of the solution viscosity $\eta$ to 
a stretched exponential $\eta_{0} 
\exp(-\alpha c^{\nu})$ for solutions studied with probe diffusion.  The Table gives the fitting parameters, polymer molecular weight, polymer:solvent system, and the reference.  $\eta_{0}$ is in 
centipoise; concentrations are converted to 
g/L except g/g marks g/g as concentration units.  Polymers and solvents as above.}
\begin{ruledtabular}
\begin{tabular}{|r|r|r|r|r|r|} \hline
$\eta_{0}$ & $\alpha$& $\nu$  & $M$(kDa) & System& Refs \\ \hline
0.89 & 0.693& 0.69 & 647  & dextran:H$_{2}$O & \cite{cush2004aDp} \\
0.89 & 0.207& 0.91 & 428  & ficoll:H$_{2}$O & \cite{cush2004aDp} \\
1.045*  & 0.0113 &0.99 &150 & dextran:H$_{2}$O & \cite{furukawa1991aDp} \\
1.00*  & 0.122 &0.72 & 40 & dextran:H$_{2}$O & \cite{furukawa1991aDp} \\
0.948 & 0.181 & 0.71 & 350 & PMMA: thf & \cite{gold1996aDp} \\
1.045 & 0.0548 & 0.972 & 350 & PMMA: dmf & \cite{gold1996aDp} \\
1.00 & 0.0554 & 0.979 & 350 & PMMA: diox & \cite{gold1996aDp} \\
0.80 & 31.9g/g & 0.848 & 70 & PBLG:DMF & \cite{gold1996bDp} \\
0.74 & 52.8g/g & 0.669 & 233 & PBLG:DMF & \cite{gold1996bDp} \\
1 & 4.41 & 0.688 & 4000 & xanthan: H$_{2}$O & \cite{koenderink2004aDp} \\
0.89 & 0.0268 & 0.91 & 50 & PAA: H$_{2}$O & \cite{lin1984aDp} \\
0.89 & 3.13 & 0.803 & 1000 & PAA: H$_{2}$O & \cite{lin1984aDp} \\
0.87 & 0.178& 0.83& 1110 & PS: dmf & \cite{onyenemezu1993aDp}\\
0.90 & 0.109 & 0.767 & 215 & PS: dmf & \cite{onyenemezu1993aDp}\\
-& 0.0141 & 0.93 & 28.5& dextran: H$_{2}$O & \cite{phillies1989bDp} \\
-& .0378& 0.81 & 46& dextran: H$_{2}$O & \cite{phillies1989bDp} \\
-& 0.0498& 0.78 & 77.2& dextran: H$_{2}$O & \cite{phillies1989bDp} \\
-& 0.0544& 0.79 & 121& dextran: H$_{2}$O & \cite{phillies1989bDp} \\
-&0.164& 0.62 & 462& dextran: H$_{2}$O & \cite{phillies1989bDp} \\
1.26& 0.189 & 0.837 & 60 & HPC: H$_{2}$O (10 C) & \cite{phillies1993aDp}\\
0.66& 0.107 & 0.977 & 60 & HPC: H$_{2}$O (39 C) & \cite{phillies1993aDp}\\
1.305& 0.167 & 1.05 & 100 & HPC: H$_{2}$O (10 C) & \cite{phillies1993aDp}\\
0.66& 0.139 & 0.999 & 100 & HPC: H$_{2}$O (39 C) & \cite{phillies1993aDp}\\
 -- & 0.462 & 0.906 & 300 & HPC: H$_{2}$O (10 C) & \cite{phillies1993aDp}\\
 -- & 0.440 & 0.845 & 300 & HPC: H$_{2}$O (39 C) & \cite{phillies1993aDp}\\
 -- & 1.21 & 0.962 & 1000 & HPC: H$_{2}$O (10 C) & \cite{phillies1993aDp}\\
 -- & 0.931 & 1.04 & 1000 & HPC: H$_{2}$O (39 C) & \cite{phillies1993aDp}\\
0.914 & 0.207 & 1 & 300 & PEO: H$_{2}$O & \cite{ullmann1985aDp}\\
0.908 & 0.0901 & 1 & 100 & PEO: H$_{2}$O & \cite{ullmann1985aDp}\\
0.933 & 0.0064 & 1 & 18 & PEO: H$_{2}$O & \cite{ullmann1985aDp}\\
0.90 & 0.0413 & 0.80 & 7.5 & PEO: H$_{2}$O & \cite{ullmann1985aDp}\\
1.016 & 0.104 & 0.90 & 110 & PAA: H$_{2}$O & \cite{yang1988aDp}\\
0.998 & 0.144 & 0.868 & 140 & PAA: H$_{2}$O & \cite{yang1988aDp}\\
1.000 & 0.409 & 0.875 & 450 & PAA: H$_{2}$O & \cite{yang1988aDp}\\
0.950 & 0.582 & 0.896 & 850 & PAA: H$_{2}$O & \cite{yang1988aDp}\\
\hline
\end{tabular}
\end{ruledtabular}
\end{table*}

\begin{table*}
\caption{\label{tab:tableiia}
Fits of the concentration dependence of the solution viscosity $\eta$ to a power law $\bar{\eta } c^{x} + \eta_{0}$, for solutions studied with probe diffusion.  The Table gives the fitting parameters, lowest concentration $c_{m}$ included in the fit, polymer molecular weight, polymer:solvent system, and the reference.  $\bar{\eta}$ and $\eta_{0}$ are in centipoise; concentrations are converted to g/L except g/g marks g/g as concentration units.}
\begin{ruledtabular}
\begin{tabular}{|r|r|r|r|r|l|r|} \hline
$\bar{\eta}$& x& $\eta_{0}$ & $c_{m}$ & $M$(kDa) & System & Refs \\ \hline
$1.33 \times 10^{-3}$ & 2.85 & $3.16 \times 10^{-3}$ & 20 & 60 & HPC: H$_{2}$O (10 C) & \cite{phillies1993aDp}\\
$1.55 \times 10^{-3}$ & 2.62 & $-2.68 \times 10^{-4}$ &  18 & 60 & HPC: H$_{2}$O (39 C) & \cite{phillies1993aDp}\\
$9.37 \times 10^{-4}$ & 3.14 & $1.13 \times 10^{-4}$ & 18& 100 & HPC: H$_{2}$O (10 C) & \cite{phillies1993aDp}\\
$8.4 \times 10^{-4}$ & 2.96 & $1.1 \times 10^{-6}$ & 18 & 100 & HPC: H$_{2}$O (39 C) & \cite{phillies1993aDp}\\
$1.02 \times 10^{-3}$ & 4.40 & $1.14 \times 10^{-4}$ & 20 & 300 & HPC: H$_{2}$O (10 C) & \cite{phillies1993aDp}\\
$2.9 \times 10^{-4}$ & 4.41 & $9.88 \times 10^{-5}$ & 20 & 300 & HPC: H$_{2}$O (39 C) & \cite{phillies1993aDp}\\
0.36 & 3.99 & 0.057 & 6.0 & 1000 & HPC: H$_{2}$O (10 C) & \cite{phillies1993aDp}\\
0.092 & 4.01 & -0.069& 4.0  & 1000 & HPC: H$_{2}$O (39 C) & \cite{phillies1993aDp}\\ \hline
\end{tabular}
\end{ruledtabular}
\end{table*}

\pagebreak

\begin{table*}[H] 
\caption{\label{tab:tableiii}  Fits of the concentration dependence of the probe rotational diffusion coefficient $D_{r}$ or rotational relaxation time $\tau_{r}$ (in $\mu$S) to 
a stretched exponential $D_{r0} 
\exp(-\alpha c^{\nu})$ or $\tau_{r0} \exp(\alpha c^{\nu})$. The Table gives the fitting parameters, matrix molecular weight or (denoted "*") radius, probe radius or (denoted "*") major axis length, polymer:solvent system, and the reference.  $D_{r0}$ is in 
/second; concentrations are converted to 
g/L except ($^{**}$) volume fraction . SS--silica spheres.
Probes include TMV-tobacco mosaic virus, FS-fluorocarbon spheres: PBLG--polybenzyl-L-glutamate. Solvents include pyr-pyridine.}
\begin{ruledtabular}
\begin{tabular}{|r|r|r|r|r|l|r|} \hline
$D_{r0}$,$\tau_{r0}$ & $\alpha$& $\nu$  & $M$ (kDa or nm) & $R$ (nm) & System& Refs \\ \hline
346 & 0.85& 0.67 & 647 &  & TMV: dextran:H$_{2}$O & \cite{cush2004aDp} \\
336 & 0.183& 0.95 & 428 &  & TMV: ficoll:H$_{2}$O & \cite{cush2004aDp} \\
0.892 & 0.257 & 0.1.56 & 4000 & 93  &FS: xanthan: H$_{2}$O  & \cite{koenderink2004aDp} \\
15.4 & 1.19** & 1 & 62* & 68*& PBLG:SS:pyr:DMF &  \cite{phalak2000aDp}\\
32.9  &13.4** &1 &62* & 128* & PBLG:SS:pyr:DMF&  \cite{phalak2000aDp}\\
51.6   &16.1**  &1 & 62*& 170*& PBLG:SS:pyr:DMF&  \cite{phalak2000aDp}\\
\hline
\end{tabular}
\end{ruledtabular}
\end{table*}

\begin{table*}[H]
\caption{\label{tab:tableiv}  Fits of the concentration dependence of the fast ($D_{f}$) and slow ($D_{s}$) diffusion coefficients to $D_{0} \exp (-\alpha c_{\nu})$ for probes having bimodal spectra. "F/S" for fast or slow. $D_{p0}$ in units $10^{-7}$ cm$^{2}$/s. $R$ as a value for $\alpha$ denotes a mode showing re-entrance.  Polymers include PAAM-polyacrylamide and NaPAA-fully neutralized poly-acrylic acid.  Probes include PSL-polystyrene latex and hema-hematite. }
\begin{ruledtabular}
\noindent \begin{tabular}{|r|r|r|r|r|r|l|r|} \hline
F/S &$D_{p0}$ & $\alpha$& $\nu$  & $M$(kDa) & $R$(nm)& System& Refs \\ \hline
F &  & R & & 3000 & 200 & PSL: PAAM: H$_{2}$O & \cite{bremmell2001aDp} \\
S & 0.235  & 5.3 &0.36 & 3000 & 200 & PSL: PAAM: H$_{2}$O & \cite{bremmell2001aDp} \\
F & 0.125 & 3.57 & 0.12 & large & 65 & NaPAM: hema: H$_{2}$O & \cite{bremmell2001aDp} \\
S & 0.177 & 1.65 & 0.51 & large & 65 & NaPAM: hema: H$_{2}$O & \cite{bremmell2001aDp} \\ \hline
\end{tabular}
\end{ruledtabular}
\end{table*}


\begin{thebibliography}{66.}


\bibitem{turner1976aDp} D. N. Turner and F. R. Hallett. A study of the diffusion of compact particles in polymer solutions using quasi-elastic light scattering. {\em Biochimica et Biophysica Acta}, {\bf 451} (1976), 305-312.

\bibitem{weitz1993aDp} D. A. Weitz and D. W. Pine. Diffusing-wave spectroscopy. In \emph{Dynamic Light Scattering}, ed.\ W. Brown, (Oxford, UK: Oxford University Press, 1993), pp.\ 652-720.

\bibitem{saxton1997aDp} M. J. Saxton and K. Jacobson. Single particle tracking: applications to membrane dynamics. {\em Annu.\ Rev.\ Biophys.\ Biomol.\ Struct.}, {\bf 26} (1997), 373-399.


\bibitem{phillies1985bDp} G. D. J. Phillies, G. S. Ullmann, K. Ullmann and T.-H. Lin. Phenomenological scaling laws for 'semidilute' macromolecule solutions  light scattering by optical probe particles. {\em J.\ Chem.\ Phys.}, {\bf 82} (1985), 5242-5246.

\bibitem{phillies2001aDp} G. D. J. Phillies and K. A. Streletzky. Microrheology of complex fluids via observation of tracer microparticles. {\em Recent Res.\ Devel.\ Phys.\ Chem.}, {\bf 5} (2001), 269-285.

\bibitem{harden2001aDp} J. L. Harden and V. Viasnoff. Recent advances in DWS-based micro-rheology. {\em Curr.\ Op.\ Coll.\ Interf,\ Sci.}, {\bf 6} (2001), 438-445.

\bibitem{solomon2001aDp} M. J. Solomon and Q. Lu. Rheology and dynamics of particles in viscoelastic media. {\em Curr.\ Op.\ Coll.\ Interf.\ Sci.}, {\bf 6} (2001), 430-437.

\bibitem{mackintosh1999aDp} F. C. MacIntosh and C. F. Schmidt, Microrheology. {\em Curr.\ Op.\ Coll.\ Interf.\  Sci.}, {\bf 4} (1999), 300-307.

\bibitem{mukhopaday2001aDp} A. Mukhopaday and S. Granick. Micro- and nanorheology. {\em Curr.\ Op.\ Coll.\ Interf.\  Sci.}, {\bf 6} (2001), 423-429.

\bibitem{tseng2002aDp} Y. Tseng, T. P. Kole, S.-H. J. Lee and D. Wirtz. Local dynamics and viscoelastic properties of cell biological systems. {\em Curr.\ Op.\ Coll.\ Interf.\ Sci.}, {\bf 8} (2002), 210-217.


\bibitem{habdas2002aDp} P. Habdas and E. R. Weeks. Video microscopy of colloidal suspensions and colloidal crystals. {\em Curr.\ Op.\ Coll.\ Interf.\ Sci.}, {\bf 7} (2002), 196-203.

\bibitem{phillies2004xDp} George D. J. Phillies. Self and tracer diffusion of polymers in
solution, arXiv:cond-mat/0403109 (3 March 2004).

\bibitem{berne1976aDp} B. J. Berne and R. Pecora. \emph{Dynamic Light Scattering} (New York, NY: Wiley, 1976), especially Chapter 5.

\bibitem{doob1942aDp} J. L. Doob.  The brownian movement and stochastic equations. \emph{Annals Math.}, {\bf 43} (1942), 351-369.

\bibitem{phillies2005aDp} G. D. J. Phillies. Interpretation of light scattering spectra in terms of particle displacements, \emph{J. Chem.\ Phys.}, {\bf 122} (2005), 224905 1-8.

\bibitem{bremmell2001aDp} K. E. Bremmell, N. Wissenden and D. E. Dunstan. Diffusing probe measurements in newtonian and elastic solutions. {\em Adv.\ Coll.\ Interf.\ Sci.}, {\bf 89-90} (2001), 141-154.

\bibitem{bremmell2002aDp} K. E. Bremmell and D. E. Dunstan. Probe diffusion measurements of polystyrene latex particles in polyelectrolyte solutions of varying ionic strength. {\em Macromolecules}, {\bf 35}  (2002), 1994-1999.
    

\bibitem{brown1987bDp} W. Brown and R. Rymden. Interaction of carboxymethylcellulose with latex spheres studied by dynamic light scattering. {\em Macromolecules}, {\bf 20} (1987), 2867-2873.

\bibitem{brown1988cDp} W. Brown and R. Rymden. Comparison of the translational diffusion of large spheres and high molecular weight coils in polymer solutions. {\em Macromolecules}, {\bf 21} (1988), 840-846.


\bibitem{bu1994aDp} Z. Bu and P. S. Russo. Diffusion of dextran in aqueous hydroxypropylcellulose. {\em Macromolecules}, {\bf 27} (1994), 1187-1194.  

\bibitem{langevin1978aDp} D. Langevin and F. Rondelez. Sedimentation of large colloidal particles through semidilute polymer solutions. {\em Polymer}, {\bf 19} (1978), 875-882.


\bibitem{busch2000aDp} N. A. Busch, T. Kim and V. A. Bloomfield. Tracer diffusion of proteins in DNA solutions.  2. Green fluorescent protein in crowded DNA solutions. {\em Macromolecules}, {\bf 33}  (2000), 5932-5937.

\bibitem{cao1997aDp} X. Cao, R. Bansil, D. Gantz, E. W. Moore, N. Niu and N. H. Afdhal. Diffusion behavior of lipid vesicles in entangled polymer solutions. {\em Biophys.\ J.}, {\bf 73} (1997), 1932-1939.

\bibitem{cheng2002aDp} Y. Cheng, R. K. Prud'homme and J. L. Thomas. Diffusion of mesoscopic probes in aqueous polymer solutions measured by fluorescence recovery after photobleaching. {\em Macromolecules}, {\bf 35} (2002), 8111-8121.

\bibitem{desmedt1994aDp} S. C. De Smedt, A. Lauwers, J. Demeester, Y. Engelborghs, G. De Mey and M. Du. Structural information on hyaluronic acid solutions as studied by probe diffusion experiments. {\em Macromolecules}, {\bf 27} (1994), 141-146.

\bibitem{desmedt1993aDp} S. C. De Smedt, P. Dekeyser, V. Ribitsch, A. Lauwers and J. Demeester. Viscoelastic and transient network properties of hyaluronic acid as a function of the concentration. {\em Biorheology}, {\bf 30} (1994), 31-42.

\bibitem{desmedt1997aDp} S. C. De Smedt, T. K. L. Meyvis, J. Demeester, P. Van Oostveldt, J. C. G. Blonk and W. E. Hennink. Diffusion of macromolecules in dextran methacrylate solutions and gels as studied by confocal scanning laser microscopy. {\em Macromolecules}, {\bf 30} (1997), 4863-4870. 

\bibitem{delfino2005aDp} I. Delfino, C. Piccolo and M. Lepore. Experimental study of short- and long-time diffusion regimes of spherical particles in carboxymethylcellulose solutions. {\em Eur.\ Polym.\ J.}, {\bf 41} (2005), 1772-1780 

\bibitem{dunstan2000aDp} D. E. Dunstan and J. Stokes. Diffusing probe measurements in polystyrene latex particles in polyelectrolyte solutions: deviations from stokes-einstein behavior. {\em Macromolecules}, {\bf 33} (2000), 193-198.


\bibitem{furukawa1991aDp} R. Furukawa, J. L. Arauz-Lara and B. R. Ware. Self-diffusion and probe diffusion in dilute and semidilute solutions of dextran. {\em Macromolecules}, {\bf 24} (1991), 599-605. 

\bibitem{gold1996aDp} D. Gold, C. Onyenemezu and W. G. Miller. Effect of solvent quality on the diffusion of polystyrene latex spheres in solutions of poly(methylmethacrylate). {\em Macromolecules}, {\bf 29} (1996), 5700-5709.
  
\bibitem{gorti1985aDp} S. Gorti and B. R. Ware. Probe diffusion in an aqueous polyelectrolyte solution. {\em J. Chem.\ Phys.}, {\bf 83} (1985), 6449-6456 .

\bibitem{lin1978aDp} S. C. Lin, W. I. Lee and J. M. Schurr. {\em Biopolymers}, {\bf 17} (1978), 1041.

\bibitem{jamieson1982aDp} A. M. Jamieson, J. G. Southwick and J. Blackwell. Dynamical behavior of xanthan polysaccharide in solution. {\em J. Polymer Sci.: Polymer Phys.\ Ed.}, {\bf 20} (1982), 1513-1524.

\bibitem{johansson1991aDp} L. Johansson, U. Skantze and J.-E. Loefroth. Diffusion and interaction in gels and solutions. 2. Experimental results on the obstruction effect. {\em Macromolecules}, {\bf 24} (1991), 6019-6023. 


\bibitem{konak1982aDp} C. Konak, B. Sedlacek and Z. Tuzar. Diffusion of block copolymer micelles in solutions of a linear polymer. {\em Makromol.\ Chem., Rapid Commun.}, {\bf 3} (1982), 91-94.

\bibitem{lin1982aDp} T.-H. Lin and G. D. J. Phillies. Translational diffusion of a
macroparticulate probe species in salt-free poly(acrylic) acid:water.
{\em  J.\ Phys.\ Chem.}, {\bf 86} (1982), 4073-4077. 

\bibitem{lin1984bDp} T.-H. Lin and G. D. J. Phillies. Probe diffusion in poly(acrylic
acid): water.  Effect of probe size. {\em Macromolecules}, {\bf 17} (1984), 1686-1691.

\bibitem{lin1984aDp} T.-H. Lin and G. D. J. Phillies. Probe diffusion in polyacrylic
acid:water --- effect of polymer molecular weight. {\em J. Coll.\ Interf.\ Sci.},
{\bf 100} (1984), 82-95. 
    
\bibitem{lin1986aDp}  T.-H. Lin. Diffusion of TiO$_{2}$ particles through a poly(ethylene oxide) melt. {\em Makromol.\ Chem.}, {\bf 187} (1986), 1189-1196.

\bibitem{mangenot2003aDp} S. Mangenot, S. Keller and J. Raedler. Transport of nucleosome core particles in semidilute DNA solutions. {\em Biophys.\ J.}, {\bf 85} (2003), 1817-1825.  

\bibitem{nehme1989aDp} O. A. Nehme, P. Johnson and A. M. Donald. Probe diffusion in poly-l-lysine solution. {\em Macromolecules}, {\bf 22} (1989), 4326-4333. 

\bibitem{onyenemezu1993aDp} C. N. Onyenemezu, D. Gold, M. Roman, and W. G. Miller. Diffusion of polystyrene latex spheres in linear polystyrene nonaqueous solutions. {\em Macromolecules}, {\bf 26} (1993), 3833-3837. 


\bibitem{phillies1985aDp} G. D. J. Phillies. Diffusion of bovine serum albumin in a neutral polymer
solution. {\em Biopolymers}, {\bf 24} (1985), 379-386.

\bibitem{phillies1989bDp} G. D. J. Phillies, J. Gong, L. Li, A. Rau, K. Zhang,  L.-P. Yu and J.
Rollings. Macroparticle diffusion in dextran solutions. {\em J. Phys.\
Chem.}, {\bf 93} (1989), 6219-6223.
    

\bibitem{phillies1989aDp} G. D. J. Phillies, T. Pirnat, M. Kiss, N. Teasdale, D. Maclung, H. Inglefield, C.
Malone, L.-P. Yu and J. Rollings. Probe diffusion in solutions of low-molecular-weight polyelectrolytes. {\em Macromolecules}, {\bf 22} (1989), 4068-4075. 

\bibitem{phillies1987aDp} G. D. J. Phillies, C. Malone, K. Ullmann, G. S. Ullmann, J. Rollings and L.-P. Yu. Probe diffusion in solutions of long-chain polyelectrolytes. {\em Macromolecules}, {\bf 20} (1987), 2280-2289.

\bibitem{phillies1997bDp} G. D. J. Phillies, M. Lacroix and J. Yambert,  Probe diffusion
in sodium polystyrene sulfonate - water: experimental determination of sphere-chain binary hydrodynamic interactions. {\em Journal of Physical Chemistry}, {\bf 101} (1997), 5124-5130. 

\bibitem{phillies1993tDp}  G. D. J.  Phillies and P. C. Kirkitelos, Higher-order hydrodynamic interactions in the calculation of polymer transport properties, {\em Journal of Polymer Science B: Polymer Physics}, {\bf 31} (1993),
1785-1797.

\bibitem{phillies1991aDp} G. D. J. Phillies, A. Saleh,  L. Li, Y. Xu, D. Rostcheck, M. Cobb
and T. Tanaka. Temperature dependence of probe diffusion in solutions
of low-molecular-weight polyelectrolytes. {\em Macromolecules},   {\bf 24} (1991),
5299-5304.

\bibitem{phillies1992cDp} G. D. J. Phillies, D. Rostcheck and S. Ahmed, Probe diffusion in intermediate-molecular-weight polyelectrolytes: temperature dependence. {\em Macromolecules}, {\bf 25} (1992), 3689-3694.

\bibitem{phillies1992eDp} G. D. J. Phillies and C. A. Quinlan.  Glass temperature effects
effects in probe diffusion in dextran solutions. {\em Macromolecules}, {\bf
25} (1992), 3110-3316. 


\bibitem{roberts2001aDp} C. Roberts, T. Cosgrove, R. G. Schmidt and G. V. Gordon. Diffusion of poly(dimethylsiloxane) mixtures with silicate nanoparticles. {\em Macromolecules}, {\bf 34} (2001), 538-543.


\bibitem{shibayama1999aDp} M. Shibayama, Y. Isaka and I. Shiwa. Dynamics of probe particles in polymer solutions and gels. {\em Macromolecules}, {\bf 32} (1999), 7086-7092.

\bibitem{shiwa1987aDp} Y. Shiwa. Hydrodynamic Screening and Diffusion in Entangled Polymer Solutions. {\em Phys.\ Rev.\ Lett.}, {\bf 58} (1987), 2102-2105.

\bibitem{ullmann1983aDp} G. S. Ullmann and G. D. J. Phillies. Implications of the failure of the Stokes-Einstein relation for measurements with QELSS of polymer adsorption by small particles. {\em Macromolecules}, {\bf 16} (1983), 1947-1949.

\bibitem{ullmann1985aDp} G. S. Ullmann, K. Ullmann, R. M. Lindner and G. D. J. Phillies. Probe diffusion of polystyrene latex spheres in poly(ethylene oxide):water. {\em J.\ Phys.\ Chem.} {\bf 89} (1985), 692-700.

\bibitem{ullmann1985bDp} K. Ullmann, G. S. Ullmann and G. D. J. Phillies. Optical probe study of a nonentangling macromolecule solution---bovine serum albumin:water. {\em J.\ Coll.\ Interf.\ Sci.}, {\bf 105} (1985), 315-324.


\bibitem{keller1971aDp} K. M. Keller, E. R. Canales and S. I. Yum.  Tracer and mutual diffusion coefficients of proteins.  {\em J. Phys.\ Chem.}, {\bf 75} (1971), 379-387.

\bibitem{kitchen1976aDp} R. G. Kitchen, N. N. Preston and J. D. Wells, {\em J. Polym.\ Sci.}, {\bf 55} (1976), 39.


\bibitem{wattenbarger1992aDp} M. R. Wattenbarger, V. A. Bloomfield, B. Zu and P. S. Russo. Tracer diffusion of proteins in DNA solutions. {\em Macromolecules}, {\bf 25} (1992), 5263-5265.

\bibitem{won1994aDp} J. Won, C. Onyenemezu, W. G. Miller and T. P. Lodge. Diffusion of spheres in entangled polymer solutions: a return to Stokes-Einstein behavior. {\em Macromolecules}, {\bf 27} (1994), 7389-7396.


\bibitem{lodge1986aDp} T. P. Lodge and L. M. Wheeler. Translational diffusion of 
linear and 3-arm-star polystyrenes in semidilute solutions of linear 
poly(vinylmethylether). {\em Macromolecules}, {\bf 19} (1986), 2983-2986.  

\bibitem{lodge1987aDp} T. P. Lodge and P. Markland. Translational diffusion of 12-arm star 
polystyrenes in dilute and concentrated poly(vinyl methyl ether) solutions. 
{\em Polymer}, {\bf 28} (1987), 1377-1384.

\bibitem{lodge1989aDp} T. P. Lodge, P. Markland and L. M. Wheeler. Tracer 
diffusion of 3-arm and 12-arm star polystyrenes in dilute, semidilute, and 
concentrated poly(vinylmethylether) solutions. {\em Macromolecules}, {\bf 22} (1989), 
3409-3418.

\bibitem{wheeler1987aDp} L. M. Wheeler, T. P. Lodge, B. Hanley and M. Tirrell.
Translational diffusion of linear polystyrenes in dilute and semidilute 
solutions of poly(vinylmethylether). {\em Macromolecules}, {\bf 20} (1987), 
1120-1129.  

\bibitem{wheeler1989aDp} L. M. Wheeler and T. P. Lodge. Tracer diffusion of 
linear polystyrenes in dilute, semidilute, and concentrated poly(vinylmethylether) solutions.  {\em Macromolecules}, {\bf 22} (1989), 3399-3408.

\bibitem{won1993aDp} J. Won and T. P. Lodge. Tracer diffusion of star-branched polystyrenes in poly(vinylmethylether) gels. {\em J. Polym.\ Sci.\ Polym.\ Phys.\ Ed.}, {\bf 31} (1993), 1897-1907. 

\bibitem{zhou1989aDp} Zhou Pu and W. Brown. Translational diffusion of large silica spheres in semidilute polymer solutions. {\em Macromolecules}, {\bf 22} (1989), 890-896. 

\bibitem{phillies1983aDp} A. C. Fernandez and G. D. J. Phillies. Temperature dependence of the
diffusion coefficient of polystyrene latex spheres. {\em Biopolymers}, {\bf 22} (1983), 593-595. 

\bibitem{phillies1981aDp} G. D. J. Phillies. Translational diffusion coefficient of macroparticles
in solvents of high viscosity. {\em J.\ Phys.\ Chem.}, {\bf 85} (1981), 2838-2843.

   
\bibitem{wiltzius1991aDp} P. Wiltzius and W. van Saarloos. Absence of increase in length scale upon approaching the glass transition in liquid glycerol. {\em J. Chem.\ Phys.}, {\bf 94} (1991), 5061-5063.

\bibitem{kiyachenko1985a} Cited by Ref. \onlinecite{wiltzius1991aDp} as Yu. F. Kiyachenko and Yu. I. Litvinov, {\em JETP Lett.}, {\bf 42} (1985), 257.

\bibitem{phillies1992dDp} G. D. J. Phillies and D. Clomenil. Lineshape and linewidth
effects in optical probe studies of glass-forming liquids. {\em J.\ Phys.\
Chem.}, {\bf 96} (1992), 4196-4200. 

\bibitem{phillies1995aDp} G. D. J. Phillies and C. A. Quinlan. Analytic structure of the
solutionlike-meltlike transition in polymer solution dynamics. {\em Macromolecules}, {\bf 28} (1995), 160-164.

\bibitem{phillies1992rDp} G. D. J. Phillies. Range of validity of the hydrodynamic scaling
model. {\em J. Phys.\ Chem.}, {\bf 96} (1992), 10061-10066. 

\bibitem{phillies1988aDp} G. D. J. Phillies. Quantitative prediction of $\alpha$ in the scaling law for self-diffusion. \emph{Macromolecules}, {\bf 21} (1988), 3101-3106.

\bibitem{brown1986aDp} W. Brown and R. Rymden. Diffusion of polystyrene latex spheres in polymer solutions studied by dynamic light scattering. {\em Macromolecules}, {\bf 19} (1986), 2942-2952.    

\bibitem{yang1988aDp} T. Yang  and A. M. Jamieson. Diffusion of latex spheres through solutions of hydroxypropylcellose in water. {\em J. Coll.\ Interf.\ Sci.}, {\bf 126} (1988), 220-230. 

\bibitem{russo1988bDp} P. S. Russo, M. Mustafa, T. Cao and L. K. Stephens. Interactions between polystyrene latex spheres and a semiflexible polymer, hydroxypropylcellulose. {\em J. Coll.\ Interf.\ Sci.}, {\bf 122} (1988), 120-137.

\bibitem{mustafa1989aDp} M. Mustafa and P. S. Russo. Nature and effects of nonexponential correlation functions in probe diffusion experiments by quasielastic light scattering. {\em J. Coll.\ Interf.\ Sci.}, {\bf 129} (1989), 240-253. 

\bibitem{phillies1993bDp} G. D. J. Phillies and D. Clomenil. Probe diffusion in polymer solutions under $\theta$ and good conditions. {\em Macromolecules}, {\bf 26} (1993), 167-170. 

\bibitem{altenberger1986aDp} A. R. Altenberger, M. Tirrell, and J. S. Dahler, {\em J. Chem.\ Phys.}, {\bf 84} (1986), 5122.

\bibitem{phillies1987dDp} G. D. J. Phillies. Dynamics of polymers in concentrated solution: the
universal scaling equation derived. \emph{Macromolecules}, {\bf 20} (1987), 558-564. 

\bibitem{phillies1988cDp} G. D. J. Phillies and P. Peczak. The ubiquity of stretched-exponential
forms in polymer dynamics. {\em Macromolecules}, {\bf 21} (1988), 214-220.

\bibitem{phillies1993aDp} G. D. J. Phillies, C. Richardson, C. A. Quinlan and S. Z. Ren. Transport in intermediate and high molecular weight hydroxypropylcellulose/water solutions. {\em Macromolecules}, {\bf 26} (1993), 6849-6858.


\bibitem{mustafa1993aDp} M. B. Mustafa, D. L. Tipton, M. D. Barkley, P. S. Russo. and F. D. Blum. Dye diffusion in isotropic and liquid crystalline aqueous (hydroxypropyl)cellulose. {\em Macromolecules}, {\bf 26} (1993), 370-378.

\bibitem{ngai1996aDp} K. L. Ngai and G. D. J. Phillies. Coupling model analysis of polymer dynamics in solution: probe diffusion and viscosity. {\em J. Chem.\ Phys.}, {\bf 105} (1996), 8385-8397.

\bibitem{ngai1994aDp} K. L. Ngai. In \emph{Disorder Effects In Relaxation Processes} ed.\ R. Richert and A. Blumen, (Berlin, Germany: Springer-Verlag (1994)).

\bibitem{phillies1997aDp} G. D. J. Phillies and M. Lacroix. Probe diffusion in hydroxypropylcellulose--water: Radius and line-shape effects in the solutionslike regime. {\em J. Phys.\ Chem.\ B}, {\bf 101} (1997), 39-47.

\bibitem{streletzky1998aDp} K. A. Streletzky and G. D. J. Phillies. Translational diffusion of small and large mesoscopic probes in hydroxypropylcellulose-water in the solutionlike regime. {\em J. Chem.\ Phys.}, {\bf 108} (1998), 2975-2988.

\bibitem{streletzky1998bDp} K. A. Streletzky and G. D. J. Phillies. Relaxational mode structure for optical probe diffusion in high molecular weight hydroxypropylcellulose. {\em J. Polym.\ Sci.\ B}, {\bf 36} (1998), 3087-3100.

\bibitem{streletzky1999aDp} K. A. Streletzky and G. D. J. Phillies. Confirmation of the reality of the viscoelastic solutionlike-meltlike transition via optical probe diffusion. {\em Macromolecules}, {\bf 32} (1999), 145-152.

\bibitem{streletzky1999cDp} K. A. Streletzky and G. D. J. Phillies. Coupling analysis of probe diffusion in high molecular weight hydroxypropylcellulose. {\em J. Phys.\ Chem.\ B}, {\bf 103} (1999), 1811-1820.

\bibitem{streletzky2000aDp} K. Streletzky and G. D. J. Phillies. Optical probe study of solution-like and melt-like solutions of high molecular weight hydroxypropylcellulose. in {\em Scattering from Polymers}, ed.\ B. S. Hsiao,
(Washington, D.C.: Am.\ Chem.\ Soc.\ Symp.\ Ser. 2000) Vol.\ 739, 297-316.

\bibitem{phillies2003aDp} G. D. J. Phillies, R. O'Connell, P. Whitford and K. A. Streletzky. Mode structure of diffusive transport in hydroxypropylcellulose:water. {\em J. Chem.\ Phys.}, {\bf 119} (2003), 9903-9913.

\bibitem{oconnell2005aDp} R. O'Connell, H. Hanson, and G. D. J. Phillies.  Neutral Polymer Slow Mode and Its Rheological Correlate.  {\em  J. Polym.\ Sci.\ B. Polym.\ Phys.}, {\bf 43} (2005), 323-333. 

\bibitem{kivelson1994aDp} S. A. Kivelson, X. Zhao, D. Kivelson, T. M. Fischer and C. M. Knobler. Frustration-limited clusters in liquids. {\em J. Chem.\ Phys.}, {\bf 101} (1994), 2391-2397.

\bibitem{camins1994aDp} B. Camins and P.S. Russo. Following polymer gelation by depolarized dynamic light scattering from optically and geometrically anisotropic latex particles. {\em Langmuir}, {\bf 10} (1994), 4053-4059.

\bibitem{cheng2003aDp} Z. Cheng and T. G. Mason. Rotational diffusion microrheology. {\em Phys.\ Rev.\ Lett.}, {\bf 90} (2003), 018304 1-4.

\bibitem{cush2004aDp} R. Cush, D. Dorman and P. S. Russo. Rotational and translational diffusion of tobacco mosaic virus in extended and globular polymer solutions. {\em Macromolecules}, {\bf 37} (2004), 9577-9584.

\bibitem{cush1997aDp} R. Cush, P. S. Russo, Z. Kucukyavuz, Z. Bu, D. Neau, D. Shih, S. Kucukyavuz and H. Ricks. Rotational and translational diffusion of a rodlike virus in random coil polymer solutions. {\em Macromolecules}, {\bf 30} (1997), 4920-4926.


\bibitem{hill1989aDp} D. A. Hill and D. S. Soane. Measurement of rotational diffusivity of rodlike molecules in amorphous polymer matrices by the dynamic kerr effect. {\em J. Polymer Science B}, {\bf 27} (1989), 2295-2320.

\bibitem{jamil1998aDp} T. Jamil and P. A. Russo. Interactions between colloidal poly(tetrafluoroethylene) latex and sodium poly(styrenesulfonate). {\em Langmuir}, {\bf 14} (1998), 264-270.

\bibitem{koenderink2004aDp} G. H Koenderink, S. Sacanna, D. G. A. L. Aarts and A. P. Philipse. Rotational and translational diffusion of fluorocarbon tracer spheres in semidilute xanthan solutions. {\em Phys.\ Rev.\ E}, {\bf 69} (2004), 021804 1-12. 

\bibitem{legoff2002aDp} L. Le Goff, O. Hallatschek, E. Frey and F. Amblard. Tracer studies on F-actin fluctuations. {\em Phys.\ Rev.\ Lett.}, {\bf 89} (2002), 258101 1-4.

\bibitem{phalak2000aDp} J. G. Phalakornkul, A. P. Gast and R. Pecora. Rotational dynamics of rodlike polymers in a rod/sphere mixture. {\em J. Chem.\ Phys.}, {\bf 112} (2000), 6487-6494. 

\bibitem{sohn1996aDp} D. Sohn, P. S. Russo, A. Davila, D. S. Poche and M. L. McLaughlin. Light scattering study of magnetic latex particles and their interaction with polyelectrolytes. {\em J. Coll. Interf.\ Sci.}, {\bf 177} (1996), 31-44.

\bibitem{apgar2000aDp} J. Apgar, Y. Tseng, E. Federov, M. B. Herwig, S. C. Almo and D. Wirtz. Multiple-particle tracking measurements of heterogeneities in solutions of actin filaments and actin bundles. {\em Biophys.\ J.}, {\bf 79} (2000), 1095-1106.

\bibitem{chen2003aDp} D. T. Chen, E. R. Weeks, J. C. Crocker, M. F. Islam, R. Verna, J. Gruber, A. J. Levine, T. C. Lubensky and A. G. Yodh. Rheological microscopy: local mechanical properties from microrheology. {\em Phys.\ Rev.\ Lett.}, {\bf 90} (2003), 108301 1-4.

\bibitem{levine2001aDp} A. J. Levine and T. C. Lubensky. Two-point microrheology and the electrostatic analogy. {\em Phys.\ Rev.\ E}, {\bf 65} (2001), 011501 1-13.

\bibitem{crocker2000aDp} J. C. Crocker, M. T. Valentine, E. R. Weeks, T. Gisler, P. D. Kaplan, A. G. Yodh and D. A. Weitz. Two-point microrheology of inhomogeneous soft materials. {\em Phys.\ Rev.\ Lett.}, {\bf 85} (2000), 888-891.

\bibitem{dichtl1999aDp} M. A. Dichtl and E. Sackmann. Colloidal probe study of short time local and long time reptational motion of semiflexible macromolecules in entangled networks. {\em New J. Physics}, {\bf 1} (1999), 18.1-18.11.

\bibitem{gardel2003aDp} M. L. Gardel, M. T. Valentine, J. C. Crocker, A. R. Bausch and D. A. Weitz. Microrheology of entangled F-actin solutions. {\em Phys.\ Rev.\ Lett.}, {\bf 91} (2003), 158302 1-4.

\bibitem{gittes1997aDp} F. Gittes, B. Schnurr, P. D. Olmstead, F. C. MacKintosh and C. F. Schmidt. Microscopic viscoelasticity: shear moduli of soft materials determined from thermal fluctuations. {\em Phys.\ Rev.\ Lett.}, {\bf 79} (1997), 3286-3289.



\bibitem{goodman2002aDp} A. Goodman, Y. Tseng and D. Wirtz. Effect of length, topology, and concentration on the microviscosity and microheterogeneity of DNA solutions. {\em J. Mol.\ Bio.}, {\bf 323} (2002), 199-215.

\bibitem{lau2003aDp} A. W. C. Lau, B. D. Hoffman, A. Davies, J. C. Crocker and T. C. Lubensky. Microrheology, stress fluctuations, and active behavior of living cells. {\em Phys.\ Rev.\ Lett.}, {\bf 91} (2003), 198101 1-4.

\bibitem{mason1997aDp} T. G. Mason, K. Ganesan, J. H. van Zanten, D. Wirtz and S. C. Kuo. Particle tracking microrheology of complex fluids. {\em Phys.\ Rev.\ Lett.}, {\bf 79} (1997), 3282-3285.

\bibitem{papag2005aDp}  A. Papagiannopolis, C. M. Ferneyhough and T. A. Waigh. The microrheology of polystyrene sulfonate combs in aqueous solution. {\em J. Chem.\ Phys.}, {\bf 123} (2005), 214904 1-10.

\bibitem{schnurr1997aDp} B. Schnurr, F. Gittes, F. C. MacKintosh, and C. F. Schmidt. Determining microscopic viscoelasticity in flexible and semiflexible polymer networks from thermal fluctuations. {\em Macromolecules}, {\bf 30} (1997), 7781-7792. 

\bibitem{tseng2001aDp} Y. Tseng and D. Wirtz. Mechanics and multiple-particle tracking microheterogeneity of $\alpha$-actinin-cross-linked actin filament networks. {\em Biophys.\ J.}, {\bf 81} (2001), 1643-1656. 

\bibitem{valentine2004aDp} M. T. Valentine, Z. E. Perlman, M. L. Gardel, J. H. Shin, P. Matsudaira, T. J. Mitchison and D. A. Weitz. Colloid surface chemistry critically affects multiple particle tracking measurements of biomaterials. {\em Biophys.\ J.}, {\bf 86} (2004), 4004-4014.

\bibitem{xu1998aDp} J. Xu, A. Palmer and D. Wirtz. Rheology and microrheology of semiflexible polymer solutions: actin filament networks. {\em Macromolecules}, {\bf 31} (1998), 6486-6492.

\bibitem{xu1998bDp} J. Xu, V. Viasnoff and D. Wirtz. Compliance of actin network filaments measured by particle-tracking microrheology and diffusing wave spectroscopy. {\em Rheol.\ Acta}, {\bf 37} (1998), 387-398.

\bibitem{xu2002aDp} J. Xu, Y. Tseng, C. J. Carriere and D. Wirtz. Microheterogeneity and microrheology of wheat gliadin suspensions studied by multiple-particle tracking. {\em Biomacromolecules}, {\bf 3} (2002), 92-99.

\bibitem{amblard1996aDp} F. Amblard, A. C. Maggs, B. Yurke, A. N. Pargellis and S. Leibler. Subdiffusion and anomalous local viscoelasticity in actin networks. {\em Phys.\ Rev.\ Lett.}, {\bf 77} (1996), 4470-4473.



\bibitem{bishop2004aDp} A. I. Bishop, T. A. Nieminen, N. R. Heckenberg and H. Rubinsztein-Dunlop. Optical microrheology using rotating laser-trapped particles. {\em Phys.\ Rev.\ Lett.}, {\bf 92} (2004), 198104 1-4.

\bibitem{hough1999aDp} L. A. Hough and H. D. Ou-Yang. A new probe for mechanical testing of nanostructures in soft materials. {\em J. Nanoparticle Research} {\bf 1} (1999), 495-499.

\bibitem{keller2001aDp} M. Keller, J. Schilling and E. Sackmann. Oscillatory magnetic bead rheometer for complex fluid microrheometry. {\em Rev.\ Sci.\ Instr.}, {\bf 72} (2001), 3626-3634 .

\bibitem{schmidt2000aDp} F. G. Schmidt, B. Hinner, and E. Sackmann. Microrheometry underestimates the values of the viscoelastic moduli in measurements on F-actin solutions compared to macrorheometry. {Phys.\ Rev.\ E}, {\bf 61} (2000), 5646-5653.

\bibitem{maggs1998aDp} A. C. Maggs, Micro-bead mechanics with actin filaments. {\em Phys.\ Rev.\ E}, {\bf 57} (1998), 2091-2094. 

\bibitem{schmidt2000bDp} F. G. Schmidt, B. Hinner, E. Sackmann and J. X. Tang. Viscoelastic properties of semiflexible filamentous bacteriophage fd.  {\em Phys.\ Rev.\ E}, {\bf 62} (2000), 5509-5517.


\bibitem{morse1998aDp} D. Morse. Viscoelasticity of concentrated isotropic solutions of semiflexible polymers. 2. Linear response. {\em Macromolecules}, {\bf 31} (1998), 7044-7067.

\bibitem{allain1986aDp} C. Allain, M. Drifford and B. Gauthier-Manuel. Diffusion of calibrated particles during the formation of a gel. {\em Polymer Communications}, {\bf 27} (1986), 177-180.

\bibitem{matsukawa1996aDp} S. Matsukawa and I. Ando. A study of self-diffusion of molecules in polymer gel by pulsed-field-gradient $^{1}$H NMR. {\em Macromolecules}, {\bf 29} (1996), 7136-7140.

\bibitem{nishio1987aDp} I. Nishio, J. C. Reina and R. Bansil. Quasielastic light scattering study of the movement of particles in gels. {\em Phys.\ Rev.\ Lett.}, {\bf 59} (1987), 684-687.

\bibitem{park1990aDp} I. H. Park, C. S. Johnson, Jr., and D. A. Gabriel. Probe diffusion in polyacrylamide gels as observed by means of holographic relaxation methods: search for a universal equation. {\em Macromolecules}, {\bf 23} (1990), 1548-1553.

\bibitem{reina1990aDp}  J. C. Reina, R. Bansil, and C. Konak. Dynamics of probe particles in polymer solutions and gels. {\em Polymer}, {\bf 31} (1990), 1038-1044.

\bibitem{suzuki1992aDp} Y. Suzuki and I. Nishio. Quasielastic-light-scattering study of the movement of particles in gels: Topological structure of pores in gels. {\em Phys.\ Rev.\ B}, {\bf 45} (1992), 4614-4619.

\bibitem{fadda2001aDp} G. C. Fadda, D. Lairez and J. Pelta. Critical behavior of gelation probed by the dynamics of latex spheres. {\em Phys.\ Rev.\ E}, {\bf 63} (2001), 061405 1-9.

\bibitem{hou1990aDp} L. Hou, F. Lanni and K. Luby-Phelps. Tracer diffusion in F-actin and ficoll mixtures. Toward a model for cytoplasm. {\em Biophys.\ J.}, {\bf 58} (1990), 31-43. 


\bibitem{lubyphelps1986aDp} K. Luby-Phelps, D. L. Taylor and F. Lanni. Probing the structure of cytoplasm. {\em J. Cell.\ Biol.}, {\bf 102} (1986), 2015-2022.

\bibitem{lubyphelps1987aDp} K. Luby-Phelps, P. E. Castle,  D. L. Taylor and F. Lanni. Hindered diffusion of inert tracer particles in the cytoplasm of mouse 3T3 cells. {\em Proc.\ Natl.\ Acad.\ Sci.}, {\bf 84} (1987), 4910-4913.

\bibitem{madonia1983aDp} F. Madonia, P. L. San Biagio, M. U. Palma, G. Schiliro', S. Musumeci and G. Russo. Photon scattering as a probe of microviscosity and channel size in gels such as sickle haemoglobin. {\em Nature}, {\bf 302} (1983), 412-415.

\bibitem{newman1989aDp} J. Newman, N. Mroczka and K. L. Schick. Dynamic light scattering measurements of the diffusion of probes in filamentatious actin solutions. {\em Biopolymers}, {\bf 28} (1989), 655-666.

\bibitem{newman1991aDp} J. Newman, G. Gukelberger, K. L. Schick and K. S. Zaner. Probe diffusion in solutions of filamentatious actin formed in the presence of gelsolin. {\em Biopolymers}, {\bf 31} (1991), 1265-1271.

\bibitem{schmidt1989aDp} C. F. Schmidt, M. Baermann, G. Isenberg, and E. Sackmann. Chain dynamics, mesh size, and diffusive transport in networks of polymerized actin.  A quasielastic light scattering and microfluorescence study. {\em Macromolecules}, {\bf 22} (1989), 3638-3649. 

\bibitem{stewart1988aDp} U. A. Stewart, M. S. Bradley, C. S. Johnson, Jr., and D. A. Gabriel. Transport of probe molecules through fibrin gels as observed by means of holographic relaxation methods. {\em Biopolymers}, {\bf 27} (1988), 173-185.

\bibitem {wong2004aDp} I. M. Wong, M. L. Gardel, D. R. Reichman, E. R. Weeks, M. T. Valentine, A. R. Bausch and D. A. Weitz. Anomalous diffusion probes microstructure dynamics of entangled F-actin networks. {\em Phys.\ Rev.\ Lett.}, {\bf 92} (2004), 178101 1-4.

\bibitem{arriodupont2000aDp} M. Arrio-Dupont, G. Foucault, M. Vacher, P. F. Devaux and S. Cribier. Translational diffusion of globular proteins in the cytoplasm of cultured muscle cells. {\em Biophys.\ J.}, {\bf 78} (2000), 901-907.

\bibitem{arriodupont1996aDp} M. Arrio-Dupont, S. Cribier, J. Foucault, P. F. Devaux and A. D'Albis. Diffusion of fluorescently labelled macrolecules in cultured muscle cells. {\em Biophys.\ J.}, {\bf 70} (1996), 2327-2332.


\bibitem{kao1993aDp} H. P. Kao, J. R. Abney and A. S. Verkman. Determinants of the translational mobility of a small solute in cell cytoplasm. {\em J. Cell Biol.}, {\bf 120} (1993), 175-184.

\bibitem{lubyphelps1993bDp} K. Luby-Phelps, S. Mujumdar, R. B. Mujumdar, L. A. Ernst, W. Galbraith and A. S. Waggoner. A novel fluorescence ratiometric method confirms the low solvent viscosity of the cytoplasm. {\em Biophys.\ J.}, {\bf 65} (1993), 236-242.

\bibitem{seksek1997aDp} O. Seksek, J. Biwersi, and A. S. Verkman. Translational diffusion of macromolecule-sized solutes in cytoplasm and nucleus. {\em J. Cell Biology}, {\bf 138} (1997), 131-142. 

\bibitem{shenoy1995aDp} V. Shenoy and J. Rosenblatt. Diffusion of macromolecules in collagen and hyaluronic acid, rigid-rod--flexible polymer, composite matrices. {\em Macromolecules}, {\bf 28} (1995), 8751-8758.

\bibitem{bellour2002aDp} M. Bellour, M. Skouri, J.-P. Munch and P. Hebraud. Brownian motion of particles embedded in a solution of giant micelles. {\em Eur.\ Phys.\ J. E}, {\bf 8} (2002), 431-436.

\bibitem{dasgupta2002aDp} B. R. Dasgupta, S.-Y. Tee, J. C. Crocker, B. J. Frisken and D. A. Weitz. Microrheology of polyethylene oxide using diffusing wave spectroscopy and single scattering. {\em Phys.\ Rev.\ E}, {\bf 65} (2002), 051505 1-10.

\bibitem{gisler1999aDp} T. Gisler and D. A. Weitz. Scaling of the microrheology of semidilute F-actin solutions. {\em Phys.\ Rev.\ Lett.}, {\bf 82} (1999), 1606-1609.


\bibitem{heinemann2004aDp} C. Heinemann, F. Cardinaux, F. Scheffold, P. Shurtenberger, F. Escher and B. Conde-Petit. Tracer Microrheology of $\gamma$-dodecalactone induced gelation of aqueous starch dispersions. {\em Carbohydrate Polymers}, {\bf 55} (2004), 155-161.

\bibitem{kao1993bDp} M. H. Kao, A. G. Yodh and D. J. Pine. Observation of brownian motion on the time scale of hydrodynamic interactions. {\em Phys.\ Rev.\ Lett.}, {\bf 70} (1993), 242-245.


\bibitem{kaplan1993aDp} P. D. Kaplan, A. G. Yodh and D. F. Townsend. Noninvasive study of gel formation in polymer-stabilized dense colloids using multiply scattered light. {\em J. Coll.\ Interf.\ Sci.}, {\bf 155} (1993), 319-324.

\bibitem{knaebel2002aDp}  A. Knaebel, R. Skouri, J. P. Munch and S. J. Candau. Structural and rheological properties of hydrophobically modified alkali-soluble emulsion solutions. {\em J. Polymer Sci.\ B}, {\bf 40} (2002), 1985-1994.

\bibitem{lu2002aDp} Q. Lu and M. J. Solomon. Probe size effects on the microrheology of associating polymer solutions. {\em Phys.\ Rev.\ E}, {\bf 66} (2002), 061504 1-11.

\bibitem{mason1995aDp} T. G. Mason and D. A. Weitz. Optical measurements of frequency-dependent linear viscoelastic moduli of complex fluids. {\em Phys.\ Rev.\ Lett.}, {\bf 74} (1995), 1250-1253.

\bibitem{narita2001aDp} T. Narita, A. Knaebel, J.-P. Munch and S. J. Candau. Microrheology of poly(vinyl alcohol) aqueous solutions and chemically cross-linked gels. {\em Macromolecules}, {\bf 34} (2001), 8224-8231.

\bibitem{nisato2000aDp} G. Nisato, P. Hebraud, J.-P. Munch and S. J. Candau. Diffusing-wave-spectroscopy investigation of latex particle motion in polymer gels. {\em Phys.\ Rev.\ E}, {\bf 61} (2000), 2879-2887.

\bibitem{palmer1998aDp} A. Palmer, J. Xu and D. Wirtz. High-Frequency Viscoelasticity of crosslinked actin filament networks measured by diffusing wave spectroscopy. {\em Rheol.\ Acta}, {\bf 37} (1998), 97-106.

\bibitem{palmer1999aDp} A. Palmer, T. G. Mason, J. Xu, S. C. Kuo and D. Wirtz. Diffusing wave spectroscopy microrheology of actin filament networks. {\em Biophys.\ J.}, {\bf 76} (1999), 1063-1071.

\bibitem{pine1988aDp} D. J. Pine, D. A. Weitz, P. M. Chaikin and E. Herbolzheimer. Diffusing-wave spectroscopy. {\em Phys.\ Rev.\ Lett.}, {\bf 60} (1988), 1134-1137. 

\bibitem{maret1987aDp} G. Maret and P. E. Wolf. Multiple light-scattering from disordered media. The effect of brownian motion of scatterers. {\em Z. Phys.\ B},  {\bf 65} (1987), 409-413. 

\bibitem{popescu2002aDp} G. Popescu, A. Dogariu and R. Rajagopalan. Spatially resolved microrheology using localized coherence volumes. {\em Phys.\ Rev.\ E}, {\bf 65} (2002), 041504 1-8.

\bibitem{popescu2001aDp} G. Popescu and A. Dogariu. Dynamic light scattering in localized coherence volumes. {\em 
Optics Letters}, {\bf 26} (2001), 551-553.

\bibitem{sohn2004aDp} I. S. Sohn, R. Rajagopalan and A. C. Dogariu. Spatially resolved microrheology through a liquid/liquid interface. {\em J. Coll.\ Interf.\ Sci.}, {\bf 269} (2004), 503-513.

\bibitem{rojas2002aDp} L. F. Rojas-Ochoa, S. Romer, F. Scheffold and P. Schurtenberger. Diffusing wave spectroscopy and small-angle neutron scattering from concentrated colloidal suspensions. {\em Phys.\ Rev.\ E}, {\bf 65} (2002), 051403 1-8. 

\bibitem{romer2000aDp} S. Romer, F. Scheffold and P. Schurtenberger. Sol-gel transition of concentrated colloidal suspensions. {\em Phys.\ Rev.\ Lett.}, {\bf 85} (2000), 4980-4983.

\bibitem{romer2001aDp} S. Romer, C. Urban, H. Bissig, A. Stradner, F. Scheffold and P. Schurtenberger. Dynamics of concentrated colloidal suspensions: diffusion, aggregation, and gelation. {\em Phil.\ Trans.\ R. Soc.\ London A}, {\bf 359} (2001), 977-984.

\bibitem{rufener1999aDp} K. Rufener, A. Palmer, J. Xu and D. Wirtz. High-frequency dynamics and microrheology of macromolecular solutions probed by diffusing wave spectroscopy: the case of concentrated solutions of F-actin. {\em J. Non-Newtonian Fluid Mech.}, {\bf 82} (1999), 303-314. 

\bibitem{vandergucht2003aDp} J. van der Gucht, N. A. M. Besseling, W. Knoben, L. Bouteiller and M. A. Cohen Stuart. Brownian particles in supramolecular polymer solutions. {\em Phys.\ Rev.\ E}, {\bf 67} (2003), 051106 1-10.

\bibitem{vanzanten2004aDp} J. H. van Zanten, S. Amin and A. A. Abdala. Brownian motion of colloidal spheres in aqueous PEO solutions. {\em Macromolecules}, {\bf 37} (2004), 3874-3880.


\bibitem{degennes1978xDp} P. G. deGennes, P. Pincus, and R. M. Velasco.  Cited by Ref.\ \onlinecite{langevin1978aDp} as 'Personal Communication'. (1978)

\bibitem{degennes1979aDp}  P. G. deGennes.  Scaling Concepts in Polymer Physics. Cornell University Press, Ithaca (1979).

\bibitem{phillies1992aDp} G. D. J. Phillies, W. Brown and P. Zhou. Chain and sphere diffusion in polyisobutylene-CHCl$_{3}$: a reanalysis.  {\em Macromolecules}, {\bf 25} (1992), 4948-4954.


\end{thebibliography}
\end{document}